\documentclass[aps,twocolumn,groupedaddress]{revtex4}
\usepackage{epsfig}
\usepackage{graphicx}
\usepackage[T1]{fontenc}
\usepackage{ae}
\usepackage{color}
\usepackage[latin1]{inputenc}
\usepackage{amssymb,amsbsy,amsmath}
\usepackage{bbm}

\newtheorem{lemma}{Lemma}
\newtheorem{defi}{Definition}
\newtheorem{prop}{Proposition}

\newtheorem{theorem}{Theorem}
\newtheorem{cor}{Corollary}
\newtheorem{ex}{Example}

\begin{document}



\title[Theory of ground states IV]
{Theory of ground states for classical Heisenberg spin
systems IV}

\author{Heinz-J\"urgen Schmidt$^1$
\footnote[1]{Correspondence should be addressed to hschmidt@uos.de}
}
\address{$^1$Universit\"at Osnabr\"uck, Fachbereich Physik,
Barbarastr. 7, D - 49069 Osnabr\"uck, Germany}


\begin{abstract}
We extend the theory of ground states of classical Heisenberg spin systems previously published to the case where the interaction
with an external magnetic field is described by a Zeeman term. The ground state problem for the Heisenberg-Zeeman Hamiltonian can be reduced
first to the relative ground state problem, and, in a second step, to the absolute ground state problem for pure Heisenberg Hamiltonians
depending on an additional Lagrange parameter. We distinguish between continuous and discontinuous reduction.
Moreover, there are various general statements about Heisenberg-Zeeman systems that will be proven under most general assumptions.
One topic is the connection between the minimal energy functions $E_{min}$ for the Heisenberg energy and $H_{min}$ for the
Heisenberg-Zeeman energy which turn out to be essentially mutual Legendre-Fenchel transforms.
This generalization of the traditional Legendre transform is especially suited to cope with situations where the function $E_{min}$ is not convex
and consequently there is a magnetization jump at a critical field. Another topic is magnetization and the
occurrence of threshold fields $B_{thr}$ and saturation fields $B_{sat}$, where we provide a general formula for the latter.
We suggest a distinction between ferromagnetic and anti-ferromagnetic systems based on the vanishing of $B_{sat}$ for the former ones.
Parabolic systems are defined in such a way that $E_{min}$ and $H_{min}$ have a particularly simple form and studied in detail. For a large class
of parabolic systems the relative ground states can be constructed from the absolute ground state by means of a so-called umbrella family.
Finally we provide a counter-example of a parabolic system where this construction is not possible.
\end{abstract}

\maketitle

\section{Introduction}\label{sec:I}
This is the forth of a series of papers devoted to the theory of ground states of finite classical Heisenberg
spin systems. The general motivation of such a theory can be found in \cite{S17a} and need
not be repeated here. Extended examples are contained in \cite{S17b}, whereas \cite{S17c} is devoted to the study of the Gram
set and the case of $N=3$ spins. The case of a Heisenberg Hamiltonian plus a Zeeman term, henceforward called `` Heisenberg-Zeeman system",
is not yet covered by \cite{S17a} -- \cite{S17c}, but it can be reduced to this theory by the following considerations. First, in section
\ref{sec:HZR}, we draw upon the well-known fact that the ground states of the Heisenberg-Zeeman system are among the relative
ground states of the pure Heisenberg system. ``Relative ground states" means states minimizing the energy under the additional constraint
of fixed magnetization or, more appropriate for our purposes, total spin length $S$.
Ground states without this additional constraints will also be called ``absolute ground states".
Second, in section \ref{sec:RH} we utilize the circumstance that the
square $S^2$ of the total spin length has essentially the form of a Heisenberg Hamiltonian. Hence the relative ground state problem
can be reduced to the absolute ground state problem of some modified Heisenberg Hamiltonian $H_\gamma$ where $\gamma$ is the
Lagrange parameter of the minimization problem with the additional constraint $S^2=\mbox{ const.}$.

Besides following this reduction program we find it in order to state and prove some general facts about Heisenberg-Zeeman systems in sections
\ref{sec:HZ} and \ref{sec:HZR}. These facts are not completely new; some of them belong to folk wisdom and
some are scattered to various places in the literature. It is a secondary aim of this paper to summarize these facts and to provide proofs under
conditions as general as possible.
One topic is the connection between the minimal Heisenberg energy $E_{min}(\mu)$ of relative ground states with magnetization $\mu$ and the minimal Heisenberg-Zeeman energy $H_{min}(b)$ depending on the magnetic field $b$. It turns out that $H_{min}$ is the negative Legendre transform of $E_{min}$,
but the usual definition of the Legendre transform is too narrow to cover those cases where the functions involved are not smooth
everywhere. These cases are of special importance since they show conspicuous features as, e.~g., magnetization plateaus and jumps.
The appropriate generalization of the Legendre transform called ``Legendre-Fenchel transform" is well-known
and has also be used in the context of statistical mechanics, e.~g., for the problem of non-equivalent ensembles, see \cite{CDFR14}.
Moreover, it has been explicitly applied to spin systems, see also \cite{TB06} -- \cite{T10}, but only w.~r.~t.~the pair of dual
variables (inverse temperature, energy). In this paper we will rather apply the Legendre-Fenchel transform to the pair of variables (magnetic field, magnetization), see subsection \ref{sec:HZLF}. Here also the 
case of a non-convex function $E_{min}$ is considered that leads to a first order phase transition of the magnetization at zero temperature. There exist examples of  spin systems where this happens, e.~g., the AF icosahedron \cite{SSSL05}, see also \cite{CT92} -- \cite{K17b} for similar findings.
However, I have decided not to include these examples in the present paper since they would deserve a separate treatment.

Another relevant theme for Heisenberg-Zeeman systems is ``saturation", that is the effect that the magnetization of the ground state reaches a maximal value if the magnetic field is equal or larger than the saturation field, see subsection \ref{sec:HZS}. This effect is well-known but I do not know whether the general formula for the saturation field has been published elsewhere.
The subject is related to the fundamental distinction between ferromagnetic and anti-ferromagnetic systems. Here we propose a definition
that is equivalent to saying that a system is ferromagnetic iff its saturation field vanishes. In order to prove some related facts about
the function $H_{min}(b)$ we utilize linear and quadratic energy bounds that have partially been published in \cite{SL03}. Magnetization plateaus are treated in
subsection \ref{sec:HZM}. The next subsection \ref{sec:HZP} concerns systems where the mentioned parabolic energy bounds are identically assumed and hence will
be called ``parabolic systems". Here we generalize results of \cite{SL03} by including the cases where the magnetization of the ground state is constant below a certain ``threshold field" $B_{thr}$.

It seems that large parts of sections \ref{sec:HZ} and \ref{sec:HZR}  are readable without having digested the theory outlined in \cite{S17a} -- \cite{S17c}.
However, section \ref{sec:RH} presupposes some notions and results of the theory of ground states that will be presented in the following section \ref{sec:T}
in condensed form. In contrast to the general theory of \cite{S17a} -- \cite{S17c} and its recapitulation in section \ref{sec:T}, for the remainder of this paper we insist on the condition that spin configurations have a dimension less or equal three.

The numerous examples are, with the exception of Example \ref{cex}, elementary ones or known from the literature and mainly serve to illustrate the preceding definitions and statements.
The begin and the end of an example will be indicated by the symbol $\clubsuit$.

\section{General definitions and results}\label{sec:T}
We consider general spin configurations ${\mathbf s}$ with ${\mathbf s}_\mu\in{\mathbbm R}^N,\;\mu=1,\ldots,N$ satisfying
\begin{equation}\label{D0}
 {\mathbf s}_\mu\cdot{\mathbf s}_\mu=1 \mbox{ for all } \mu=1,\ldots,N,
\end{equation}
and denote by ${\mathcal P}^N$ the phase space of all such configurations.
Any ${\mathbf s}\in{\mathcal P}^N$ can be represented by its ``Gram matrix" $G$ with entries
\begin{equation}\label{D1}
  G_{\mu\nu}={\mathbf s}_\mu\cdot{\mathbf s}_\nu,\;\mu,\nu=1,\ldots,N.
\end{equation}

The dimension $\dim{\mathbf s}$ of ${\mathbf s}\in{\mathcal P}^N$ will be identified with the rank of $G({\mathbf s})$.
Two spin configurations have the same Gram matrix iff they are equivalent w.~r.~t.~a global rotation/reflection $R\in O(N)$.
A spin configuration ${\mathbf s}\in{\mathcal P}^N$ of dimension $n<N$ can be represented by vectors of ${\mathbbm R}^n$ upon
a suitable  rotation/reflection $R\in O(N)$ and the natural embedding ${\mathbbm R}^n\subset {\mathbbm R}^N$.
Let ${\mathcal G}={\mathcal G}_N$ denote the convex set of all Gram matrices, i.~e.~, of all $N\times N$-matrices  $G$  that are positively semi-definite
and satisfy $G_{\mu\mu}=1$ for all $\mu=1,\ldots,N$.
Let
\begin{equation}\label{D2}
 H_0({\mathbf s})=\sum_{\mu,\nu=1}^N J_{\mu\nu}\,{\mathbf s}_\mu\cdot{\mathbf s}_\nu=\mbox{ Tr} \left({\mathbbm J}\,G\right)
\end{equation}
be the Heisenberg Hamiltonian of the spin system where ${\mathbbm J}$ denotes the symmetric  $N\times N$-matrix with entries $J_{\mu\nu}$.
The mean row sum of ${\mathbbm J}$ will be denoted by $j$ such that
\begin{equation}\label{D2j}
 \sum_{\mu,\nu=1}^{N}J_{\mu\nu}=N\,j
\end{equation}
holds. The diagonal entries of ${\mathbbm J}$ can be arbitrary real numbers satisfying ${\mbox Tr } {\mathbbm J}=0$. This is a kind
of ``gauge freedom" that does not change the Hamiltonian (\ref{D2}). If $\check{\mathbf s}$ is a ground state of $H_0$, i.~e.~,
realizing its global minimum $e_0$, there is a unique ``ground state gauge" of ${\mathbbm J}$, denoted by ${\mathbbm J}^{(g)}$,  satisfying the eigenvalue equation
\begin{equation}\label{D2a}
 \sum_{\nu=1}^{N}{\mathbbm J}^{(g)}_{\mu\nu}\,\check{\mathbf s}_\nu=j_{min}^{(g)}\, \check{\mathbf s}_\mu,
 \; \mu=1,\ldots,N\;,
 \end{equation}
and consequently $e_0=j_{min}^{(g)}\,N$, where $j_{min}^{(g)}$ denotes the lowest eigenvalue of ${\mathbbm J}^{(g)}$.
For the convenience of the reader we will repeat the derivation of (\ref{D2a}) from \cite{S17a}.

The condition that the ground state $\check{\mathbf s}$ minimizes the energy (\ref{D2}) under the $N$ constraints (\ref{D0}) implies
the following ``stationary state equation" (SSE):
\begin{equation}\label{D7}
\sum_{\nu=1}^N J_{\mu\nu}\check{\mathbf s}_\nu = - \kappa_\mu\,\check{\mathbf s}_\mu,\quad \mu=1,\ldots,N
\;.
\end{equation}
Here the $\kappa_\mu$  are the Lagrange parameters due to the constraints (\ref{D0}).
Let us rewrite (\ref{D7}) in the following way:
\begin{equation}\label{D8}
\sum_{\nu=1}^N J_{\mu\nu}\check{\mathbf s}_\nu = (\bar{\kappa}- \kappa_\mu)\,\check{\mathbf s}_\mu-\bar{\kappa}\,\check{\mathbf s}_\mu
=-\lambda_\mu\,\check{\mathbf s}_\mu-\bar{\kappa}\,\check{\mathbf s}_\mu
\;,
\end{equation}
where we have introduced the mean value of the Lagrange parameters
\begin{equation}\label{D9}
\bar{\kappa}\equiv \frac{1}{N}\sum_{\mu=1}^N\,\kappa_\mu
\;,
\end{equation}
and the deviations from the mean value
\begin{equation}\label{D10}
\lambda_\mu\equiv \kappa_\mu-\bar{\kappa},\;\mu=1,\ldots,N
\;,
\end{equation}
such that
\begin{equation}\label{D11}
\sum_{\mu=1}^N\lambda_\mu=0
\;.
\end{equation}
Defining
\begin{equation}\label{D12}
{\mathbbm J}^{(g)}_{\mu\nu}=J_{\mu\nu}+\lambda_\mu\,\delta_{\mu\nu},\quad \mu,\nu=1,\ldots N
\;,
\end{equation}
renders (\ref{D8}) in the form of the eigenvalue equation
\begin{equation}\label{D13}
\sum_{\nu=1}^N {\mathbbm J}^{(g)}_{\mu\nu}\,\check{\mathbf s}_{\nu} = -\bar{\kappa}\,\check{\mathbf s}_{\mu}
\;,
\end{equation}
which is identical with (\ref{D2a}) if $j_{min}^{(g)}=-\bar{\kappa}$. For the proof of the latter equation we refer the reader to \cite{S17a}.\\

All Gram matrices of ground states, i.~e., satisfying $\mbox{ Tr} \left({\mathbbm J}\,G\right)=e_0$,
are of the form
\begin{equation}\label{D3}
G=W\,\Delta\,W^\top\;,
\end{equation}
where $W$ is some $N\times M$-matrix the columns of which span the eigenspace of ${\mathbbm J}^{(g)}$ corresponding to its lowest eigenvalue
$j_{min}^{(g)}$ and $\Delta$ is some positively semi-definite $M\times M$-matrix that is a solution of the ``additional degeneracy equation" (ADE)
\begin{equation}\label{D4}
\left(W\,\Delta\,W^\top\right)_{\mu\mu}=1\mbox{    for all   }\mu=1,\ldots,N.
\end{equation}
The convex set of solutions $\Delta\ge 0$ of the ADE is denoted by ${\mathcal S}_{ADE}$. It is affinely isomorphic to the face
of all Gram matrices $G$ satisfying $\mbox{ Tr} \left({\mathbbm J}\,G\right)=e_0$.

Further we recall some results for the special case $N=3$, see \cite{S17c}. The Gram matrices $G\in{\mathcal G}_3$ have the form
\begin{equation}\label{ST2}
 G=\left(
\begin{array}{ccc}
1& u&v\\
u&1&w\\
v&w&1
\end{array}
\right)
\;,
\end{equation}
such that the Heisenberg Hamiltonian can be written as
\begin{equation}\label{ST2a}
H_0(u,v,w)=J_1\,w+J_2\,v+J_3\,u\;,
\end{equation}

The Gram matrix of co-planar ground states of $H_0$  can be obtained from
\begin{eqnarray}\label{ST7a}
u&=& \frac{{J_1}}{2 {J_2}} \left(\frac{{J_2}^2}{{J_3}^2}-1\right)-\frac{{J_2}}{2 {J_1}},\\
\label{ST7b}
v&=& \frac{{J_3}}{2 {J_1}} \left(\frac{{J_1}^2}{{J_2}^2}-1\right)-\frac{{J_1}}{2 {J_3}},\\
\label{ST7c}
w&=&\frac{{J_2}}{2 {J_3}} \left(\frac{{J_3}^2}{{J_1}^2}-1\right)-\frac{{J_3}}{2 {J_2}}\;.
\end{eqnarray}

\section{Ground states of Heisenberg-Zeeman spin systems}\label{sec:HZ}

As mentioned in the Introduction, in this section we consider $n$-dimensional spin configurations with $n\le 3$.
This is in contrast to \cite{S17a} -- \cite{S17c} where the dimension of the spin configuration is left open.
The corresponding restricted phase space will be denoted by ${\mathcal P}^{\le3}$. Spin configurations with
$\dim {\mathbf s}=1$ will be called ``collinear" or ``Ising states"; those configurations with $\dim {\mathbf s}=2$
are denoted as ``co-planar states".
Ising states will occasionally be marked by strings of up/down arrows as, e.~g., $\uparrow\downarrow\ldots\uparrow$.

We are looking for ground states of spin systems with a Hamiltonian of the form
\begin{equation}\label{HZ1}
H(\mathbf{s})=\sum_{\mu,\nu=1}^N J_{\mu \nu}\,\mathbf{s}_\mu\cdot \mathbf{s}_\nu-{\mathbf B}\cdot {\mathbf S}
\equiv H_0(\mathbf{s})-{\mathbf B}\cdot {\mathbf S}
\;,
\end{equation}
where
\begin{equation}\label{HZ2}
{\mathbf S}\equiv \sum_{\mu=1}^N {\mathbf s}_\mu
\end{equation}
denotes the total spin vector and ${\mathbf B}=B\,{\mathbf e}\in{\mathbbm R}^{3}$ the dimensionless magnetic field.
We consider the unit vector ${\mathbf e}$ as fixed and $B\ge 0$ as variable.
If ${\mathbf s}$ is a ground state of (\ref{HZ1}) then it follows that $ {\mathbf S}$ points into the
direction of the unit vector ${\mathbf e}$, and hence ${\mathbf S}=M\,{\mathbf e},\; M\ge 0$.
Otherwise one could perform a global rotation $R\in O(3)$ of ${\mathbf s}$ such that ${\mathbf S}=M\,{\mathbf e},\; M\ge 0$ holds.
This rotation will not change the pure Heisenberg energy $H_0({\mathbf s})$ and definitely lower the Zeeman term $-{\mathbf B}\cdot {\mathbf S}$.
The latter is not possible since, by assumption, ${\mathbf s}$ was already a ground state before the rotation.
Hence any ground state of (\ref{HZ1}) satisfies
\begin{equation}\label{HZ3}
{\mathbf S}=M\,{\mathbf e},\; M\ge 0
\;.
\end{equation}
$M=M({\mathbf s})$ will be called the ``magnetization" of the ground state ${\mathbf s}$.
Moreover, it is sensible to restrict the total phase space ${\mathcal P}^{\le3}$ of spin configurations to the subset
of configurations satisfying (\ref{HZ3}), as far as the ground state problem is concerned. Hence we define
\begin{defi}\label{defPe+}
 \begin{equation}\label{Pe+}
   {\mathcal P}_{\mathbf e}^+\equiv \{{\mathbf s}\in{\mathcal P}^{\le3}\left|  {\mathbf S}=M\,{\mathbf e},\;M\ge 0  \right. \}\;.
 \end{equation}
\end{defi}
By restricting spin configurations to ${\mathcal P}_{\mathbf e}^+$ it is possible to rewrite the Hamiltonian (\ref{HZ1}) as
\begin{equation}\label{HZ4}
 H(\mathbf{s})= H_0(\mathbf{s})-B\,M({\mathbf s})=H_0(\mathbf{s})-B\,||{\mathbf S}||,\quad {\mathbf s}\in {\mathcal P}_{\mathbf e}^+
 \;.
\end{equation}
For ground states ${\mathbf s}$ of (\ref{HZ1}) the two notions of ``magnetization $M({\mathbf s})$" and ``total spin length $||{\mathbf S}||$" coincide.
However, we prefer to use the term ``magnetization" because of its physical appeal.
If, for given $B\ge 0$, the spin configuration $\check{\mathbf s}\in {\mathcal P}_{\mathbf e}^+$ is a ground state of (\ref{HZ4}) we will denote its energy as
\begin{equation}\label{HZ5}
  H_{min}(B)\equiv  H_0(\check{\mathbf s})-B\,M(\check{\mathbf s})
  \;.
\end{equation}
Obviously, the value $H_{min}(B)$ does not depend on the choice of the ground state $\check{\mathbf s}$ and hence
(\ref{HZ5}) defines a real function $H_{min}:{\mathbbm R}^+\longrightarrow{\mathbbm R}$.

So far, the field $B$ and the magnetization $M({\mathbf s}),\;{\mathbf s}\in{\mathcal P}_{\mathbf e}^+$ are, by definition,
non-negative. For mathematical reasons, mainly in order to facilitate the application of the Legendre-Fenchel transform,
it is convenient to extend the function $b\mapsto H_{min}(b)$ to an even function defined on the whole real axis. This
suggests to also introduce negative fields and negative magnetization. Hence we extend the set  ${\mathcal P}_{\mathbf e}^+$
to the larger set
\begin{defi}\label{defPe}
 \begin{equation}\label{Pe}
   {\mathcal P}_{\mathbf e}\equiv \{{\mathbf s}\in{\mathcal P}^{\le3}\left|  {\mathbf S}=M\,{\mathbf e},\;M\in{\mathbbm R}  \right. \}\;,
 \end{equation}
\end{defi}
and re-define the magnetization function
\begin{eqnarray}\nonumber
 M&:& {\mathcal P}_{\mathbf e}\longrightarrow{\mathbbm R}\\
 \label{Pe1}
 M({\mathbf s})&\equiv& {\mathbf S}\cdot {\mathbf e}\;.
\end{eqnarray}
Hence $ M({\mathbf s})=||{\mathbf S}||\ge 0$ if ${\mathbf S}$ is parallel to ${\mathbf e}$ and
$ M({\mathbf s})=-||{\mathbf S}||\le 0$ if ${\mathbf S}$ is anti-parallel to ${\mathbf e}$. The equation
\begin{equation}\label{Pe2}
  H(\mathbf{s})= H_0(\mathbf{s})-B\,M({\mathbf s})
\end{equation}
then holds for all $B\in{\mathbbm R}$ and all ${\mathbf s}\in{\mathcal P}_{\mathbf e}$.
Similarly, (\ref{HZ5}) holds for all  $\check{\mathbf s}\in{\mathcal P}_{\mathbf e}$
that minimize the energy  (\ref{Pe2}) for given $B\in{\mathbbm R}$.

As a first general property of the function $H_{min}$ we state:
\begin{prop}\label{prop1}
$H_{min}:{\mathbbm R}\longrightarrow {\mathbbm R}$ is a concave and continuous function.
\end{prop}
Recall that a concave function $f$ can be defined by the property that its subgraph $\Sigma_f\equiv \{(x,y)\}\left| y\le f(x)\right.\}$
is a convex subset of ${\mathbbm R}^2$.\\
\noindent{\bf Proof:}\\
For each ${\mathbf s}\in {\mathcal P}_{\mathbf e}$ define the affine function $L_{\mathbf s}:{\mathbbm R}\longrightarrow{\mathbbm R}$ by
\begin{equation}\label{HZ6}
L_{\mathbf s}(B)=H_0({\mathbf s})-B\,M({\mathbf s})
\;.
\end{equation}
Obviously, $H_{min}\le L_{\mathbf s}$ and $H_{min}(B)= L_{\check{\mathbf s}}(B)$ iff  $\check{\mathbf s}$ is a ground state of (\ref{Pe2})
for some $B\in{\mathbbm R}$.
Hence
$\Sigma_{H_{min}} = \bigcap_{{\mathbf s}\in {\mathcal P}_{\mathbf e}} \Sigma_{L_{ \mathbf s}}$,
and this set is convex since it is an intersection of convex sets.

For the remaining claim we note that $-H_{min}$ is a convex function defined on the whole real axis, and hence continuous, see \cite{R97}, Cor.~10.1.
\hfill$\Box$\\

\vspace{5mm}
\begin{ex} \label{ex1}
 The AF dimer ($N=2$)\\
\begin{center}
\unitlength1cm
\begin{picture}(4,1)(0,0)
\thicklines
 \put(0.5,0.5){\circle{0.5}}
  \put(2.5,0.5){\circle{0.5}}
 \textcolor{red}{ \put(0.75,0.5){\line(1,0){1.5}}}
\end{picture}
\end{center}
\end{ex}
\vspace{5mm}

\begin{figure}[ht]
  \centering
    \includegraphics[width=1.0\linewidth]{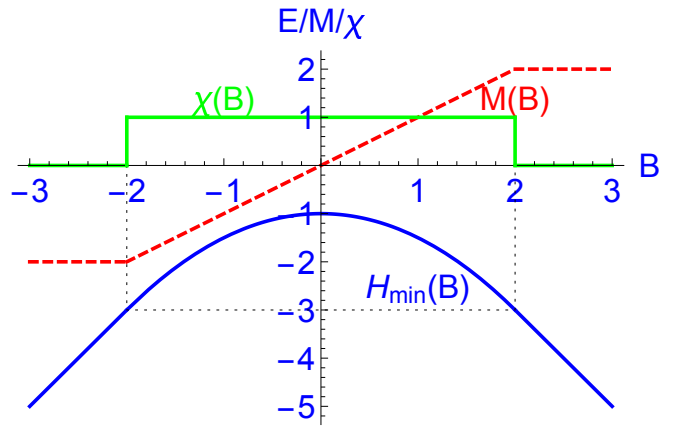}
  \caption[Example AF Dimer]
  {The minimal  energy $H_{min}$, the magnetization ${\sf M}$ and the susceptibility $\chi$ as a function of the magnetic field $B$ for the AF dimer.
  At the saturation field $|B|=B_{sat}=2$   (dotted line) there occurs a phase transition of $2$nd order.
  }
  \label{FHZ1a}
\end{figure}

$\clubsuit$ For the sake of illustration we will consider the AF spin dimer as an elementary example. It has the Hamiltonian
\begin{eqnarray}\label{HZ7a}
H&=&{\mathbf s}_1\cdot{\mathbf s}_2-\mathbf{B}\cdot\left( {\mathbf s}_1 +{\mathbf s}_2\right)\\
\label{HZ7b}
&=&\frac{1}{2}\left( {\mathbf S}-{\mathbf B}\right)^2-\left(1+\frac{1}{2}B^2\right)\;.
\end{eqnarray}
For $B=0$ its ground state is of the form $\uparrow\downarrow$, whereas for large $B$ the ferromagnetic ground state $\uparrow\uparrow$
will have the lowest energy. For intermediate values of $B$ we have a competition between parallel and anti-parallel
alignment and one wonders, how the system's compromise would look like.
The solution can be obtained by elementary considerations. Due to the ``completing squares" trick in (\ref{HZ7b})
it is clear that the ground states are exactly those minimizing the distance $|| {\mathbf S}-{\mathbf B}||$.
For $0\le B\le 2$ this is possible by choosing $ {\mathbf S}={\mathbf B}$. Consequently, $H_{min}(B)=-1-\frac{1}{2}B^2$
in this case. On the other hand, for $B>2$ the distance $|| {\mathbf S}-{\mathbf B}||$ is minimized by the Ising ground state
${\mathbf s}_1={\mathbf s}_2 ={\mathbf e}$ or, in the arrow notation, ${\mathbf s}=\uparrow\uparrow$. In this case,
$H_{min}(B)= 1-2 {\mathbf B}\cdot {\mathbf e}=1-2 B$, and hence
\begin{equation}\label{HZ8}
 H_{min}(B)=\left\{\begin{array}{r@{\quad:\quad}l}
-1-\frac{1}{2}B^2& 0\le |B|\le 2 ,\\
 1-2 |B|&2<|B| ,
  \end{array} \right.
\end{equation}
where we have written $H_{min}$ as an even function of $B$. Clearly, $H_{min}$ is a concave function, see Figure \ref{FHZ1a},
where also the magnetization ${\sf M}(B)$ and the susceptibility $\chi(B)$  have been displayed that will be defined later.
Obviously, the saturation field has the value $B_{sat}=2$.

\hfill$\clubsuit$\\

\section{Ground states of Heisenberg-Zeeman spin systems and relative ground states}\label{sec:HZR}
\subsection{Legendre-Fenchel transform}\label{sec:HZLF}

Returning to the general case we assume that, for given $B$, a ground state $\check{\mathbf s}$ realizes the minimal energy
$H_{min}(B)$. Hence the line in ${\mathbbm R}^2$ given by the graph of the affine function
$L_{\check{\mathbf s}}(b)=H_0({\check{\mathbf s}})-b\,M(\check{\mathbf s})$, cp.~ (\ref{HZ6}),
intersects the graph of $H_{min}$ (at least) at the point $(B,H_{min}(B))$. We set $\mu=M(\check{\mathbf s})$ and argue that the minimum
$H_{min}(B)$ can already be obtained by varying ${\mathbf s}$ not over ${\mathcal P}_{\mathbf e}$ but over the smaller
set
\begin{equation}\label{HZ9}
 {\mathcal P}_{{\mathbf e},\mu}\equiv \{{\mathbf s}\in{\mathcal P}_{\mathbf e}\left| M({\mathbf s})=\mu\right.\}
 \;.
\end{equation}
The reason is simply that, by construction, the state $\check{\mathbf s}$ realizing the minimal energy
$H_{min}(B)$ is contained in ${\mathcal P}_{{\mathbf e},\mu}$. But if we vary  ${\mathbf s}$ over ${\mathcal P}_{{\mathbf e},\mu}$
the Zeeman term becomes $-B\,\mu$ and thus will be constant. Hence the minimum $H_{min}(B)$ is obtained exactly for those states where
$H_0({\mathbf s})$ will be minimized for ${\mathbf s}\in {\mathcal P}_{{\mathbf e},\mu}$. These states will be called ``relative ground states";
they are ground states of the pure Heisenberg Hamiltonian $H_0$ under the constraint
$M({\mathbf s})=\mu,\;{\mathbf s}\in{\mathcal P}_{{\mathbf e},\mu}$. Hence we have shown that
the ground states of (\ref{HZ1}) are among the relative ground states of $H_0$.
The ground state problem for Heisenberg-Zeeman Hamiltonians can thus be reduced to the relative ground state problem for pure Heisenberg Hamiltonians.
However, it does not follow that {\it all} relative ground states of $H_0$ are ground states of $H$. We will later present a couple of counter-examples.

It is necessary to introduce some more notation. Let
\begin{eqnarray}\label{HZ10a}
\widetilde{E}&\equiv& \{(\mu,E)\left| \exists\, {\mathbf s}\in {\mathcal P}_{{\mathbf e},\mu} \mbox{ such that } E=H_0({\mathbf s})\right.\}\\
\nonumber
&=& \{(\mu,E)\left| \exists\, {\mathbf s}\in {\mathcal P}^{\le 3} \mbox{ such that } E=H_0({\mathbf s})\right.\\
\label{HZ10aa}
&& \mbox{ and } {\mathbf S}^2=\mu^2\}
\end{eqnarray}
and $E_{min}:[-N,N]\longrightarrow{\mathbbm R}$ denote the function
\begin{equation}\label{HZ10b}
E_{min}(\mu)\equiv \mbox{ Min }\{ H_0({\mathbf s})\left| {\mathbf s}\in  {\mathcal P}_{{\mathbf e},\mu}\right.\}
\;.
\end{equation}
Obviously, $E_{min}$ is an even function and the graph of $E_{min}$ is a subset of $\tilde{E}$.
For later purposes we note the following
\begin{lemma}\label{lemma1}
  $\tilde{E}$ is a compact subset of ${\mathbbm R}^2$.
\end{lemma}
\noindent{\bf Proof:}\\
We will consider the restricted Gram set
\begin{equation}\label{lemma1a}
{\mathcal G}^{\le 3}\equiv \{G\in{\mathcal G}\left| \mbox{ rank }(G)\le 3 \right.\}
\;.
\end{equation}
It is obviously bounded and also closed since ${\mathcal G}$ is closed and the condition $ \mbox{ rank }(G)\le 3$
can be reformulated by the vanishing of all minors of $G$ of order $4$. Hence ${\mathcal G}^{\le 3}$ is compact
and we will prove the claim by showing that  $\tilde{E}$ is essentially the image of ${\mathcal G}^{\le 3}$ under a continuous map.
For the second component of $(\mu,E)$ this is clear since $E=\mbox{ Tr }\left( {\mathbbm J}\,G\right)$.
For the first component and $\mu\ge 0$ it is clear that $\mu = M({\mathbf s})= ||{\mathbf S}||=
\left( \sum_{\mu,\nu=1}^{N}{\mathbf s}_\mu\cdot{\mathbf s}_\nu\right)^{1/2}=
\sqrt{ \mbox{ Tr }\left( {\mathbf 1}\,G\right)}$,
where $ {\mathbf 1}$ is the $N\times N$-matrix completely filled with $1$. Hence $\tilde{E}\cap \{(\mu,E)\left| \mu\ge 0\right.\}$
is the image of  ${\mathcal G}^{\le 3}$ under the continuous map
$G\mapsto (\mbox{ Tr }\left( {\mathbbm J}\,G\right),\sqrt{ \mbox{ Tr }\left( {\mathbf 1}\,G\right)})$ and hence compact.
Analogously, one shows that
$\tilde{E}\cap \{(\mu,E)\left| \mu\le 0\right.\}$ is compact and hence $\tilde{E}$ is compact as the union of two compact sets.     \hfill$\Box$\\

The above arguments showing the reduction to the relative ground state problem can be sharpened to prove that
the function $H_{min}$ will be the negative ``Legendre-Fenchel transform" of the function $E_{min}$ or,
equivalently, of the set $\tilde{E}$. Some explanations will
be in order. The Legendre transform is well-known in physics by its applications in mechanics and thermodynamics.
Recall that the Hamiltonian $H$ of a mechanical system can be obtained as the Legendre transform of the corresponding Lagrangian $L$,
in symbols $H={\mathcal L}\left( L\right)$.
Consider the simple case $L(v)=\frac{m}{2}v^2 -V(q)$ with self-explaining notation and define $p(v)\equiv \frac{\partial L}{\partial v}=m\,v$
with the inverse $v(p)=\frac{1}{m}p$. Then $H(p)\equiv p\, v(p)-L(v(p))= \frac{p^2}{2m}+V(q)$. The Legendre transform has the nice
geometric interpretation that $H(p)$ is the negative intersection of the tangent to the graph of $L(v)$  having the slope $p$ with the vertical coordinate axis.
For the applications to thermodynamics including phase transitions
it turns out that the notion of Legendre transform is too narrow and needs to be generalized to the ``Legendre-Fenchel transform" or
``convex conjugate", see \cite{T84}, \cite{R97}, and \cite{T05} for a short introduction. This generalization essentially consists of replacing the tangent to the graph
of a function by a ``supporting line". In this way the assumptions of smoothness of the function  to be transformed
can be weakened. Moreover, the definition can be extended to the Legendre-Fenchel transform of rather general subsets of ${\mathbbm R}^2$.

Before explaining the details of the Legendre-Fenchel transform we will first treat the smooth case where $H_{min}$ can be obtained as
the negative Legendre transform of $E_{min}$ in the traditional way. We will use the abbreviations $E_{min}=E$ and $H_{min}=H$ in the following Proposition
but retain the meaning of these functions.
\begin{prop}\label{propL}
Let $E:[-N,N]\longrightarrow {\mathbbm R}$ be an even function that is twice continuously differentiable and satisfies
\begin{equation}\label{propL1}
  \frac{d^2\,E(\mu)}{d\mu^2} >0
  \;,
\end{equation}
thus being strictly convex. Especially, the limits $\lim_{\mu\rightarrow \pm N}\frac{d\,E(\mu)}{d\,\mu}=\pm B_0$ exist.
Then $b\mapsto H(b)$ is the negative Legendre transform of $\mu\mapsto E(\mu)$.
\end{prop}
\noindent{\bf Proof:}\\
It follows that ${\sf B}(\mu)\equiv\frac{dE}{d\mu}$ is continuous and strictly monotonically increasing for all $-N\le \mu \le N$.
Hence ${\sf B}$ has an inverse ${\sf M}: [-B_0,B_0]\longrightarrow [-N,N]$. The Legendre transform of $E$ is $b\mapsto{\sf M}(b)\,b-E({\sf M}(b))$.
For fixed $b\in[-B_0,B_0]$ we consider the function
$h_b(\mu)\equiv E(\mu) -\mu\,b$. Its derivative $\frac{d\,h_b}{d\,\mu}=\frac{d\,E}{d\,\mu}-b$ vanishes exactly for $b={\sf B}(\mu)$
and hence for $\mu={\sf M}(b)$.  Since $\frac{d^2\,h_b}{d\,\mu^2}=\frac{d^2\,E}{d\,\mu^2}>0$ there is a unique and global minimum
of $h_b$ at $\mu={\sf M}(b)$ if it can be excluded that a minimum occurs at the boundary of $[-N,N]$ without vanishing derivative of $h_b$.
To show the latter let us assume that a minimum of $h_b$ at $\mu=N$ exists such that $\left.\frac{d\,h_b}{d\,\mu}\right|_{\mu=N}={\sf B}(N)-b<0$.
This implies $b>{\sf B}(N)=B_0$ which contradicts $b\in[-B_0,B_0]$. The case $\mu=-N$ is analogous.

Next consider, for given $-B_0\le b\le B_0$, a ground state $\check{{\mathbf s}}\in{\mathcal P}_{\mathbf e}$ of the Heisenberg-Zeeman Hamiltonian with minimal energy $H(b)$ and magnetization $\mu=M(\check{\mathbf s})$. Hence $H(b)$ is also the minimum of the Heisenberg-Zeeman energy of all states
${\mathbf s}\in {\mathcal P}_{{\mathbf e},\mu}$. Since  for fixed $\mu$ and $b$ the term $-\mu\,b$ is constant,
the minimum is attained for states realizing the minimum of the pure Heisenberg energy, i.~e., $H(b)=E(\mu)-\mu\,b$.

Now let $\mu$ vary over the whole domain $[-N,N]$. It follows that  $H(b)$ is the global minimum of $E(\mu)-\mu\,b=h_b(\mu)$.
According to the above consideration this minimum is attained at $\mu={\sf M}(b)$. Hence
$H(b)=E({\sf M}(b))-{\sf M}(b)\,b$ for all $-B_0\le b\le B_0$, which is the negative Legendre transform of $E$.
\hfill$\Box$\\

In the smooth case the magnetization function is the negative derivative of $H_{min}$ :
\begin{cor}\label{corL}
 Under the conditions of Proposition \ref{propL} let the magnetization function ${\sf M}: [-B_0,B_0]\longrightarrow [-N,N]$ be defined as
 the inverse function of ${\sf B}=\frac{\partial E}{\partial \mu}$, cp.~the preceding proof.
 Then
 \begin{equation}\label{corL1}
  \frac{\partial H_{min}(b)}{\partial b}=-{\sf M}(b)\;.
 \end{equation}
\end{cor}
\noindent{\bf Proof:}\\
This follows from
\begin{eqnarray}
   \label{corL2a}
   \frac{\partial H(b)}{\partial b} &=& \frac{\partial}{\partial b}\left(E({\sf M}(b))- {\sf M}(b)\,b\right)\\
    \label{corL2b}
    &=& \left.\frac{\partial E}{\partial \mu}\right|_{\mu= {\sf M}(b)}\,\frac{\partial {\sf M}}{\partial b}
    -\frac{\partial {\sf M}}{\partial b}\,b-{\sf M}(b)\\
    &=& {\sf B}({\sf M}(b))\,\frac{\partial {\sf M}}{\partial b}  -\frac{\partial {\sf M}}{\partial b}\,b-{\sf M}(b)\\
     \label{corL2c}
    &=&b\, \frac{\partial {\sf M}}{\partial b}  -\frac{\partial {\sf M}}{\partial b}\,b-{\sf M}(b)\\
     \label{corL2d}
    &=& -{\sf M}(b)\;.
\end{eqnarray}
\hfill$\Box$\\

We return to the general case and will illustrate the pertaining definitions in connection with the Legendre-Fenchel transform
for the function $\mu\mapsto E_{min}(\mu)$ and the set $\tilde{E}$.
A line in the $(\mu,E)$-plane given by the equation $E=E_0 +b\,\mu$ is said to ``support $E_{min}$ at $\mu_0$" iff it satisfies
\begin{eqnarray}\label{HZ11a}
 E_{min}(\mu_0)&=&E_0 +b\,\mu_0,\\
 \nonumber
  \mbox{and}&&\\
 \label{HZ11b}
  E_{min}(\mu)&\ge& E_0 +b\,\mu \mbox{  for all }-N\le \mu\le N
 \;.
\end{eqnarray}
If $E_{min}$ is a convex function and differentiable in the neighborhood of $\mu_0$ then its tangent at $\mu_0$ will be
the only supporting line of  $E_{min}$ at $\mu_0$.

Let $f:[-N,N]\longrightarrow {\mathbbm R}$ be a function bounded from below. Then the Legendre-Fenchel transform\\
${\mathcal L}\left(f\right):{\mathbbm R}\longrightarrow{\mathbbm R}$ of  $f$ will be defined by
\begin{equation}\label{HZ12}
{\mathcal L}\left(f\right)(b)\equiv \sup_{\mu\in[-N,N]}\left( b\,\mu-f(\mu)\right),\mbox{ for all }b\in{\mathbbm R}.
\end{equation}
The assumption that $f$ is bounded from below assures that the supremum in (\ref{HZ12}) exists.

\begin{figure}[ht]
  \centering
    \includegraphics[width=1.0\linewidth]{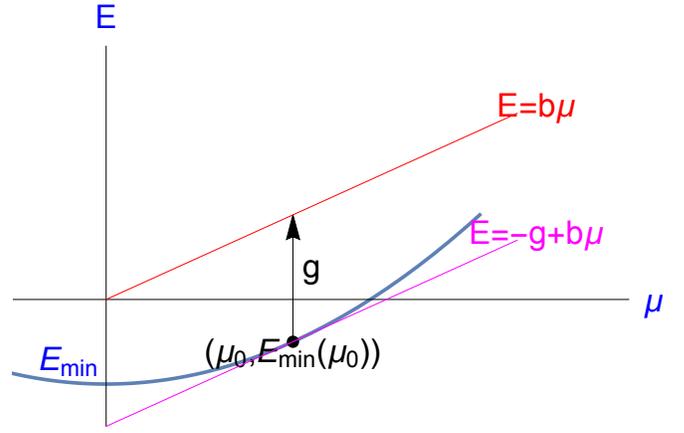}
  \caption[Example LF transform]
  {Illustration of the Legendre-Fenchel transform ${\mathcal L}\left( E_{min}\right)$.
  The (red) line $E= b\,\mu$  with slope $b$ has the maximal vertical distance $g={\mathcal L}\left( E_{min}\right)(b)$
  to the graph of $E_{min}$  at the point $(\mu_0,E_{min}(\mu_0))$. The correspondingly shifted (magenta) line $E= -g+b\,\mu$
  supports $E_{min}$ at $\mu_0$.
  }
  \label{FHZ2}
\end{figure}

The relation of this definition to the concept of supporting lines will be explained for the case $f=E_{min}$, see Figure \ref{FHZ2}:
Let us assume that the supremum in (\ref{HZ12}) will be attained at some (not necessarily unique) point
$(\mu_0,E_{min}(\mu_0))$.  Let $g\equiv b\,\mu_0-E_{min}(\mu_0)$ and consider the line $\ell$ given by the equation $E=-g+b\,\mu$.
Then $\ell$ will support $E_{min}$ at $\mu_0$. To show the latter, we first note that $E_{min}(\mu_0)=-g +b\,\mu_0$ by the definition
of $g$. Further, since $g$ is a supremum, we conclude $g\ge b\,\mu -E_{min}(\mu)$ for all $-N\le\mu\le N$. This implies
$E_{min}(\mu)\ge -g +b\,\mu$ for all $-N\le\mu\le N$ and completes the proof that $\ell$ supports $E_{min}$ at $\mu_0$.

Analogously to (\ref{HZ12}) the Legendre-Fenchel transform of a set $\widehat{E}\subset {\mathbbm R}^2$ bounded from below will be defined by
\begin{equation}\label{HZ13}
{\mathcal L}(\widehat{E})(b)\equiv \sup_{(\mu,E)\in\widehat{E}}\left( b\,\mu-E\right),\mbox{ for all }b\in{\mathbbm R}.
\end{equation}
Both definitions yield the same function when applied to $E_{min}$ and $\widetilde{E}$:
\begin{lemma}\label{lemma2}
$ {\mathcal L}(\tilde{E})={\mathcal L}\left( E_{min}\right)$.
\end{lemma}

\noindent{\bf Proof:}\\
Let $b\in{\mathbbm R}$ be arbitrary. Since $\tilde{E}$ is compact, see Lemma \ref{lemma1}, the supremum
${\mathcal L}(\tilde{E})(b)= \sup_{(\mu,E)\in\tilde{E}}\left( b\,\mu-E\right)$ is assumed at some (not necessarily unique)
point $(\mu_0,E_0)\in \tilde{E}$. It follows that $E_0$ is the minimum of the set $\{ E \left| (\mu_0,E)\in \tilde{E}\right. \}$
and hence $E_0=E_{min}(\mu_0)$. \\We conclude\\
${\mathcal L}(\tilde{E})(b)= b\,\mu_0-E_0\in\{ b\,\mu-E_{min}(\mu)\left| -N\le \mu \le N\right.\}$
and hence ${\mathcal L}(\tilde{E})(b)$ is bounded by the supremum of the set $\{ b\,\mu-E_{min}(\mu)\left| -N\le \mu \le N\right.\}$,
which is ${\mathcal L}\left( E_{min}\right)(b)$.

Conversely, ${\mathcal L}\left( E_{min}\right)(b)\le {\mathcal L}(\tilde{E})(b)$ since the first supremum is taken over the graph
of $E_{min}$ which is a subset of $\tilde{E}$.
 \hfill$\Box$\\

 The proof immediately implies:
 \begin{cor}\label{cor1}
The  supremum of the set\\ $\{ b\,\mu-E_{min}(\mu)\left| -N\le \mu \le N\right.\}$ is attained for all $b\in{\mathbbm R}$.
 \end{cor}

 The main result of this subsection is the following:
\begin{theorem}\label{theorem1}
 $H_{min}=-{\mathcal L}\left( E_{min}\right)$.
\end{theorem}

\noindent{\bf Proof:}\\
The statement is equivalent to
\begin{equation}\label{th1}
 H_{min}(B)=\inf_{\mu\in[-N,N]}\left(E_{min}(\mu)-B\,\mu\right),\mbox{ for all }B\in{\mathbbm R}.
\end{equation}
This holds since the search for an infimum, or, equivalently (using Corollary \ref{cor1}), minimum of  $H_0(\mathbf{s})-B\,M({\mathbf s})$
can be decomposed into two steps: First, we fix $\mu=M({\mathbf s})$ and look for a minimum of $H_0(\mathbf{s})$
within the set ${\mathbf s}\in{\mathcal P}_{{\mathbf e},\mu}$. This gives the result
$\mbox{Min }\{H_0(\mathbf{s})\left|\mathbf{s}\in{\mathcal P}_{{\mathbf e},\mu}\right.\} =E_{min}(\mu)$.
In the second step we minimize $E_{min}(\mu)-B\,\mu$ over $-N\le \mu\le N$. This yields (\ref{th1}).
\hfill$\Box$\\

Since the Legendre-Fenchel transform clearly reverses functional inequalities $\le$, see also \cite{T84}  6.3(a), we conclude
\begin{lemma}\label{corT1}
The negative Legendre-Fenchel transform is monotone, i.~e.~, if $f_1\le f_2$ then $-{\mathcal L}(f_1)\le -{\mathcal L}(f_2)$ .
\end{lemma}

At this point we would like to define the magnetization of ground states of (\ref{HZ1}) as a function  ${\sf M}(B)$.
However, there are examples where the points $(\mu_0,E_0)\in\tilde{E}$ realized by ground states are not unique for given $B$ and
thus the magnetization function would be multi-valued.
Hence we will rather define a ``magnetization graph" ${\mathcal M}$ in the following way:
\begin{defi}\label{defM1}
\begin{equation}\label{HZ14}
 {\mathcal M}\equiv\left\{(b,\mu_0) \left|\exists\,E_0: \sup_{(\mu,E)\in\tilde{E}}\left( b\,\mu-E\right)= b\,\mu_0-E_0\right.\right\}.
\end{equation}
\end{defi}

An equivalent definition that more directly refers to the ground states is the following:
\begin{defi}\label{defM2}
\begin{eqnarray}\nonumber
(B,\mu)\in{\mathcal M}& \Leftrightarrow& \mbox{ there exists a ground state } {\mathbf s}\in{\mathcal P}_{\mathbf e} \mbox{ of (\ref{Pe2})}\\
\label{HZ15}
&&\mbox{ such that } \mu=M({\mathbf s}).
\end{eqnarray}
\end{defi}

The magnetization graph is odd, i.~e., it satisfies $(B,\mu)\in{\mathcal M} \Leftrightarrow (-B,-\mu)\in{\mathcal M}$.

The magnetization increases with $B$. Instead of proving the monotonic increase of the magnetization function we have to resort to
the following formulation:
\begin{lemma}\label{lemma3}
  If $(B_1,\mu_1),\,(B_2,\mu_2)\in {\mathcal M}$ and $ B_1 <B_2$ then $\mu_1\le \mu_2$.
\end{lemma}

\noindent{\bf Proof:}\\
In the case of a smooth magnetization function ${\sf M}(B)$ its monotonic increase would follow immediately from
$H_{min}(B)$ being concave and ${\sf M}(B)=-\frac{\partial H_{min}}{\partial B}$, see Corollary \ref{corL}. It is plausible that this also
holds in the limit of a non-smooth magnetization graph ${\mathcal M}$ but it seems difficult to make this idea rigorous.
Hence we proceed with a direct proof of the lemma.

According to the Definition \ref{defM2} of ${\mathcal M}$ let ${\mathbf s}^1$ and ${\mathbf s}^2$ be ground states corresponding to the points
$(B_1,\mu_1),\,(B_2,\mu_2)\in {\mathcal M}$ with minimal energies $E_1$ and $E_2$, resp.~. Then we conclude
\begin{eqnarray}
\label{HZ16a}
  E_1 &=& H_0({\mathbf s}^1)-B_1\,M({\mathbf s}^1)\\
 \label{HZ16b}
 E_2 &=& H_0({\mathbf s}^2)-B_2\,M({\mathbf s}^2)\\
 \label{HZ16c}
   E_1 &\le& H_0({\mathbf s}^2)-B_1\,M({\mathbf s}^2) \equiv E_1'\\
   \label{HZ16d}
   E_2 &\le& H_0({\mathbf s}^1)-B_2\,M({\mathbf s}^1) \equiv E_2'\\
   \label{HZ16e}
  M({\mathbf s}^1) &=& \frac{E_1-E_2'}{B_2-B_1}\le  \frac{E_1'-E_2}{B_2-B_1}=M({\mathbf s}^2)
 ,
\end{eqnarray}
which proves $\mu_1\le \mu_2$. Here  (\ref{HZ16c}) holds since ${\mathbf s}^1$ was assumed to be a ground state
at $B=B_1$, analogously for (\ref{HZ16d}) and ${\mathbf s}^2$. Moreover, (\ref{HZ16e}) follows from (\ref{HZ16a}) -- (\ref{HZ16d})
and $B_1<B_2$.    \hfill$\Box$\\

In all examples that we have investigated the magnetization graph consists of parts that are
graphs of smooth functions ${\sf M}_j(B)$ and possible ``jumps"
of height $h_i$ at $B_i,\;i=1,\ldots,K$.
In these ``smooth" cases, but not in general, we would define the susceptibility $\chi(B)$ piecewise
as the  derivative of the magnetization function
plus a sum of $\delta$-functions
\begin{equation}\label{HZ15}
 \chi(B)=\frac{\partial {\sf M}_j}{\partial B}+\sum_{i=1}^{K}h_i\,\delta(B-B_i)
 \;.
\end{equation}
In the smooth case the magnetization is an odd function of $B$ and hence $\chi(B)$ will be an even one. Moreover,
Lemma \ref{lemma3} implies that the magnetization is monotonically increasing and hence $\chi(B)\ge0$ for all $B\in{\mathbbm R}$.
In other words, classical spin systems are necessarily paramagnetic at $T=0$.
\cite{foot1}
 \\

$\clubsuit$ We will illustrate the preceding definitions and results for the above elementary example of the AF dimer.
The general spin configuration ${\mathbf s}\in{\mathcal P}_{\mathbf e}$ will be of the form
\begin{equation}\label{HZ17}
 {\mathbf s}_1={\cos\alpha \choose \sin\alpha},\;{\mathbf s}_2={\cos\alpha \choose -\sin\alpha},\;0\le\alpha\le \pi,
\end{equation}
if ${\mathbf e}={1\choose 0}$. Hence $M({\mathbf s})=2 \cos\alpha$ and
$H_0({\mathbf s})=\cos^2\alpha-\sin^2\alpha=2\,\cos^2\alpha-1=\frac{1}{2}\,M({\mathbf s})^2-1$.
This implies that $\tilde{E}$ is identical to the graph of the function
\begin{equation}\label{HZ18}
E_{min}(\mu)=\frac{1}{2}\,\mu^2-1
\;.
\end{equation}
This is compatible with Lemma \ref{lemma2} but will be rarely  satisfied in other examples.
In the interval $-2<\mu<2$ the function $E_{min}$ is smooth and hence the Legendre transform can be calculated in the
traditional way by $B(\mu)=\frac{\partial E_{min}}{\partial \mu}=\mu$, hence ${\sf M}(B)=B$ and $H_{min}(B)=-{\mathcal L}\left(E_{min}\right)(B)=
E_{min}({\sf M}(B))-B\,{\sf M}(B)=\frac{1}{2}\,B^2-1-B^2=-1-\frac{1}{2}B^2$ which complies with (\ref{HZ8}) for $-2<\mu=B<2$.

For $B\ge 2$ all supporting lines of $E_{min}$ with slope $B$ pass through the point $(\mu,E)=(2,1)$ and hence are given by the equation
$E=1-2\,B+B\,\mu$. This implies  $H_{min}(B)=-{\mathcal L}\left(E_{min}\right)(B)=1-2\,B$ for $B\ge 2$.
Together with the result for $B\le 2$ that can be analogously calculated as $H_{min}(B)=1+2\,B$, this complies with
(\ref{HZ8}) for $|B|\ge 2$.

Finally we calculate the magnetization function ${\sf M}(B)$ as
\begin{equation}\label{HZ19}
{\sf M}(B)= \left\{\begin{array}{r@{\quad:\quad}l}
B&  |B|<2 ,\\
2&B\ge 2 ,\\
-2 & B\le 2,
  \end{array} \right.
\end{equation}
and consequently the susceptibility function $\chi(B)$ as
\begin{equation}\label{HZ20}
\chi(B)= \left\{\begin{array}{r@{\quad:\quad}l}
1&  |B|<2 ,\\
0&|B|\ge 2 ,
  \end{array} \right.
\end{equation}
see Figure \ref{FHZ1a}.
According to the Ehrenfest classification the AF dimer hence undergoes a phase transition of second order at $|B|=2$ and $T=0$.
But this is a kind of trivial phase transition rarely mentioned since it will occur for all AF Heisenberg spin systems as we will see
below.
\hfill$\clubsuit$\\

\vspace{5mm}
\begin{ex} \label{ex2}
 Non-convex $E_{min}$
\end{ex}
\vspace{5mm}

\begin{figure}[ht]
  \centering
    \includegraphics[width=1.0\linewidth]{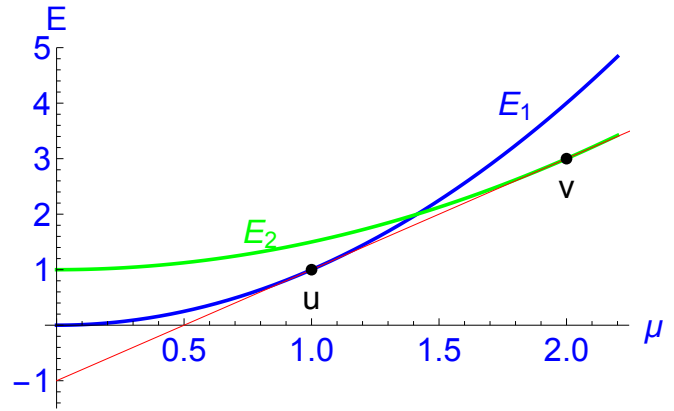}
  \caption[Non-convex]
  {Illustration of a non-convex function $E_{min}$ that is the minimum of two convex functions $E_1$ (blue) and $E_2$ (green).
  The graph of $E_{min}$ has the convex closure generated by the segment of the common tangent (red line) with slope $b=2$ between the points
  $u=(1,1)$ and $v=(2,3)$.
  }
  \label{FHZ12}
\end{figure}

$\clubsuit$ In order to justify the use of the generalization of the Legendre transform to the Legendre-Fenchel transform we will
consider another possible case where a phase transition occurs, this time of first order. The realization of this case by concrete spin
systems, see \cite{SSSL05} -- \cite{K17b}, will not be considered in this article; here we confine ourselves to a toy example.
As already mentioned, it is not necessary
that all relative ground states are ground states of the Heisenberg-Zeeman system. The reason is that the function
$E_{min}$ need not be convex in general and hence there will be points in the graph of $E_{min}$ that will never
be touched by supporting lines. Consider a case where $E_{min}$ is the minimum of two convex functions $E_1$ and $E_2$ without being convex itself,
namely
\begin{equation}\label{HZ21}
E_{min}=\mbox{ Min }\{E_1,E_2\},\quad E_1(\mu)=\mu^2,\quad E_2(\mu)=1+\frac{1}{2}\mu^2,
\end{equation}
see Figure \ref{FHZ12}. The definition domain of $E_{min}$ is some interval $[-N,N]$ that will play no role in the following.
The two functions $E_1,\;E_2$ intersect at $(\mu_0, e_0)=(\sqrt{2},2)$ and have a common tangent with slope $b=2$ connecting the points
$u=(1,1)$ and $v=(2,3)$. This tangent generates the convex hull of the epigraph of $E_{min}$.
We try to formally calculate the (negative) Legendre transforms $H_1$ and $H_2$ of $E_1$ and $E_2$, resp.:
$B=\frac{\partial E_1}{\partial \mu}=2\,\mu$ hence $\mu(B)=\frac{B}{2}$ and
$H_1(B)= E_1(\mu(B))-B\,\mu(B)=\frac{1}{4}B^2-\frac{1}{2}B^2=-\frac{1}{4}B^2$.
Analogously, $B=\frac{\partial E_2}{\partial \mu}=\mu$ hence $\mu(B)=B$ and
$H_2(B)= E_2(\mu(B))-B\,\mu(B)=1+\frac{1}{2}B^2-B^2=1-\frac{1}{2}B^2$. It is not clear which $H_i$ we should
take as the (negative) Legendre transform. It is plausible, analogously as for $E_{min}$,
to choose the minimum of both function, $H_{min}=\mbox{ Min }\{H_1,H_2\}$, which is concave, but this would be an {\it ad hoc} choice.

\begin{figure}[ht]
  \centering
    \includegraphics[width=1.0\linewidth]{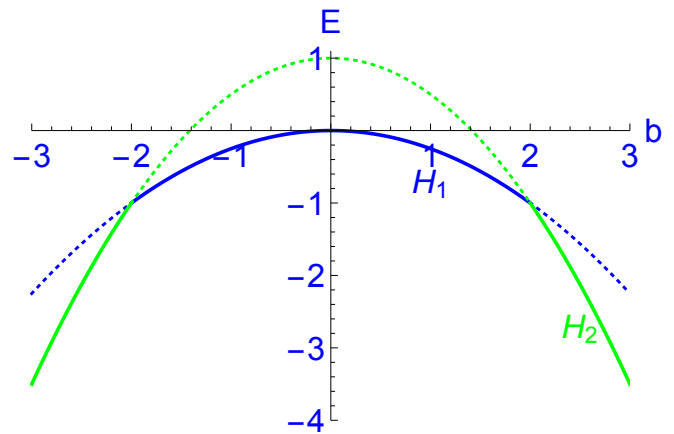}
  \caption[Hmin]
  {The negative Legendre-Fenchel transform $H_{min}(b)$ of $E_{min}(\mu)$. It is the minimum of the two Legendre transforms
  $H_1(b)$ and $H_2(b)$ of $E_1(\mu)$ and $E_2(\mu)$, resp., see Figure \ref{FHZ12}. At $|b|=2$ the graph of $H_{min}$ shows
  a kink and admits there various supporting lines with slopes $\mu$ such that $1\le|\mu|\le 2$.
  }
  \label{FHZ14}
\end{figure}

\begin{figure}[ht]
  \centering
    \includegraphics[width=1.0\linewidth]{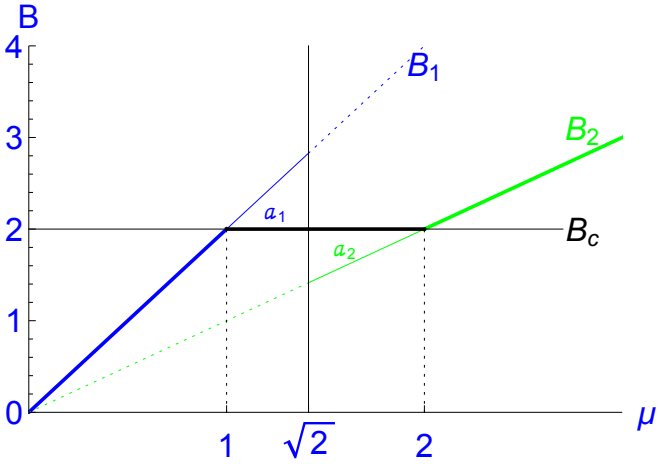}
  \caption[Maxwell]
  {The inverse magnetization functions $B_i(\mu),\;i=1,2$ for the non-convex $E_{min}$ of Figure \ref{FHZ12} and the critical line $B=B_c$. The thick
  line is the physical function according to the Legendre-Fenchel transform. The critical line can also be obtained by the postulate that the
  two areas ${\textit{ a}}_1$ and ${ \textit{a}}_2$ are equal, the so-called Maxwell construction.
  }
  \label{FHZ13}
\end{figure}

The Legendre-Fenchel approach clarifies the situation without any hand-waving.
The supporting lines of $E_{min}$ with non-negative slope fall into three classes:
Those which are tangents to the graph of $E_{min}$ at $0\le\mu<1$ and have a slope $0\le b<2$, or those tangents at $\mu>2$
and slope $b>2$, and the third class consisting of the single
supporting line which is a common tangent of $E_1$ and $E_2$ (the red line in Figure \ref{FHZ12}) having the critical slope $b=B_c=2$.
The latter touches the graph of $E_{min}$ at the two points $u$ and $v$, see  Figure \ref{FHZ12}, and hence illustrates the above remarks that the
intersection of the supporting line and the graph of $E_{min}$ need not occur at a unique $\mu$. The negative Legendre-Fenchel transform
of $E_{min}$ is hence $H_{min}(b)=H_1(b)$ for $|b|<2$ and $H_{min}(b)=H_2(b)$ for $|b|>2$, see Figure \ref{FHZ14}.
For $b=2$ the supporting line intersects
the $E$-axis at $E=-1$, see Figure \ref{FHZ12}, and hence $H_{min}(2)=-1$ which is the intersection of $H_1$ and $H_2$.

It is further interesting to consider the inverse magnetization curve $B={\sf B}(\mu)\Leftrightarrow \mu={\sf M}(B)$.
 We first formally calculate the two functions ${\sf B}_i$ corresponding to $E_i,\;i=1,2$
 according to ${\sf B}_i(\mu)=\frac{\partial E_i}{\partial \mu}$. The result
is ${\sf B}_1(\mu)=2\mu$ and ${\sf B}_2(\mu)=\mu$. From the Legendre-Fenchel approach it is clear which parts of these functions belong
to physical values: ${\sf B}_1(\mu)=2\mu$ corresponds to the first class of supporting lines with $0\le \mu <1$,
${\sf B}_2(\mu)=\mu$ corresponds to the second class of supporting lines for $2 <\mu \le N$, and the constant function ${\sf B}_3(\mu)=B_c=2$
corresponds to the common tangent for $1\le \mu \le 2$, see Figure \ref{FHZ13}. The traditional justification of $B_c=2$
is the so-called Maxwell construction, namely the postulate that the two areas ${\textit{ a}}_1$ and ${ \textit{a}}_2$ bounded
by ${\sf B}_i(\mu),\;i=1,2$, ${\sf B}_3(\mu)=B_c=2$, and $\mu=\mu_0=\sqrt{2}$ are equal, see Figure \ref{FHZ13}. We will show that
the Maxwell construction follows from the condition of a common tangent
\begin{equation}\label{HZ22}
  \frac{E_2(v_1)-E_1(u_1)}{v_1-u_1}=B_c
  \;,
\end{equation}
see Figure \ref{FHZ12}, and the equations ${\sf B}_i(\mu)=\frac{\partial {E_i}}{\partial \mu},\;i=1,2$. The latter implies
\begin{eqnarray}
 \nonumber
  E_2(v_1)-E_1(u_1) &=& \left( E_2(v_1)-E_2(\mu_0)\right)\\
  \label{HZ23a}
  &&+\left(E_1(\mu_0)-E_1(u_1))\right) \\
 \nonumber
   &=& \int_{\mu_0}^{v_1} {\sf B}_2(\mu)d\mu +\int_{u_1}^{\mu_0} {\sf B}_1(\mu)d\mu\\
    \label{HZ23b}
   &&\\
   &\stackrel{(\ref{HZ22})}{=}& B_c \left(v_1-u_1 \right)\\
  \label{HZ23c}
  &=&B_c \left(v_1-\mu_0 \right)+B_c \left(\mu_0-u_1 \right)
  \end{eqnarray}
and hence
\begin{equation}\label{FZ24}
{\textit{ a}}_1=\int_{u_1}^{\mu_0} \left({\sf B}_1(\mu) -B_c\right)d\mu =\int_{\mu_0}^{v_1} \left(B_c-{\sf B}_2(\mu)\right)d\mu ={\textit{ a}}_2.
\end{equation}

We close our toy example by the remark that its magnetization graph (the inverse graph of Figure \ref{FHZ13}) shows
a magnetization jump at $B=B_c=2$ from $\mu=1$ to $\mu=2$ and hence a phase transition of first order.
\hfill$\clubsuit$\\

It follows from the general theory of Legendre-Fenchel transforms that $g={\mathcal L}(f)$ is a closed convex function,
see \cite{R97}, theorem 12.2. For our purposes a function $g$ will be defined as ``closed" iff the epigraph
$\Sigma^g\equiv \{ (x,y)\left| y\ge g(x)\right.\}$ of $g$ is closed.
For the comparison with the definition in \cite{R97} we refer to
theorem 7.1.(c) of \cite{R97} and stress that all functions considered in this paper have finite values and hence are
``proper functions" in the sense of \cite{R97}.
 Moreover, ${\mathcal L}^2(f)=f$ iff $f$ is a closed, convex
function, see \cite{T84} 6.15. This has the consequence that in the above example of a non-convex $E_{min}$, the  Legendre-Fenchel transform
of $-H_{min}$ will not return $E_{min}$ but a suitable defined convex envelope $E_{min}^{(co)}$ of $E_{min}$.
In the following we will provide the details of the definition of $E_{min}^{(co)}$ such that ${\mathcal L}^2(E_{min})=E_{min}^{(co)}$.

First, recall that $\widetilde{E}$ is the compact set of all possible points $(\mu,E)$ such that $\mu=M({\mathbf s})$, and
$E=H_0({\mathbf s})$ where ${\mathbf s}\in {\mathcal P}^{\le 3}$, see Lemma \ref{lemma1}. Let $\widetilde{E}^{(co)}$ denote
the convex hull of $\widetilde{E}$.
\begin{lemma}\label{lemmaE}
$\widetilde{E}^{(co)}$ is a compact subset of ${\mathbbm R}^2$.
\end{lemma}
\noindent{\bf Proof:}\\
$\widetilde{E}^{(co)}$ is the image of the compact set $\widetilde{E}\times \widetilde{E}\times [0,1]$
under the continuous map $A:\widetilde{E}\times \widetilde{E}\times [0,1]\longrightarrow \widetilde{E}^{(co)}$
given by $A(x,y,\lambda)\equiv \lambda x+(1-\lambda)y$, and hence compact.  \hfill$\Box$

\noindent Then we define $E_{min}^{(co)}:[-N,N]\longrightarrow {\mathbbm R}$ by
\begin{defi}\label{defC}
\begin{eqnarray}
\label{defCa}
 E_{min}^{(co)}(\mu) &\equiv& \inf\; \{E\left|  (\mu,E)\in \widetilde{E}^{(co)}\right.\} \\
  \label{defCb}
   &=& \mbox{ Min}\;  \{E\left|  (\mu,E)\in  \widetilde{E}^{(co)}\right.\}\;,
\end{eqnarray}
where the infimum will be attained and hence can be replaced by the minimum since $\widetilde{E}^{(co)}$ is compact.
\end{defi}
$E_{min}^{(co)}$ will be the convex envelope of $E_{min}$ we are seeking for. To verify this we have to prove the following
\begin{lemma}\label{lemmaC}
 $E_{min}^{(co)}$ is a closed, convex function.
\end{lemma}
\noindent{\bf Proof:}\\
We define
\begin{equation}\label{lemmaC1}
 \bar{E}\equiv \{(\mu,E)\left|  \exists E_0: (\mu,E_0)\in \widetilde{E}^{(co)} \mbox{ and } E\ge E_0   \right.\}\;,
\end{equation}
and will show that $\bar{E}$ is a closed, convex set. To prove the latter we consider $(\mu_1,E_1),(\mu_2,E_2)\in\bar{E}$
and $\lambda\in[0,1]$.
There exist $e_1, e_2$ such that $(\mu_i,e_i)\in \widetilde{E}^{(co)}$ and $E_i\ge e_i$ for $i=1,2$.
Let $\mu=\lambda\,\mu_1+(1-\lambda)\,\mu_2$ and $E=\lambda\,E_1+(1-\lambda)\,E_2$, further $e=\lambda\,e_1+(1-\lambda)\,e_2$.
It follows that $(\mu,E)\in\widetilde{E}^{(co)}$ since $\widetilde{E}^{(co)}$ is convex. Moreover,
$E=\lambda\,E_1+(1-\lambda)\,E_2 \ge \lambda\,e_1+(1-\lambda)\,e_2 =e$, hence $(\mu,E)\in\bar{E}$ and $\bar{E}$ is convex.

In order to prove that $\bar{E}$ is closed we first recall that $\mu=\pm N$ and $(\mu,E)\in \widetilde{E}$ implies $E=j\,N$
since  $\mu=M({\mathbf s})=\pm N$ is only realized by the ferromagnetic ground state ${\mathbf s}={\mathbf f}$. The same
conclusion holds for the convex hull of $\widetilde{E}$, namely $\mu=\pm N$ and $(\mu,E)\in \widetilde{E}^{(co)}$ implies $E=j\,N$.
Let $C$ be the epigraph $C=\Sigma^c$ of the constant function $c:[-N,N]\longrightarrow{\mathbbm R},\; c(\mu)=j\,N$.
Obviously, $C$ is closed and the graph of $c$,  $H_c=\{(\mu,j\,N)\left| -N\le \mu\le N\right.\}$ is contained in  $\widetilde{E}^{(co)}$.
We will show that $\bar{E}=\widetilde{E}^{(co)}\cup C$. Obviously,
$\widetilde{E}^{(co)}\subset \bar{E}$ and $C\subset \bar{E}$ hence $\widetilde{E}^{(co)}\cup C\subset \bar{E}$.
Conversely, let $(\mu,E)\in\bar{E}$. If $E\ge j\,N$ then $(\mu,E)\in C$. If $E< j\,N$ then, according to the
definition of $\bar{E}$, there exists an $E_0$ such that $(\mu,E_0)\in \widetilde{E}^{(co)}$ and $j\,N> E\ge E_0$.
Hence $(\mu,E)$ is a point of the line segment between $(\mu,j\,N)\in H_c\subset \widetilde{E}^{(co)}$ and $(\mu,E_0)\in \widetilde{E}^{(co)}$
and thus $(\mu,E)\in\widetilde{E}^{(co)}$ by means of the convexity of $\widetilde{E}^{(co)}$.
Summarizing,  $\bar{E}= \widetilde{E}^{(co)}\cup C$ and hence $\bar{E}$ is closed as the union of two closed sets.

Finally we will show that $\bar{E}$ is the epigraph of  $E_{min}^{(co)}$ which would complete the proof of the lemma.
This follows from the equivalences
\begin{eqnarray}
\label{lemmaC2}
&& (\mu,e)\in \Sigma^{E_{min}^{(co)}} \\
   &\Leftrightarrow& e\ge E_{min}^{(co)}(\mu) \\
   &\stackrel{(\ref{defCb})}{\Leftrightarrow}& e \ge \mbox{Min }\{E\left| (\mu,E)\in \widetilde{E}^{(co)}\right.\}\\
   &\Leftrightarrow& \exists E_0: (\mu,E_0)\in\widetilde{E}^{(co)} \mbox{ and } e\ge E_0\\
    &\stackrel{(\ref{lemmaC1})}{\Leftrightarrow}& (\mu,e)\in \bar{E}\;.
\end{eqnarray}
 \hfill$\Box$\\

\noindent With the preceding definition of $E_{min}^{(co)}$ the following holds:
\begin{prop}\label{propL2}
 ${\mathcal L}^2(E_{min})=E_{min}^{(co)}$\;.
\end{prop}
\noindent{\bf Proof:}
The proposition follows from
${\mathcal L}(E_{min})={\mathcal L}(\widetilde{E})= {\mathcal L}(\widetilde{E}^{(co)})={\mathcal L}(\bar{E})={\mathcal L}(E_{min}^{(co)})$
and the above-mentioned fact that ${\mathcal L}^2(f)=f$ iff $f$ is a closed, convex function. \hfill$\Box$\\

Examples show that $E_{min}$ is not monotonically increasing for $0\le \mu \le N$ in general, see, e.~g., Example \ref{ex3} below.
However, this holds for the convex envelope $E_{min}^{(co)}$ and can even be sharpened to strict monotonicity in a suitable restricted domain:
\begin{prop}\label{propmono}
 Let $\check{\mu}$ be the maximal magnetization of all absolute ground states of the Heisenberg Hamiltonian $H_0$,
 such that $E_{min}^{(co)}(\mu)=e_0$ for all $-\check{\mu}\le \mu \le \check{\mu}$.
 Then it follows that $\mu\mapsto E_{min}^{(co)}(\mu)$ is strictly monotonically increasing for  $\check{\mu}\le \mu\le N$.
\end{prop}
\noindent{\bf Proof:}\\
Let $\check{\mu}\le \mu_1 <\mu_2 \le N$ such that $\mu_1=\lambda\,\check{\mu}+ (1-\lambda)\,\mu_2$ for some $0<\lambda \le 1$.
According to the definition of $\check{\mu}$ and $\mu_2>\check{\mu}$ we have  $e_2\equiv E_{min}^{(co)}(\mu_2)>e_0$.
By convexity of $E_{min}^{(co)}$ we conclude
\begin{eqnarray}
\label{propmono1}
  e_1 &\equiv& E_{min}^{(co)}(\mu_1)\le \lambda\, E_{min}^{(co)}(\check{\mu})+(1-\lambda)\, E_{min}^{(co)}(\mu_2) \\
   &=& \lambda\,e_0+(1-\lambda)\,e_2=e_2 -\lambda\,(e_2-e_0) \\
   &<& e_2\;.
\end{eqnarray}
This proves the strict monotonicity of $E_{min}^{(co)}$ in the domain $[\check{\mu},N ]$.  \hfill$\Box$\\
From $E_{min}(\pm N)=E_{min}^{(co)}(\pm N)=N\,j$ it follows that
\begin{cor}\label{cormono}
 $E_{min}^{(co)}(\mu) < N\,j$ for all $|\mu|\in (\check{\mu},N)$\;.
\end{cor}

It is plausible and follows from the Definition \ref{defferro} below that for ferromagnetic systems we have $\check{\mu}=N$ and hence the domain $(\check{\mu},N )$ where $E_{min}^{(co)}$ is strictly monotonically increasing becomes empty. Hence Proposition \ref{propmono}
is only meaningful for AF systems.

\subsection{Energy bounds and saturation}\label{sec:HZS}

We have seen in the Example 1 that the Heisenberg-Zeeman spin system assumes the ferromagnetic ground state
$\uparrow\uparrow$ if the magnetic field $B$ surpasses a certain value $B_{sat}$, the ``saturation field".
It is plausible that this happens for all Heisenberg-Zeeman systems and we will prove this and generally calculate
the saturation field in the present subsection. As a tool for the proof we will use certain general inequalities for
the energy of Heisenberg-Zeeman systems that are also interesting in their own right.

As we have mentioned in Section \ref{sec:T} and explained in more details in \cite{S17a} the real symmetric matrix of
coupling constants $J_{\mu\nu}$ is not uniquely determined by the Hamiltonian $H_0({\mathbf s})$. The addition of
diagonal values $\lambda_\mu$ such that $\sum_{\mu=1}^{N}\lambda_\mu=0$ does not change $H_0({\mathbf s})$. The choice of the
$\lambda_\mu$ has been called a ``gauge" in \cite{S17a} and the notation ${\mathbbm J}({\boldsymbol\lambda})$
explicitly emphasizes the gauge dependence of the matrix of coupling constants.
Recall that the eigenvalues of ${\mathbbm J}({\boldsymbol\lambda})$ in general non-trivially depend on the gauge.
The most important gauge has hitherto been the ``ground state gauge" (\ref{D2a}).
In this subsection we will make use of another gauge, called ``homogeneous gauge", see also \cite{SL03}, which results
in constant row sums of the homogeneously gauged matrix ${\mathbbm J}^{(h)}$. Given a symmetric matrix
$N\times N$-matrix $\widetilde{J}$
with vanishing trace we define
\begin{eqnarray}
\label{HZS1a}
  \tilde{\jmath}_\mu &=& \sum_{\nu=1}^{N}\widetilde{J}_{\mu\nu},\;\mu=1,\ldots,N,\\
  \label{HZS1b}
  \tilde{\jmath} &=& \frac{1}{N}\sum_{\mu=1}^{N}\tilde{\jmath}_\mu,\\
  \label{HZS1c}
  \lambda_\mu &=& \tilde{\jmath}-\tilde{\jmath}_\mu,\;\mu=1,\ldots,N,\\
  \label{HZS1d}
  J_{\mu\nu}&=& \widetilde{J}_{\mu\nu}-\lambda_\mu\,\delta_{\mu\nu},\;\mu,\nu=1,\ldots,N.
\end{eqnarray}
The matrix with entries (\ref{HZS1d}) will  be denoted by ${\mathbbm J}^{(h)}$.
It has the property that its row sum
\begin{equation}\label{HZS2}
j\equiv \sum_{\mu=1}^{N}J_{\mu\nu}
\end{equation}
will be independent of $\nu$ and equals the mean row sum of $\widetilde{J}$, i.~e., $j=\tilde{\jmath}$. \cite{foot2}\\

$\clubsuit$ Consider, for example, the $3\times 3$-matrix
\begin{equation}\label{HZS3}
  \widetilde{J}=\left(\begin{array}{ccc}
 0 & J_3 & J_2 \\
 J_3 &0 &J_1\\
 J_2& J_1 & 0 \\
\end{array}\right).
\end{equation}
Its mean row sum is $\tilde{\jmath}=\frac{2}{3}\left( J_1+J_2+J_3\right)$. If we subtract in each diagonal element
the actual row sum and add the mean row sum, we obtain the homogeneously gauged matrix
\begin{equation}\label{HZS4}
\mathbbm{J}^{(h)}=\left(\begin{array}{ccc}
 \frac{2J_1-J_2-J_3}{3} & J_3 & J_2 \\
 J_3 &\frac{2J_2-J_1-J_3}{3} &J_1\\
 J_2& J_1 & \frac{2J_3-J_1-J_2}{3} \\
\end{array}\right)
\end{equation}
with vanishing trace and the constant row sum $j=\tilde{\jmath}$. \hfill$\clubsuit$\\

In the remainder of this subsection we will always assume that the matrix $\mathbbm{J}=\mathbbm{J}^{(h)}$ is homogeneously
gauged with constant row sum $j$. It follows that the constant vector ${\mathbf f}=(1,1,\ldots,1)\in{\mathbbm R}^N$
will be an eigenvector of ${\mathbbm J}$ with eigenvalue $j$, i.~e.,
\begin{equation}\label{HZS5}
  {\mathbbm J}\,{\mathbf f}=j\,{\mathbf f}
  \;.
\end{equation}

 We will first obtain the following linear upper bound:
\begin{prop}\label{propLI}
 For all $B\in {\mathbbm R}$ there holds
 \begin{equation}\label{HZS5}
   H_{min}(B)\le N(j-|B|)
   \;.
 \end{equation}
\end{prop}

\noindent{\bf Proof:}\\
It suffices to prove the claim for $B\ge 0$.
We note that the eigenvector ${\mathbf f}=(1,1,\ldots,1)\in{\mathbbm R}^N$ of  ${\mathbbm J}$ can also be
viewed as the spin configuration representing the ferromagnetic ground state that has the maximal magnetization $M({\mathbf f})=N$. Hence
\begin{eqnarray}
\label{HZS6a}
   H_{min}(B) &\le& H_0({\mathbf f}) - M({\mathbf f})\,B\\
 &=& {\mathbf f}\cdot {\mathbbm J}\,{\mathbf f}- N\,B\\
 &\stackrel{(\ref{HZS5})}{=}& j\, {\mathbf f}\cdot {\mathbf f}- N\,B =  N(j-B)
 \;.
\end{eqnarray}
 \hfill$\Box$\\

The bounding line given by the equation\\
$E= N(j-B)$ will also be referred to as the ``ferromagnetic line" due to the nature of the
state ${\mathbf f}$. Next we will prove a parabolic lower bound for  $H_{min}$. Let $j_{min}^{(h)}$ denote the lowest eigenvalue
of ${\mathbbm J}$, where the superscript ``$h$" reminds us that ${\mathbbm J}$ is homogeneously gauged. Thus, by definition,
$j_{min}^{(h)}\le j$, but for the lower bound we need the stronger condition $j_{min}^{(h)}< j$:
\begin{theorem}\label{theorem2}
If $j_{min}^{(h)}< j$ then the following holds for all $B\in {\mathbbm R}$ :
\begin{equation}\label{HZS6}
   H_{min}(B)\ge H_{bound}(B)\equiv j_{min}^{(h)}\,N-\frac{N\,B^2}{4\left(j-j_{min}^{(h)}\right)}
   \;.
 \end{equation}
\end{theorem}
\noindent{\bf Proof A:}\\
Rayleigh's principle yields
\begin{equation}\label{HZS7}
 \sum_{\mu,\nu=1}^{N}{\mathbbm J}_{\mu\nu}x_\mu x_\nu \ge j_{min}^{(h)}\sum_{\mu=1}^{N}x_\mu^2
 \;,
\end{equation}
for any vector ${\mathbf x}=(x_1,x_2,\ldots,x_N)\in {\mathbbm R}^N$. Choosing
$x_\mu={\mathbf s}_\mu^{(i)}-\beta\,{\mathbf B}^{(i)}$, where $\beta\in{\mathbbm R}$ and ${\mathbf s}\in{\mathcal P}_{\mathbf e}$ are arbitrary,
and summing over $i=1,2,3$ yields
\begin{eqnarray}\label{HZS8a}
g&\equiv& \sum_{\mu,\nu=1}^{N}{\mathbbm J}_{\mu\nu} \left({\mathbf s}_\mu-\beta\,{\mathbf B}\right)\cdot \left({\mathbf s}_\nu-\beta\,{\mathbf B}\right)\\
\label{HZS8b}
&\stackrel{(\ref{HZS7})}{\ge}& j_{min}^{(h)}\sum_{\mu=1}^{N}\left({\mathbf s}_\mu-\beta\,{\mathbf B}\right)^2
\;.
\end{eqnarray}
Expanding the dot product in (\ref{HZS8a}) and (\ref{HZS8b}) gives
\begin{eqnarray}\nonumber
g&=& \sum_{\mu,\nu=1}^{N}{\mathbbm J}_{\mu\nu} {\mathbf s}_\mu\cdot{\mathbf s}_\nu\\
\label{HZS9a}
&&-2\beta{\mathbf B}\cdot\sum_{\mu,\nu=1}^{N} {\mathbbm J}_{\mu\nu} {\mathbf s}_\mu + \beta^2 B^2 j N\\
\label{HZS9b}
&\ge& j_{min}^{(h)}\left( N+N\beta^2 B^2 +2 \beta\, {\mathbf B}\cdot {\mathbf S}\right)
\;.
\end{eqnarray}
Using $\sum_{\mu,\nu=1}^{N}{\mathbbm J}_{\mu\nu} {\mathbf s}_\mu\cdot{\mathbf s}_\nu=H_0({\mathbf s})$
and $\sum_{\mu,\nu=1}^{N}{\mathbbm J}_{\mu\nu} {\mathbf s}_\mu=j\,{\mathbf S}$
we rewrite the inequality (\ref{HZS9a}), (\ref{HZS9b}) as
\begin{eqnarray}
\nonumber
  H_0({\mathbf s})-2\beta  (j-j_{min}^{(h)}){\mathbf B}\cdot {\mathbf S} &\ge&  \\
  \label{HZS10a}
  j_{min}^{(h)}N\left( 1+\beta^2B^2\right)-j N \beta^2 B^2. &&
\end{eqnarray}
The l.~h.~s.~of (\ref{HZS10a}) just equals\\ $H_0({\mathbf s}) -{\mathbf B}\cdot {\mathbf S}=H_0({\mathbf s}) -B\, M({\mathbf s})=H({\mathbf s}) $
if we choose
\begin{equation}\label{HZS11}
 \beta=\frac{1}{2\left( j- j_{min}^{(h)}\right)}.
\end{equation}
Then  (\ref{HZS10a}) simplifies to
\begin{eqnarray}\label{HZS12a}
 H({\mathbf s})&\ge&  j_{min}^{(h)}N-\left( j- j_{min}^{(h)}\right)N\beta^2 B^2\\
 \label{HZS12b}
 &=&j_{min}^{(h)}\,N-\frac{N\,B^2}{4\left(j-j_{min}^{(h)}\right)}
\;,
\end{eqnarray}
which says that, for given $B$, $j_{min}^{(h)}\,N-\frac{N\,B^2}{4\left(j-j_{min}^{(h)}\right)}$ is a lower bound
of the energy of the Heisenberg-Zeeman system for arbitrary ${\mathbf s}\in{\mathcal P}_{\mathbf e}$. 
Hence it is also a lower bound of the minimum $H_{min}(B)$, which yields (\ref{HZS6}).
\hfill$\Box$\\

\noindent{\bf Proof B:}\\
It will be illustrative to sketch an alternative proof of Theorem \ref{theorem2}. It is based on the inequality
\begin{equation}\label{ProofB1}
  E_{min}(\mu)\ge E_{bound}(\mu)\equiv j_{min}^{(h)}\,N +\frac{j-j_{min}^{(h)}}{N}\mu^2\;,
\end{equation}
that holds for all $-N\le\mu\le N$, see \cite{SL03} Theorem $1$. Its proof can be found in \cite{SL03} and need not be repeated here.
Since $E_{bound}$ is smooth we may calculate its negative Legendre-Fenchel transform $H_1$ in the following way:
\begin{eqnarray}
\label{ProofB2a}
 b(\mu) &\equiv& \frac{\partial E_{bound}}{\partial \mu} = 2\,\frac{j-j_{min}^{(h)}}{N}\mu,\\
 \label{ProofB2b}
\Rightarrow \mu(b) &=& \frac{N}{2\left(j-j_{min}^{(h)}\right)}b,\\
  H_1(b) &=& E_{bound}\left(\mu(b) \right)-\mu(b)\,b\\
\nonumber
 &=&  j_{min}^{(h)}N + \frac{j-j_{min}^{(h)}}{N} \left( \frac{N\,b}{2\left(j-j_{min}^{(h)}\right)} \right)^2 \\
 \label{ProofB2c}
 && - \,\frac{N}{2\left(j-j_{min}^{(h)}\right)}b^2\\
\nonumber
 &=&  j_{min}^{(h)}N  - \frac{N}{4\left(j-j_{min}^{(h)}\right)}b^2=H_{bound}(b).\\
  \label{ProofB2d}
 &&
 \end{eqnarray}
 The monotonicity of the negative Legendre-Fenchel transform, see Lemma \ref{corT1}, then implies
 $H_{min}\ge H_1=H_{bound}$ which is (\ref{HZS6}).   \hfill$\Box$\\

$\clubsuit$ For the above example of the AF dimer the matrix
${\mathbbm J}=\frac{1}{2}\left(\begin{array}{cc}
 0 &1 \\
 1 &0
\end{array}\right)$
is already homogeneously gauged and $j=\frac{1}{2}$ whereas $j_{min}^{(h)}=-\frac{1}{2}$. It follows that (\ref{HZS6})
assumes the form $H_{min}(B)\ge -1 -\frac{1}{2}B^2$. Comparison with (\ref{HZ8}) shows that this lower bound is assumed
for $|B|\le 2$ whereas for $|B|\ge 2$ the upper bound
$N(j-|B|)=2(\frac{1}{2}-|B|)=1-2|B|$, see (\ref{HZS5}), is assumed.
\hfill$\clubsuit$\\

Generally, the ferromagnetic line given by\\ $E=N(j-B)$ will be the tangent to the lower  parabola given by
$H_{bound}(B)=j_{min}^{(h)}\,N-\frac{N\,B^2}{4\left(j-j_{min}^{(h)}\right)}$ at the point
\begin{equation}\label{HZS13}
(B_0,E_0)=\left(2\left(j-j_{min}^{(h)}\right),N\left(2j_{min}^{(h)}-j \right)\right)
\;.
\end{equation}
One may ask whether the ferromagnetic line is always assumed by $H_{min}(B)$ for $B\ge B_0=2\left(j-j_{min}^{(h)}\right)$
and whether $B_0$ is the smallest value of $B$ where this happens. In this case it would be legitimate to call  $B_0$ the ``saturation field".
Moreover, what happens in the case $j=j_{min}^{(h)}$ where $B_0$ vanishes?
The following proposition answers these questions.
\begin{prop}\label{propBsat}
 (i) If $j=j_{min}^{(h)}$ then
 \begin{equation}\label{propBsat1}
   H_{min}(B)=N(j-|B|)\mbox{ for all } B\in{\mathbbm R}
   \;.
 \end{equation}
 (ii) If $j>j_{min}^{(h)}$ then
 \begin{equation}\label{propBsat2}
   H_{min}(B)=N(j-|B|)\mbox{ for all } |B|\ge B_0
   \;.
 \end{equation}
 (iii) If $j>j_{min}^{(h)}$ then
 \begin{equation}\label{propBsat3}
   H_{min}(B)<N(j-|B|)\mbox{ for all } |B|< B_0
   \;.
 \end{equation}
\end{prop}

\noindent{\bf Proof:}\\
(i) By applying Rayleigh's principle one shows analogously as in the proof A of Theorem \ref{theorem2} that
for all ${\mathbf s}\in{\mathcal P}_{\mathbf e}$:
\begin{equation}\label{HZS14}
  H_0({\mathbf s})\ge N\,j_{min}^{(h)}=N\,j
  \;,
\end{equation}
and hence
\begin{equation}\label{HZS15}
  H_0({\mathbf s})-M({\mathbf s})\,B\ge N\,j_{min}^{(h)}-M({\mathbf s})\,B\ge N(j-|B|)
  \;,
\end{equation}
since $|M({\mathbf s})|\le N$. Hence $H_{min}(B)\ge N(j-|B|)$.
The converse inequality holds by Proposition \ref{propLI}.\\

(ii) In order to derive a contradiction we assume that (\ref{propBsat2}) does not hold.
It suffices to consider the case where there exists a $B\ge B_0$ and a spin configuration ${\mathbf s}\in{\mathcal P}_{\mathbf e}$
such that
\begin{equation}\label{HZS16}
  H_0({\mathbf s})-M({\mathbf s})\,B < N(j-B)
  \;.
\end{equation}
${\mathbf s}$ cannot be the ferromagnetic ground state ${\mathbf f}$, hence
\begin{equation}\label{HZS16a}
 M({\mathbf s})<N
 \;.
\end{equation}
Adding the obvious inequality
\begin{equation}\label{HZS17}
  H_0({\mathbf s})-M({\mathbf s})\,B_0 \ge H_{min}(B_0)=N(j-B_0)
\end{equation}
and the negative of (\ref{HZS16})
\begin{equation}\label{HZS18}
  -H_0({\mathbf s})+M({\mathbf s})\,B > N(B-j)
  \;,
\end{equation}
we obtain
\begin{equation}\label{HZS19}
 M({\mathbf s})\left(B-B_0 \right)> N(B-B_0)
 \;,
\end{equation}
which, by virtue of $B\ge B_0$, contradicts (\ref{HZS16a}).\\

(iii)
It suffices to consider the case $B>0$.
Let ${\mathbf x}\in{\mathbbm R}^N$ be an eigenvector of ${\mathbbm J}$ corresponding to the eigenvalue
$j_{min}^{(h)}$. According to $j_{min}^{(h)}<j$, ${\mathbf x}$ will be orthogonal to the eigenvector
${\mathbf f}=(1,1,\ldots,1)$ of ${\mathbbm J}$ corresponding to the eigenvalue $j$, i.~e.,
\begin{equation}\label{HZS20}
 \sum_{\mu=1}^N {x}_\mu=0.
\end{equation}

Let $t\mapsto{\mathbf s}(t)$ be the parametrization of a smooth curve in ${\mathcal P}^{\le 3}$
defined on an open interval $-\epsilon<t<\epsilon$ that satisfies
\begin{equation}\label{HZS21}
 {\mathbf s}_\mu(0)={\mathbf e}_3\equiv \left(\begin{array}{c}
                                                0 \\
                                                0 \\
                                                1
                                              \end{array}\right),\;
 \dot{\mathbf s}_\mu(0)=                 \left(\begin{array}{c}
                                                0 \\
                                                {x}_\mu \\
                                                0
                                              \end{array}\right)
\end{equation}
for all $\mu=1,\ldots,N$.
Differentiating ${\mathbf s}_\mu(t)\cdot {\mathbf s}_\mu(t)=1$ we obtain
\begin{equation}\label{HZS22a}
\dot{\mathbf s}_\mu(t)\cdot {\mathbf s}_\mu(t)=0,
\end{equation}
and
\begin{equation}\label{HZS22b}
\ddot{\mathbf s}_\mu(t)\cdot {\mathbf s}_\mu(t)+\dot{\mathbf s}_\mu(t)\cdot \dot{\mathbf s}_\mu(t)=0
\end{equation}
for all $\mu=1,\ldots,N$. We consider the Taylor expansion of $H({\mathbf s}(t))$ at $t=0$ up to terms of second order:
\begin{eqnarray}
\label{HZS23a}
  H({\mathbf s}(t)) &=& \sum_{\mu,\nu=1}^{N} {\mathbbm J}_{\mu\nu}{\mathbf s}_\mu(t)\cdot{\mathbf s}_\nu(t)
  -  \sum_{\mu=1}^{N}{\mathbf s}_\mu(t)\cdot{\mathbf B}\\
  \label{HZS23b}
&\equiv& H^{(0)}+t\,H^{(1)}+\frac{t^2}{2}\,H^{(2)}+{\mathcal O}(t^3).\\
\label{HZS23c}
H^{(0)}&=&\sum_{\mu,\nu=1}^{N} {\mathbbm J}_{\mu\nu}{\mathbf s}_\mu(0)\cdot{\mathbf s}_\nu(0)
  -  \sum_{\mu=1}^{N}{\mathbf s}_\mu(0)\cdot{\mathbf B}\\
\label{HZS23d}
&=&\sum_{\mu,\nu=1}^{N}{\mathbbm J}_{\mu\nu}{\mathbf e}_3\cdot{\mathbf e}_3-B \sum_{\mu=1}^{N}{\mathbf e}_3\cdot{\mathbf e}_3\\
\label{HZS23e}
&=& N(j-B)\;.\\
\nonumber
H^{(1)}&=&2\sum_{\mu,\nu=1}^{N} {\mathbbm J}_{\mu\nu}\dot{\mathbf s}_\mu(0)\cdot{\mathbf s}_\nu(0)
  -  B\,\sum_{\mu=1}^{N}\dot{\mathbf s}_\mu(0)\cdot{\mathbf e}_3\\
  &&\\
\label{HZS23f}
&=&0 \mbox{ by (\ref{HZS21}).}
\end{eqnarray}
\begin{eqnarray}
\nonumber
H^{(2)}&=&2\sum_{\mu,\nu=1}^{N} {\mathbbm J}_{\mu\nu}\left(\ddot{\mathbf s}_\mu(0)\cdot{\mathbf s}_\nu(0)+
\dot{\mathbf s}_\mu(0)\cdot\dot{\mathbf s}_\nu(0)\right)\\
\label{HZS23g}
&&-\sum_{\mu=1}^N\ddot{\mathbf s}_\mu(0)\cdot{\mathbf B}\\
&\stackrel{(\ref{HZS21})}{=}&2\sum_{\mu,\nu=1}^{N} {\mathbbm J}_{\mu\nu}\left(\ddot{\mathbf s}_\mu(0)\cdot{\mathbf s}_\mu(0)
+x_\mu\,x_\nu\right)\\
&&-B\sum_{\mu=1}^N\ddot{\mathbf s}_\mu(0)\cdot{\mathbf e}_3\\
\nonumber
&\stackrel{(\ref{HZS22b})}{=}&
2\sum_{\mu,\nu=1}^{N} {\mathbbm J}_{\mu\nu}\left(-\dot{\mathbf s}_\mu(0)\cdot\dot{\mathbf s}_\mu(0)+x_\mu\,x_\nu\right)\\
\label{HZS23h}
&&+B\sum_{\mu=1}^N\dot{\mathbf s}_\mu(0)\cdot\dot{\mathbf s}_\mu(0)\\
\nonumber
&\stackrel{(\ref{HZS21})}{=}& -2\sum_{\mu,\nu=1}^{N} {\mathbbm J}_{\mu\nu}x_\mu^2+2\,j_{min}^{(h)} \sum_{\mu=1}^N x_\mu^2+B\,\sum_{\mu=1}^N x_\mu^2\\
\label{HZS23i}
&&\\
&=& \left( B+2\left(j_{min}^{(h)}-j\right)\right)\sum_{\mu=1}^N x_\mu^2\\
\label{HZS23j}
&\stackrel{(\ref{HZS13})}{=}& \left(B-B_0\right)\sum_{\mu=1}^N x_\mu^2\;<0\;,
\end{eqnarray}
since $B<B_0$ and $\sum_{\mu=1}^N x_\mu^2>0$. In (\ref{HZS23i}) and (\ref{HZS23j}) we have used the above assumption ${\mathbbm J}{\mathbf x}=j_{min}^{(h)}{\mathbf x}$.\\
Summarizing, the zeroth order of $H({\mathbf s}(t))$ at $t=0$ equals the ferromagnetic line, the first order vanishes and the second
order is negative for $B<B_0$. Hence, for sufficiently small $t$,
\begin{equation}\label{HZS22}
 H({\mathbf s}(t))< N(j-B)
 \;,
\end{equation}
and thus also
\begin{equation}\label{HZS23}
 H_{min}(B)< N(j-B)
 \;.
\end{equation}
This completes the proof of (iii). \hfill$\Box$\\

Proposition \ref{propBsat} (ii) and (iii) justify the following definition of the saturation field \cite{foot3}:
\begin{defi}\label{defsat}
\begin{equation}\label{HZS24}
 B_{sat}\equiv 2\left(j-j_{min}^{(h)}\right)
 \;,
\end{equation}
\end{defi}
such that, according to (\ref{HZS13}),
\begin{equation}\label{HZS24a}
 H_{min}\left(B_{sat}\right)=N\left(2j_{min}^{(h)}-j \right)
 \;.
\end{equation}

Moreover, \ref{propBsat} (i) shows that for all spin systems satisfying $j=j_{min}^{(h)}$ or, equivalently, $ B_{sat}=0$,
the graph of  $ H_{min}$ is completely given by the ferromagnetic lines $E(B)=N(j-|B|)$ and hence rather uninteresting.
For these systems $\pm{\mathbf f}$ will be the ground state for all $B\in{\mathbbm R}$.
Hence the following definition appears sensible:
\begin{defi}\label{defferro}
We will denote a Heisenberg-Zeeman system as ``ferromagnetic" iff $j=j_{min}^{(h)}$ and as ``anti-ferromagnetic" iff $j>j_{min}^{(h)}$.
\end{defi}
In the literature the term ``ferromagnetic" is sometimes reserved to denote the case where all $J_{\mu\nu}\le 0,\;\mu\neq \nu$.
In this case our condition $j=j_{min}^{(h)}$ follows by means of the Frobenius-Perron theorem, see, e.~g., \cite{N76}.

It will be interesting to compare the condition $j=j_{min}^{(h)}$ characterizing ferromagnetic systems with the
inequality
\begin{equation}\label{HZS25}
 J_1+J_2+J_3 \le -||{\mathbf J}||\equiv-||(J_1,J_2,J_3)||\;,
\end{equation}
characterizing the ferromagnetic cone of ${\mathbf J}$-vectors in the case of $N=3$, that was derived in \cite{S17c}. The coupling constants
$J_1,J_2,J_3$
are those appearing in (\ref{HZS3}) except for a factor $2$, which is irrelevant since (\ref{HZS25}) is homogeneous in the $J_i,\;i=1,2,3$.
\begin{lemma}\label{lemma4}
 In the case of $N=3$, $j=j_{min}^{(h)}$ is equivalent to (\ref{HZS25}).
\end{lemma}

\noindent{\bf Proof:}\\
The condition $j=j_{min}^{(h)}$ is equivalent to the statement $\widehat{\mathbbm J}\equiv{\mathbbm J}-j\,{\mathbbm 1}\ge 0$. The latter matrix
$\widehat{\mathbbm J}$ is obtained from $\widetilde{J}$, the matrix of coupling coefficients in the zero diagonal gauge, by subtracting the actual row sum from each diagonal
element. In the case $N=3$, $\widehat{\mathbbm J}$ hence assumes the form
\begin{equation}\label{HZS26}
\widehat{\mathbbm J}=\left(\begin{array}{ccc}
 -J_2-J_3  & J_3 & J_2 \\
 J_3 &-J_1-J_3 &J_1\\
 J_2& J_1 & -J_1-J_2  \\
\end{array}\right).
\end{equation}
By Sylvester's criterion, $\widehat{\mathbbm J}\ge 0$ iff all principal minors of $\widehat{\mathbbm J}$ are $\ge 0$.
Since $\det \widehat{\mathbbm J}=0$ and all principal minors of order two have the same value, the condition $j=j_{min}^{(h)}$ will
be equivalent to the conjunction of the inequalities
\begin{eqnarray}
\label{HZS27a}
  -J_2-J_3 &\ge& 0, \\
  \label{HZS27b}
  -J_1-J_3 &\ge& 0, \\
  \label{HZS27c}
   -J_1-J_2  &\ge& 0, \\
  \label{HZS27d}
   J_1 J_2+J_2 J_3 +J_1 J_3 &\ge& 0\;.
\end{eqnarray}
In order to show that  $\widehat{\mathbbm J}\ge 0$ implies (\ref{HZS25}) we note that
(\ref{HZS27a}) -- (\ref{HZS27c})  imply $J_1+J_2+J_3\le 0$. Under this condition, (\ref{HZS25}) is equivalent to
$\left(J_1+J_2+J_3\right)^2  \ge  ||{\mathbf J}||^2$ and, further, to
(\ref{HZS27d}) since
\begin{eqnarray}
 \label{HZS28a}
 \left(J_1+J_2+J_3\right)^2  &\ge & ||{\mathbf J}||^2 \\
 \nonumber
 \Leftrightarrow\quad J_1^2+J_2^2+J_3^2&+&2(J_1 J_2+J_2 J_3+J_1 J_3)\\
 \label{HZS28b}
 &\ge& J_1^2+J_2^2+J_3^2 \\
 \label{HZS28c}
 \Leftrightarrow\quad J_1 J_2+J_2 J_3+J_1 J_3 &\ge&0.
\end{eqnarray}

Conversely, (\ref{HZS25}) implies $J_1+J_2\le -||{\mathbf J}||-J_3\le 0$, where the last inequality is trivial
for $J_3\ge 0$ and follows from the triangle inequality if $J_3\le 0$. Hence $-J_1-J_2\ge 0$ and the  other two
inequalities (\ref{HZS27a}) and (\ref{HZS27b}) follow analogously. The remaining part of the proof is analogous
to the above considerations.
\hfill$\Box$\\

As another consequence of the preceding results we note the following
\begin{prop}\label{propINT}
Let the magnetization graph coincide with the graph of a smooth magnetization function ${\sf M}(b)$  except for
a finite set of arguments such that $b\mapsto {\sf M}(b)$ is integrable. Then
\begin{equation}\label{INT1}
\int_{0}^{B_{sat}}{\sf M}(b)\,db = e_0-N\left(2\,j_{min}^{(h)}-j\right)\;,
 \end{equation}
 where $e_0\equiv H_{min}(0)$ is the absolute ground state energy.
\end{prop}

\noindent{\bf Proof:}\\
We first consider the integral for an open interval $(b_1,b_2)$ where ${\sf M}(b)$ is smooth and hence the
negative derivative of a smooth part of $H_{min }(b)$. Hence
$\int_{b_1}^ {b_2}{\sf M}(b)\,db= H_{min }(b_1)-H_{min }(b_2)$. By adding the integrals over the whole domain
$[0,B_{sat}]$ we obtain
$\int_{0}^{B_{sat}}{\sf M}(b)\,db = H_{min}(0)-H_{min}(B_{sat})= e_0-N\left(2\,j_{min}^{(h)}-j\right)$\;,
using (\ref{HZS24a}) and the continuity of $H_{min}$.
\hfill$\Box$\\

\subsection{Magnetization plateaus}\label{sec:HZM}
Recall that in general the  Legendre-Fenchel transform
of $-H_{min}$ will not return $E_{min}$ but its convex envelope  $E_{min}^{(co)}$, see Definition \ref{defC}.
Nevertheless, we have a kind of duality between $\mu\mapsto E_{min}(\mu)$ and $b\mapsto H_{min}(b)$ that can be utilized
for various purposes. For example, a kink in the graph of $H_{min}$, see Figure \ref{FHZ14}, is connected with a linear part
in (the convex envelope of) $E_{min}$ , see Figure \ref{FHZ12}, and a magnetization jump, see Figure \ref{FHZ13}. By duality, a kink in the graph
of $E_{min}$ corresponds to a linear part of $H_{min}$, and, interchanging the role of $\mu$ and $b$, to a magnetization plateau
in the magnetization graph ${\mathcal M}$. A first example of this we have already encountered, namely the ferromagnetic line
assumed by $H_{min}$ for $B\ge B_{sat}$, corresponding to the ``kink" of $E_{min}$ at $\mu=N$
and the magnetization plateau for $B\ge B_{sat}$. However, this example is
somewhat degenerate and less interesting since it occurs for every AF system. Hence we will provide another, still elementary example
for a magnetization plateau.

\vspace{5mm}
\begin{ex}\label{ex3} The AF $3$-chain ($N=3$)\\
\begin{center}
\unitlength1cm
\begin{picture}(8,1)(0,0)
\thicklines
 \put(0.5,0.5){\circle{0.5}}
  \put(2.5,0.5){\circle{0.5}}
  \put(4.5,0.5){\circle{0.5}}
  \textcolor{red}{ \put(0.75,0.5){\line(1,0){1.5}}
   \put(2.75,0.5){\line(1,0){1.5}}}
\end{picture}
\end{center}
\end{ex}
\vspace{5mm}

$\clubsuit$
\begin{figure}[ht]
  \centering
    \includegraphics[width=1.0\linewidth]{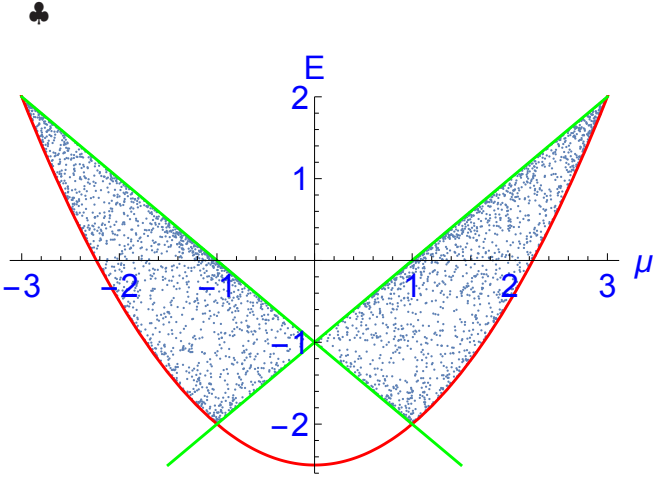}
  \caption[Emin]
  {The set of possible values of $(\mu,E)=\left( \pm \sqrt{{\mathbf S}^2},H_0({\mathbf s})\right)$ for the AF $3$-chain.
  The points correspond to randomly chosen values for $\alpha$ and $\beta$ in (\ref{HZM3a}) and (\ref{HZM3b});
  the bounding (green) lines and the (red) parabola are given by (\ref{HZM4}) and (\ref{HZM5}).
  The function  $E_{min}(\mu)$ is locally given by one of the linear or quadratic bounds.
  }
  \label{FHZM1}
\end{figure}

The anti-ferromagnetic $3$-chain will be defined by the Hamiltonian
\begin{equation}\label{HZM1}
 H({\mathbf s})={\mathbf s}_1\cdot{\mathbf s}_3+{\mathbf s}_2\cdot{\mathbf s}_3-{\mathbf S}\cdot{\mathbf B}
 \;.
 \end{equation}
We first calculate $E_{min}(\mu)$. According to the results of \cite{S17c} for $N=3$ it suffices to consider co-planar
spin configurations. Due to rotational degeneracy the general co-planar spin configuration can be assumed to be of the form
\begin{equation}\label{HZM2}
 {\mathbf s}_3={1\choose 0},\; {\mathbf s}_1={\cos\alpha\choose \sin\alpha},\; {\mathbf s}_2={\cos\beta\choose \sin\beta}\;,
\end{equation}
where $\alpha\in[0,\pi],\,\beta\in[0,2\pi]$. It follows that
\begin{eqnarray}
\nonumber
  M({\mathbf s}) &=&\pm \sqrt{3+2 \cos \alpha+2 \cos (\alpha -\beta )+2 \cos \beta }, \\
  \label{HZM3a}
  &&\\
  \label{HZM3b}
  H_0({\mathbf s})&=& \cos\alpha+\cos\beta\;.
\end{eqnarray}
The set of points with coordinates $\left(M(\alpha,\beta),H_0(\alpha,\beta)\right)$ exhausts the set $\widetilde{E}$ introduced in (\ref{HZ10a}),
if $(\alpha,\,\beta)$ runs through the set $[0,\pi]\,\times\,[0,2\pi]$. The boundary of $\widetilde{E}$ is contained
in the set of solutions of ${\sf Jac}(\alpha,\beta)=0$, where ${\sf Jac}$ denotes the Jacobian of the transformation
$(\alpha,\beta)\mapsto (M^2(\alpha,\beta),H_0(\alpha,\beta))$. After some elementary calculations we obtain
${\sf Jac}=2 \sin (\alpha -\beta ) (\sin \alpha +\sin \beta )$ and hence ${\sf Jac}(\alpha,\beta)=0$ has the two
solutions $\alpha=\beta$ and $\alpha=-\beta$. The first one leads to $\mu=M(\alpha,\alpha)=\pm \sqrt{4 \cos \alpha +5}$
and $E=H_0(\alpha,\alpha)=2\cos\alpha$, and hence to the parabola
\begin{equation}\label{HZM4}
 E=\frac{1}{2}\left(\mu^2 - 5\right)
 \;.
\end{equation}
The second solution leads to $\mu=\pm M(\alpha,-\alpha)=\pm (1+ 2\cos \alpha )$
and $E=H_0(\alpha,-\alpha)=2\cos\alpha$, and hence to the two lines
\begin{equation}\label{HZM5}
E_{\pm}=\pm\mu- 1
 \;.
\end{equation}

In Figure \ref{FHZM1} we have displayed a number of values $(\mu,E)$ obtained numerically by randomly choosing $\alpha$ and $\beta$
together with the bounding lines (\ref{HZM5}) and the parabola (\ref{HZM4}).
From this it is obvious that the function $E_{min}$ is given by
\begin{equation}\label{HZM6}
 E_{min}(\mu)=\left\{\begin{array}{r@{\quad:\quad}l}
  -1-|\mu|&0\le |\mu|\le 1\;,\\
\frac{1}{2} \left(\mu ^2-5\right)& 1\le |\mu|\le 3\;.
  \end{array} \right.
\end{equation}

There are, up to the reflection $\mu\mapsto -\mu$,
three prominent points of $\widetilde{E}$: The point with coordinates $(\mu=0,E=-1)$ corresponding to the coplanar spin state
with mutual angles of $120^\circ$, the point $(\mu=1,E=-2)$ corresponding to the total Ising ground state $\downarrow\downarrow\uparrow$,
and the point $(\mu=3,E=2)$  corresponding to the ferromagnetic ground state $\uparrow\uparrow\uparrow$.
$\widetilde{E}$ is not convex; the relative ground states ${\mathbf s}$ with $-1<M({\mathbf s})<1$ will never become
ground states of the Heisenberg-Zeeman Hamiltonian since they cannot be reached by supporting lines. Hence there will be
a magnetization jump from $\mu=-1$ to $\mu=1$ at $B=0$.

Moreover, the point $(\mu=1,E=-2)$ is a kink of $E_{min}$, even if $E_{min}$ is replaced by its convex envelope  $E_{min}^{(co)}$.
This leads to a magnetization plateau: The magnetization has the constant value $\mu=1$ if $B$ varies from $B_1=0$
to $B_2=1$. The latter value is obtained from \\
$B_2=\lim_{\mu\downarrow 1}\frac{\partial E_{min}}{\partial \mu}=
\lim_{\mu\downarrow 1}\frac{\partial }{\partial \mu} \frac{1}{2}\left(\mu ^2-5\right)=1$.

$H_{min}$ is obtained as the negative Legendre-Fenchel transform of $E_{min}$. Note that for $1< |\mu|<3$
this can be calculated as the Legendre transform in the traditional way:
$b=\frac{\partial }{\partial \mu} E_{min}(\mu)=\frac{\partial }{\partial \mu} \frac{1}{2} \left(\mu ^2-5\right)=\mu(b)$
and hence
$H_{min}(b)=E_{min}(\mu(b))-\mu(b)\,b=\frac{1}{2} \left(b^2-5\right)-b^2=-\frac{1}{2}\left(5+b^2\right)$.
We state the complete result:
\begin{equation}\label{HZM7}
H_{min}(b)=\left\{\begin{array}{r@{\quad:\quad}l}
-2-|b|& |b|\le 1\;,\\
 -\frac{1}{2}\left(5+b^2\right)&1\le |b|\le 3\;,\\
  2-3|b|&3\le |b|\;.
  \end{array} \right.
\end{equation}

\begin{figure}[ht]
  \centering
    \includegraphics[width=1.0\linewidth]{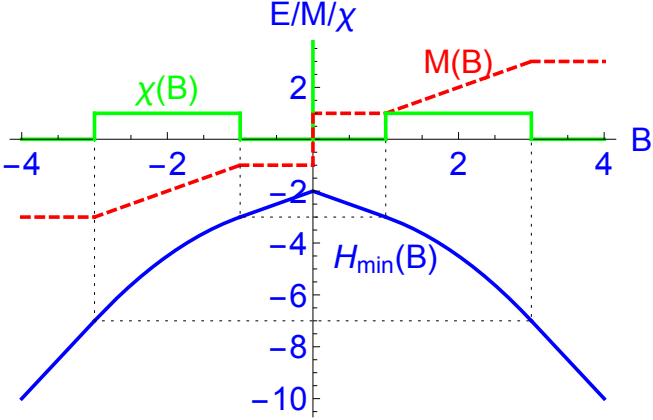}
  \caption[EMchi]
  {The three functions $H_{min}(B),{\sf M}(B)$ and $\chi(B)$ for the AF $3$-chain. Note the magnetization plateaus for $0<|B|<1$.
  }
  \label{FEMchi}
\end{figure}

This result implies the value of the saturation field being $B_{sat}=3$. It will be in order to check the definition (\ref{HZS24}).
To this end we consider the homogeneously gauged ${\mathbbm J}$-matrix, cp.~ (\ref{HZS4}),
\begin{equation}\label{HZM8}
{\mathbbm J}^{(h)}=\frac{1}{2} \left(\begin{array}{ccc}
 \frac{1}{3}  &0 & 1 \\
 0 & \frac{1}{3} &1\\
 1& 1 & - \frac{2}{3}  \\
\end{array}\right)\;,
\end{equation}
and calculate its eigenvalues $j_{min}^{(h)}=-\frac{5}{6},\;j_2=\frac{1}{6},\;j=\frac{2}{3}$.
Hence $B_{sat}=2\left(j- j_{min}^{(h)}\right)=3$ which confirms the above finding.
Note further that $H_{min}(B)$ assumes its lower bound (\ref{HZS6}) for $1\le |B|\le 3$.
The complete results for $H_{min}(B)$, magnetization ${\sf M}(B)$ and susceptibility $\chi(B)$ are represented in Figure \ref{FEMchi}.

We may check (\ref{INT1}) for the AF $3$-chain: Its r.~h.~s.~reads $\int_{0}^{B_{sat}}{\sf M}(b)\,db=\int_{0}^{1}1\,db + \int_{1}^{3}b\,db
=1+\frac{1}{2}\left( 3^2-1^2\right)=5$.  The l.~h.~s.~of (\ref{INT1}) is
$e_0-N\left( 2 j_{min}^{(h)}-j\right)= -2 -3\left( -2 \times\frac{5}{6}-\frac{2}{3}\right)=-2+7=5$ and hence (\ref{INT1}) is satisfied.

\hfill$\clubsuit$

\subsection{Parabolicity}\label{sec:HZP}

We will consider the case where the lower parabolic bound (\ref{HZS6}) is identically assumed for a maximal domain
and will call those systems ``parabolic". We already know that this cannot happen for
$|B|>B_{sat}$ where $H_{min}(B)$ is given by the ferromagnetic lines (\ref{propBsat2}).
An additional restriction has to be considered for small $B$ as we will show in the following.

In the above Example $3$ of the AF $3$-chain the absolute ground state $\check{\mathbf s}$ has a non-zero magnetization ${\sf M}(0+)=1$.
Even if a small magnetic field $B$ is applied this ground state and its magnetization remains unchanged. In this example the largest value of $B$
such that ${\sf M}(B)={\sf M}(0+)=1$ is $B=1$. This value will be called the ``threshold field" $B_{thr}$.

Generally, we will define $B_{thr}$ in the following way.
Let $\check{\mathbf s}$ be an absolute ground state of the Heisenberg-Zeeman system. If there are various absolute ground states
(except trivial degeneracy) we consider an $\check{\mathbf s}$  with maximal magnetization $\check{\mu}=M(\check{\mathbf s})$.
Then we define the line $L$ in the $(B,E)$-plane by the equation
\begin{equation}\label{HZP1}
E=H_0(\check{\mathbf s})-B\,M(\check{\mathbf s})
\;.
\end{equation}
The intersection $L_0$ of $L$ with the (sub)graph of $H_{min}$ is a closed convex subset of $L$, hence a closed interval or a single point.
We define
\begin{defi}\label{defthr}
\begin{eqnarray}\label{HZP2a}
 B_{thr}&\equiv& \mbox{ Max }\{ B\left| (B,E)\in L_0 \right. \}\\
 \nonumber
 &=& \mbox{ Max }\{ B\left| H_{min}(B)=H_0(\check{\mathbf s})-B\,M(\check{\mathbf s})\right.\}.\\
 \label{HZP2b}
\end{eqnarray}
\end{defi}

In the Example \ref{ex3} the interval $L_0$ will be the linear part of $B_{min}$ between $B=0$ and $B=1$ and the above
definition correctly yields $B_{thr}=1$. Even if there exists an absolute ground state with non-zero magnetization it
may happen that $B_{thr}=0$ if $E_{min}$ is smooth in the neighborhood of $\check{\mu}=M(\check{\mathbf s})$,
see the Examples \ref{ex3} and \ref{exPN+1} below.

In the case $B_{thr}>0$ there will be a magnetization plateau and
$H_{min}(B)=H_0(\check{\mathbf s})-|B| \,M(\check{\mathbf s})$ for $-B_{thr}\le B \le B_{thr}$. Hence the inequality
(\ref{HZS6}) cannot be replaced by an equality in the open interval $(-B_{thr},B_{thr})$, similarly as for $|B|>B_{sat}$.
The Example \ref{ex3} is typical in this respect. These considerations lead to the following
\begin{defi}\label{defP}
 Consider an anti-ferromagnetic Heisenberg-Zeeman system
 and an absolute ground state $\check{\mathbf s}$ with maximal magnetization $\check{\mu}=M(\check{\mathbf s})$ and
 corresponding threshold field $B_{thr}$.
 Then this system will be called ``parabolic" iff
 one of the following equivalent conditions is satisfied:\\
 (i) For all $B\in{\mathbbm R}$ such that
 $B_{thr}\le|B|\le B_{sat}$ there holds
 \begin{equation}\label{HZP3a}
 H_{min}(B)=H_{bound}(B)=j_{min}^{(h)}\,N-\frac{N\,B^2}{4\left(j-j_{min}^{(h)}\right)}\;.
 \end{equation}
 \\
 (ii) For all $\mu\in{\mathbbm R}$ such that
 $\check{\mu}\le |\mu| \le N$ there holds
 \begin{equation}\label{HZP3b}
 E_{min}(\mu)=E_{bound}(\mu)=j_{min}^{(h)}\,N+\frac{j-j_{min}^{(h)}}{N}\mu^2\;.
 \end{equation}
\end{defi}
The equivalence of (\ref{HZP3a}) and (\ref{HZP3b}) follows since $E_{bound}$ and $-H_{bound}$ are mutual
Legendre-Fenchel transforms in the corresponding domains $B_{thr}\le B \le B_{sat}$ and $\check{\mu}\le \mu\le N$.

According to this definition the above Examples \ref{ex1} and \ref{ex3} are parabolic.
Note that without the restriction  $B_{thr}\le|B|$ in (\ref{HZP3a}) the
Example \ref{ex3} would not be parabolic although the parabolic lower bound is assumed for  $B_{thr}\le|B|\le B_{sat}$.
The restriction to anti-ferromagnetic systems is sensible since, according to our Definition \ref{defferro},
ferromagnetic systems satisfy $B_{sat}=0$.\\

A global characterization of parabolic systems is not possible at the moment although we will present a couple
of general results. We remark that a constructive proof of parabolicity can be given in many cases as follows:
Whenever we have found a family of states ${\mathbf s}(B)$, where $B_{thr}\le B \le B_{sat}$, such that the
Heisenberg-Zeeman energy $H({\mathbf s}(B))$ is given by $H_{bound}$ it follows that
$H_{min}(B)=H({\mathbf s}(B))$ and hence the system is parabolic since $H_{bound}$ is a lower
bound of $H_{min}(B))$.

It directly follows from the definition (\ref{HZP3a}) that a parabolic system has a linear magnetization function
\begin{equation}\label{HZP4}
 {\sf M}(B)=-\frac{\partial H_{min}(B)}{\partial B}=
 \frac{N}{2\left(j-j_{min}^{(h)}\right)}\,B
\end{equation}
and hence a constant susceptibility for $B_{thr}\le |B| \le B_{sat}$. Thus for all parabolic systems the magnetization
functions look the same and the AF $3$-chain is typical, see Figure \ref{FEMchi}. Though it may happen that $B_{thr}=0$
and hence (\ref{HZP4}) holds for all $-B_{sat}\le B\le B_{sat}$, see Figure \ref{FHZ1a} for a typical example.

We will proceed with an example and two counter-examples.\\

\vspace{5mm}
\begin{ex}\label{exPN} The $N$-pantahedron\\
\begin{center}
\begin{figure}[ht]
  \centering
    \includegraphics[width=0.3\linewidth]{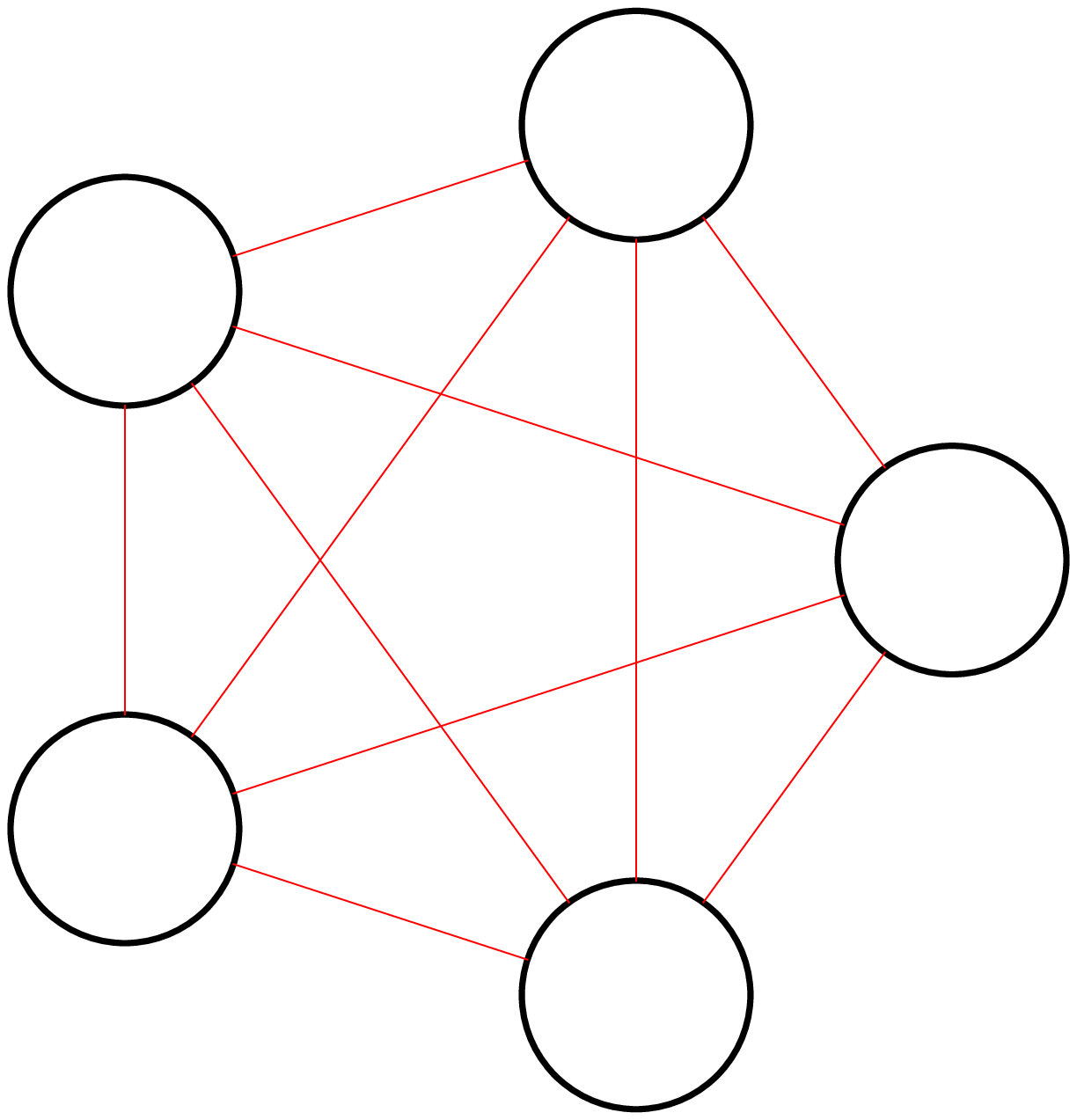}
  \label{GP1}
\end{figure}
\end{center}
\end{ex}
\vspace{5mm}

$\clubsuit$ The $N$-pantahedron or uniformly coupled AF system is defined by the Heisenberg Hamiltonian
\begin{equation}\label{PN1}
H_0({\mathbf s})=\sum_{1\le\mu<\nu\le N}{\mathbf s}_\mu \cdot {\mathbf s}_\nu = \frac{1}{2}\left({\mathbbm S}^2({\mathbf s})-N \right)
\;,
\end{equation}
where
\begin{equation}\label{PN2}
  {\mathbbm S}^2({\mathbf s})\equiv \sum_{1\le\mu,\nu\le N}{\mathbf s}_\mu \cdot {\mathbf s}_\nu =
  \left( \sum_{\mu=1}^{N}{\mathbf s}_\mu\right)^2 = {\mathbf S}^2
\end{equation}
is the total spin square function. Recall that in the presence of a magnetic field ${\mathbf B}=B\,{\mathbf e},\;B>0$ it is possible
to choose ${\mathbf S}$ parallel to ${\mathbf e}$ by means of a suitable rotation and hence ${\mathbf S}^2=M({\mathbf s})^2$.
Thus every  ${\mathbf s}\in{\mathcal P}_{\mathbf e}$ realizes a relative ground state of $H_0$ such that
\begin{equation}\label{PN3}
E_{min}(M({\mathbf s}))=H_0({\mathbf s})\;,
\end{equation}
and hence, using (\ref{PN1}),
\begin{equation}\label{PN4}
E_{min}(\mu)=-\frac{N}{2}+\frac{1}{2}\,\mu^2\;.
\end{equation}
The ${\mathbbm J}$-matrix corresponding to the Hamiltonian (\ref{PN1}) has the entries
\begin{equation}\label{PN5}
 {\mathbbm J}_{\lambda\nu}= \left\{\begin{array}{r@{\quad:\quad}l}
0& \lambda=\nu ,\\
 \frac{1}{2}&\lambda\neq \mu\;,
  \end{array} \right.
\end{equation}
where $1\le \lambda,\nu \le N$. It is already homogeneously gauged and has the eigenvalues with muliplicity
\begin{eqnarray}
\label{PN6a}
  j_{min}^{(h)} &=& -\frac{1}{2}\quad (\mbox{mult.}=N-1)\;,\\
 \label{PN6b}
  j &=&  \frac{N-1}{2}\;(\mbox{mult.}=1)\;,
\end{eqnarray}
in accordance with $\mbox{Tr }{\mathbbm J}=0$.
Thus $E_{min}(\mu)$ assumes the lower bound
\begin{equation}\label{PN7}
  E_{bound}=j_{min}^{(h)}\,N+\frac{j-j_{min}^{(h)}}{N}\,\mu^2= -\frac{N}{2}+\frac{1}{2}\mu^2
  \;,
\end{equation}
and hence the $N$-pantahedron is parabolic.\hfill$\clubsuit$\\

\vspace{5mm}
\begin{ex}\label{exPN+1} The $N$-pantahedron plus one extra spin\\
\begin{center}
\begin{figure}[ht]
  \centering
    \includegraphics[width=0.3\linewidth]{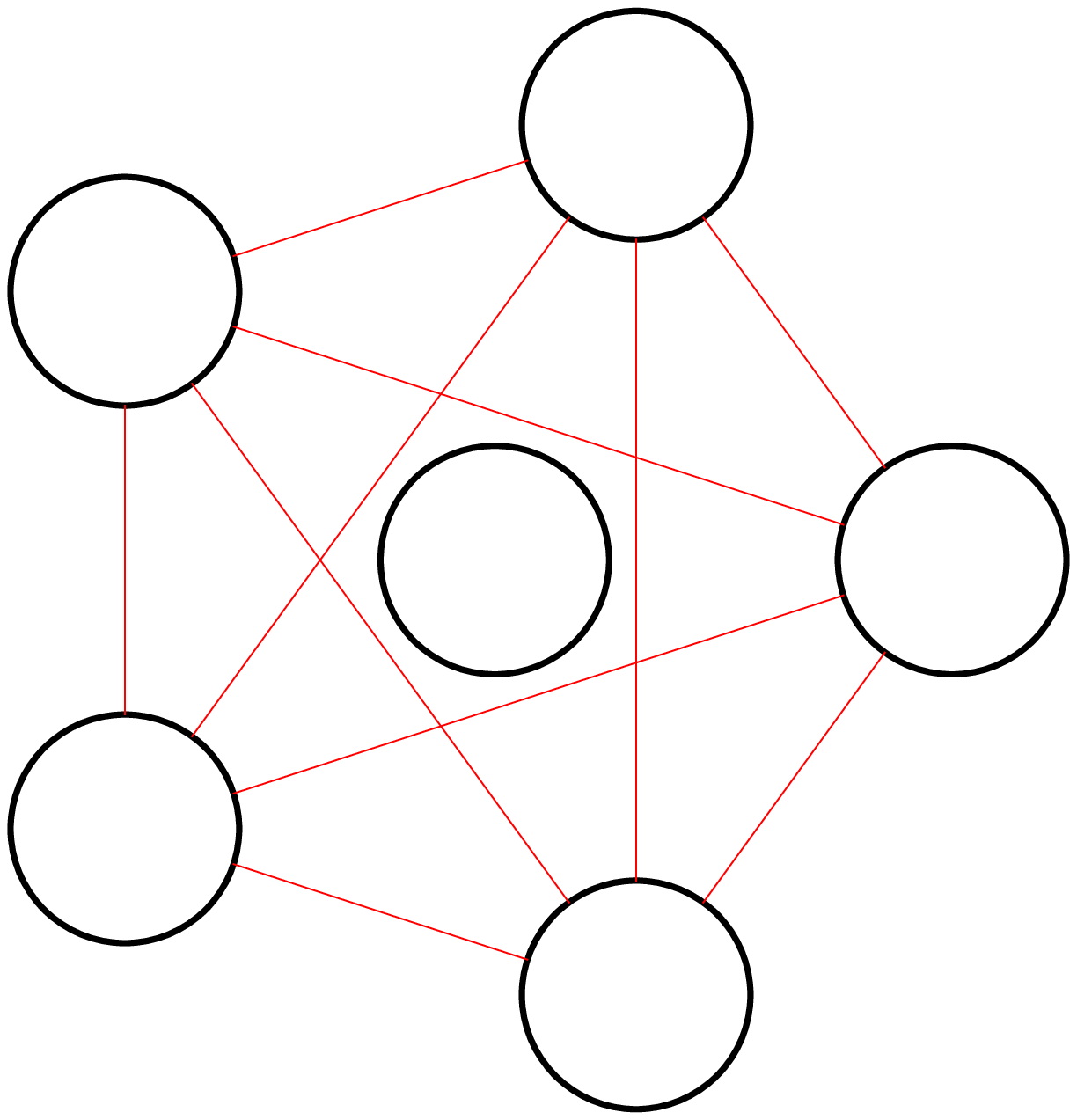}
  \label{GPP1}
\end{figure}
\end{center}
\end{ex}
\vspace{5mm}

$\clubsuit$ 
\begin{figure}[ht]
  \centering
    \includegraphics[width=1.0\linewidth]{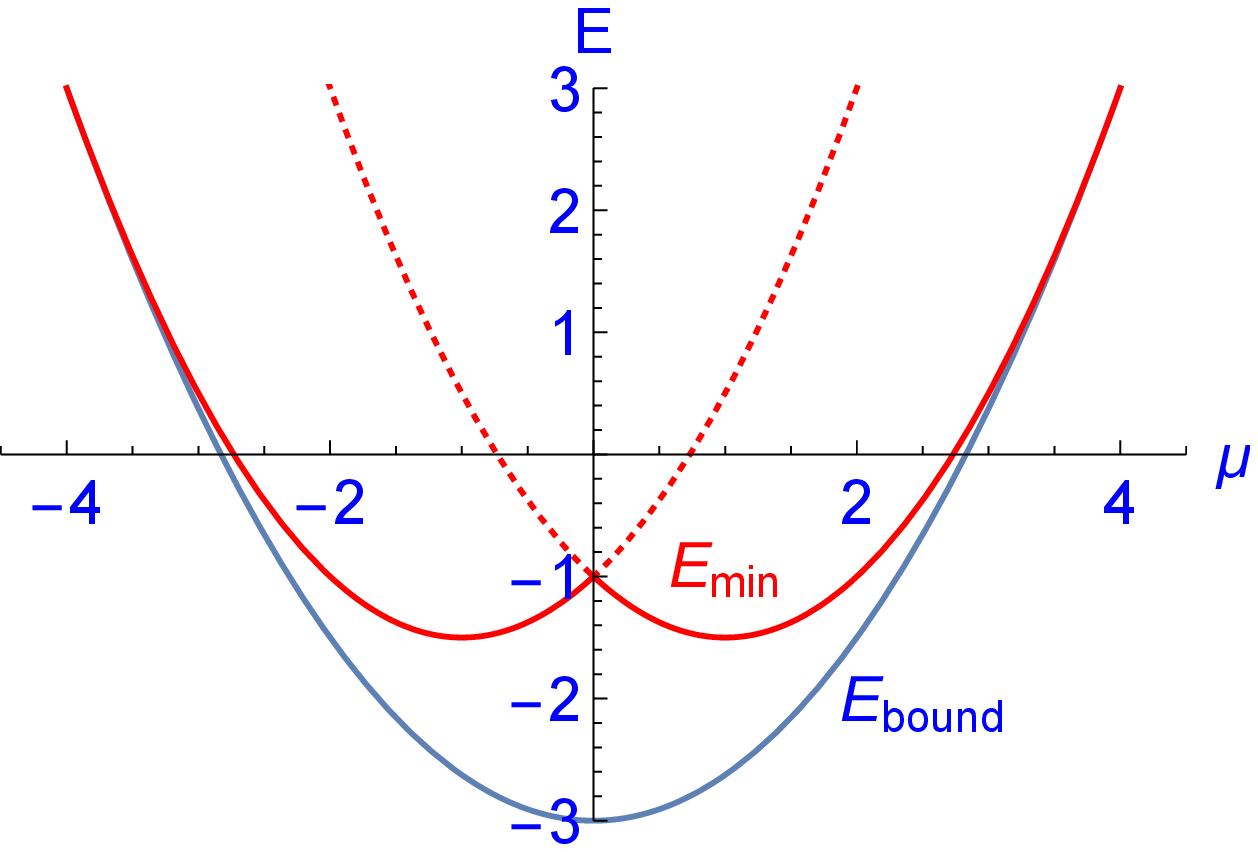}
  \caption[alpha]
  {The minimal energy $E_{min}(\mu)$ of the $3$-pantahedron with one extra spin and its lower parabola $E_{bound}(\mu)$. Since
  $E_{min}(\mu)>E_{bound}(\mu)$ for $|\mu|<N=4$ the system is not parabolic.
  }
  \label{FHZP2}
\end{figure}

This example would hardly appear in real applications but it nicely illustrates some aspects of the definition of parabolicity.
We consider a system with $N+1$ spins but the Hamiltonian (\ref{PN1}) of an $N$-pantahedron.
This means that the $(N+1)$th spin is not coupled to the other spins and does not contribute to the Heisenberg energy
of the system. But it contributes to its magnetization and hence destroys the parabolicity of the system
as we will see in a moment.

Let $\mu$ be the total magnetization of the system and ${\mu}_\ast$  the magnetization of the first $N$ spins,
such that the Heisenberg energy of the system will be $E= -\frac{N}{2}+\frac{1}{2}{\mu}_\ast^2$.
For given $\mu>0$ the minimal ${\mu}_\ast$ will be ${\mu}_\ast=\mu-1$, and hence, keeping
in mind that $E_{min}$ is an even function,
\begin{equation}\label{PN+1a}
 E_{min}(\mu)= \left\{\begin{array}{r@{\quad:\quad}l}
 -\frac{N}{2}+\frac{1}{2}(\mu-1)^2& \mu\ge 0\\
-\frac{N}{2}+\frac{1}{2}(\mu+1)^2& \mu\le 0\;,
  \end{array} \right.
\end{equation}
see Figure \ref{FHZP2}.

In order to calculate the bounding parabola of the system note that its ${\mathbbm J}$-matrix
is that of the $N$-pantahedron augmented by a zero row and a zero column.
Hence its row sum $j_\nu\equiv \sum_{\lambda=1}^{N+1}{\mathbbm J}_{\lambda\nu}$ amounts to
\begin{equation}\label{PN+1b}
 j_\nu= \left\{\begin{array}{r@{\quad:\quad}l}
 -\frac{N-1}{2}& \nu\le N\\
0&  \nu= N+1\;,
  \end{array} \right.
\end{equation}
and its mean row sum is
\begin{equation}\label{PN+1c}
  j=\frac{N(N-1)}{2(N+1)}\;.
\end{equation}
It follows that the diagonal elements of the homogeneously gauged ${\mathbbm J}$-matrix will be
\begin{equation}\label{PN+1d}
  {\mathbbm J}^{(h)}_{\nu\nu}=\left\{\begin{array}{r@{\quad:\quad}l}
-\frac{N-1}{2(N+1)}& \nu\le N\;,\\
\frac{N(N-1)}{2(N+1)}& \nu=N+1\;.
  \end{array} \right.
\end{equation}
Consequently the eigenvalues with multiplicity of ${\mathbbm J}^{(h)}$ are
\begin{eqnarray}
\label{PN+1e}
 j&=& \frac{N(N-1)}{2(N+1)},\quad (\mbox{mult.}= 2)\\
\label{PN+1f}
  j_{min}^{(h)} &=&- \frac{N}{N+1},\quad (\mbox{mult.}= N-1)\;,
\end{eqnarray}
in accordance with $\mbox{Tr }{\mathbbm J}^{(h)}=0$. The bounding parabola is hence given by
\begin{equation}\label{PN+1g}
E_{bound}(\mu)=-N +\frac{N}{2(N+1)}\,\mu^2\;,
\end{equation}
and intersects the graph of $E_{min}(\mu)$ only at $|\mu|=N+1$, see Figure \ref{FHZP2}.
Summarizing, the $N$-pantahedron plus one extra spin is not parabolic.\hfill$\clubsuit$\\

\vspace{5mm}
\begin{ex}\label{exCN} The AF $N$-chain\\
\begin{center}
\unitlength1cm
\begin{picture}(10,1)(0,0)
\thicklines
 \put(0.5,0.5){\circle{0.5}}
  \put(2.5,0.5){\circle{0.5}}
  \put(4.5,0.5){\circle{0.5}}
  \put(6.5,0.5){\circle{0.5}}
  \put(8.5,0.5){\circle{0.5}} 
   \textcolor{red}{\put(5.25,0.5){$\ldots$} \put(0.75,0.5){\line(1,0){1.5}}
   \put(2.75,0.5){\line(1,0){1.5}}\put(6.75,0.5){\line(1,0){1.5}}}
\end{picture}
\end{center}
\end{ex}
\vspace{5mm}
 $\clubsuit$ We have seen that the AF $3$-chain and the AF $2$-chain are parabolic, where the latter is identical with the AF dimer.
 Hence one could conjecture that all AF $N$-chains are parabolic too, but this is not the case as we will show.

 The homogeneously gauged ${\mathbbm J}$-matrix for the  AF $N$-chain has the form
 \begin{equation}\label{exCN1}
  {\mathbbm J}^{(h)}=\left(\begin{array}{ccccccc}
                 \frac{1}{2}-\frac{1}{N} & \frac{1}{2} & 0 & \ldots &  & \ldots & 0 \\
                 \frac{1}{2} &-\frac{1}{N} & \frac{1}{2} & \ddots &  & \ldots & 0 \\
                 0 &  \ddots &  \ddots & \ddots &  \ddots & & \vdots \\
                 \vdots &  \ddots &  \ddots &  \ddots &  \ddots &  \ddots & \vdots \\
                  &  &  \ddots &  \ddots &  \ddots &  \ddots & 0 \\
                 \vdots &  &  &  \ddots & \frac{1}{2} & -\frac{1}{N} & \frac{1}{2} \\
                 0 & \ldots &  & \ldots & 0 & \frac{1}{2} & \frac{1}{2}-\frac{1}{N}
               \end{array}\right).
 \end{equation}
By computer-algebraic means one can easily verify that its eigenvalues are
\begin{equation}\label{exCN2}
 j_\nu= \cos\left( \frac{\nu\pi}{N}\right)-\frac{1}{N},\quad \nu=0,\ldots,N-1\;,
\end{equation}
corresponding to eigenvectors $\varphi^{(\nu)}$ with components
\begin{equation}\label{exCN2}
 \varphi^{(\nu)}_\lambda = \cos\left(-\frac{\nu\pi}{2N} +\frac{\nu\lambda\pi}{N}\right),\quad \lambda=1,\ldots,N.
\end{equation}
Especially, $j=j_0=1-\frac{1}{N}$ and $j_{min}^{(h)}=\cos\frac{(N-1)\pi}{N}-\frac{1}{N}$ and hence the lower parabolic bound has the
form
\begin{equation}\label{exCN3}
 E_{bound}(\mu)=-1-N \cos \left(\frac{\pi }{N}\right)+\frac{\left(1+\cos \left(\frac{\pi
   }{N}\right)\right) }{N}\mu ^2.
\end{equation}
For even $N$ the Ising ground state $\uparrow\downarrow\ldots\downarrow$ has magnetization $\mu=0$ and the ground state
energy $e_0=-N+1$. The parabolic bound gives $E_{bound}(0)=-1-N \cos \left(\frac{\pi }{N}\right)<-1-N+\frac{\pi^2}{2 N}<-N+1=e_0$
for $N>\frac{\pi^2}{4}=2.4674\ldots$.

For odd $N$ the Ising ground state $\uparrow\downarrow\ldots\uparrow$ has magnetization $\mu=1$ and the ground state energy
energy $e_0=-N+1$. The parabolic bound gives $E_{bound}(1)=-\frac{N-1}{N} \left(1+(1+N) \cos \frac{\pi }{N}\right)
<-1-N+\frac{4+\pi ^2}{2 N}<-N+1=e_0$
for $N>1+\frac{\pi^2}{4}=3.4674\ldots$.

Hence the AF $N$-chain is not parabolic for $N>3$.
\hfill$\clubsuit$\\

Next we consider a large class of parabolic systems where the relative ground states can be constructed from the
absolute ground state.
\begin{theorem}\label{theoremP}
  We assume that an anti-ferromagnetic Heisenberg-Zeeman system  has an absolute co-planar or Ising ground state
  $\check{\mathbf s}$ with energy $e_0$, maximal magnetization $\check{\mu}=M(\check{\mathbf s})$ and threshold field
  $B_{thr}$  satisfying one of the following equivalent conditions:
  \begin{eqnarray}
 \label{HZPTa}
   e_0-\check{\mu}\,B_{thr} &=&j_{min}^{(h)} N-\frac{N}{4\left(j- j_{min}^{(h)}\right)}B_{thr}^2\;,
 \end{eqnarray}
 or
 \begin{eqnarray}
 \label{HZPTb}
   e_0 &=&j_{min}^{(h)}\, N+\frac{j- j_{min}^{(h)}}{N}\,\check{\mu}^2\;.
 \end{eqnarray}
   Then the system will be parabolic.
\end{theorem}
\noindent{\bf Proof:}\\
Let us first assume (\ref{HZPTa}).
This equation  says that the line $L$ defined in (\ref{HZP1}) intersects the graph of $H_{min}$ and
the bounding parabola $P$ given by (\ref{HZS6}) at the point $(B_{thr}, e_0-\check{\mu}\,B_{thr})$. It follows
that $L$ must be tangent to $P$ since otherwise $L$ would lie below $P$ somewhere. This implies the
identity of the (negative) slopes of $L$ and $P$ at  $B=B_{thr}$, namely
\begin{equation}\label{HZPslopes}
 \check{\mu}=\frac{N}{2\left(j-j_{min}^{(h)} \right)}\,B_{thr}
 \;.
 \end{equation}

Next we consider the one-parameter family of spin configurations
\begin{equation}\label{HZP4a}
 \tilde{\mathbf s}_\nu(\alpha)\equiv(\sqrt{1-\alpha^2}\,\check{\mathbf s}_\nu, \alpha),\;\nu=1,\ldots,N,\,
  \mbox{ and } 0\le \alpha\le 1
  \;.
\end{equation}
This generalizes the ``umbrella construction" in \cite{SL03} that was implicitly assuming $\check{\mu}=0$ and hence the  family (\ref{HZP4a}) will
also be referred to as the ``umbrella family".

\begin{figure}[ht]
  \centering
    \includegraphics[width=1.0\linewidth]{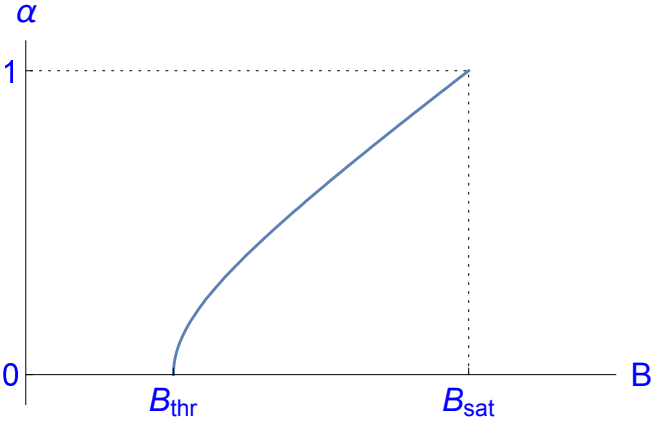}
  \caption[alpha]
  {The function $\alpha(B)$ according to (\ref{HZP8}) interpolating between $\alpha(B_{thr})=0$ and $\alpha(B_{sat})=1$.
  }
  \label{FHZP1}
\end{figure}

It satisfies $S^2\equiv||\sum_{\mu=1}^{N}\tilde{\mathbf s}_\mu(\alpha)||^2 = (1-\alpha^2)\check{\mu}^2+\alpha^2 N^2$.
Let ${\mathbf s}(\alpha)$ be a suitable rotated spin configuration such that $M({\mathbf s}(\alpha))=S$.
For the calculation of the energy we will utilize  $H_0({\mathbf s}(\alpha))=H_0(\tilde{\mathbf s}(\alpha))$:
\begin{eqnarray}
\label{HZP5a}
  {\sf H}(\alpha) &\equiv& H({\mathbf s}(\alpha))=H_0(\tilde{\mathbf s}(\alpha))-M({\mathbf s}) B \\
  \label{HZP5b}
     &=& \sum_{\mu,\nu=1}^{N} {J}_{\mu\nu}\left(\left(1-\alpha^2\right)\check{\mathbf s}_\mu\cdot\check{\mathbf s}_\nu+\alpha^2 \right)-S B\\
     \nonumber
   &=&\left(1-\alpha^2\right)e_0+N\,j\,\alpha^2-\sqrt{(1-\alpha^2)\check{\mu}^2+\alpha^2 N^2}B.\\
   \label{HZP5c}
   &&
\end{eqnarray}
We want to show that the energy of the one-parameter family ${\mathbf s}(\alpha)$ assumes the bounding parabola, i.~e.,
\begin{equation}\label{HZP6}
  {\sf H}(\alpha)=H_{bound}(B)=j_{min}^{(h)} N-\frac{N}{4\left(j- j_{min}^{(h)}\right)}B^2
  \;.
\end{equation}
We insert (\ref{HZP5c}) into (\ref{HZP6}) and consider the result as an equation of the form $f(\alpha,B)=0$.
This equation will be solved for the square root in (\ref{HZP5c}) and, by squaring both sides,
transformed into a bi-quadratic equation for $\alpha$. By squaring both sides of an equation  we have enlarged
its solution set and hence have to additionally check the final result. The solution(s) of the bi-quadratic
equation will be rewritten using the following equations that follow from (\ref{HZPTa}), (\ref{HZPslopes}), and (\ref{HZS24}):
\begin{eqnarray}
\label{HZP6a}
  \check{\mu} &=&\frac{B_{thr}}{B_{sat}}\, N \;,\\
  \label{HZP6b}
  j &=& \frac{1}{2} \left(-\frac{B_{thr}^2}{B_{sat}}+B_{sat}+\frac{2 e_0}{N}\right) \;,\\
  \label{HZP6c}
  j_{min}^{(h)} &=& \frac{e_0}{N}-\frac{B_{thr}^2}{2 B_{sat}}\;.
\end{eqnarray}

After some manipulations  we obtain
\begin{equation}\label{HZP7}
 B^2-\alpha ^2 B_{sat}^2+\left(\alpha ^2-1\right)  B_{thr}^2=0
 \;,
\end{equation}
which is the equation of a hyperbola. The relevant branch of it is given by the solution
\begin{equation}\label{HZP8}
 \alpha(B)=\sqrt{\frac{B^2-B_{thr}^2}{B_{sat}^2-B_{thr}^2}}
 \;,
\end{equation}
that interpolates between $\alpha(B_{thr})=0$ and $\alpha(B_{sat})=1$, see Figure \ref{FHZP1}.
It is straight forward to verify  that the one-parameter family ${\mathbf s}(\alpha(B))$ assumes the bounding parabola
(\ref{HZS6}) in the interval $B_{thr}\le B \le B_{sat}$ and hence the Heisenberg-Zeeman system is parabolic.
In particular, (\ref{HZPTb}) follows.

Alternatively, we may assume (\ref{HZPTb}) and  define
\begin{equation}\label{HZP9}
 \alpha(\mu)=\sqrt{\frac{\mu^2-\check{\mu}^2}{N^2-\check{\mu}^2}}\;,
\end{equation}
and, after some steps using (\ref{HZPTb}), prove
\begin{equation}\label{HZP10}
  H_0({\mathbf s}(\alpha(\mu)))=E_{bound}(\mu)=j_{min}^{(h)}N+\frac{j-j_{min}^{(h)}}{N}\mu^2
\end{equation}
for all $\check{\mu}\le \mu\le N$. This proves the parabolicity of the Heisenberg-Zeeman system in the second case.
In particular, (\ref{HZPTa}) follows.
\hfill$\Box$\\

In the special case of $B_{thr}=0$ the condition (\ref{HZPTa}) reduces to $e_0=j_{min}^{(h)}\, N$ which can be easily checked
in many cases where the system has a co-planar or Ising absolute ground state. We provide an example, see also \cite{SL03}.\\
\pagebreak

\vspace{5mm}
\begin{ex}\label{exR} The AF $N$-ring\\
\begin{center}
\begin{figure}[ht]
  \centering
    \includegraphics[width=0.3\linewidth]{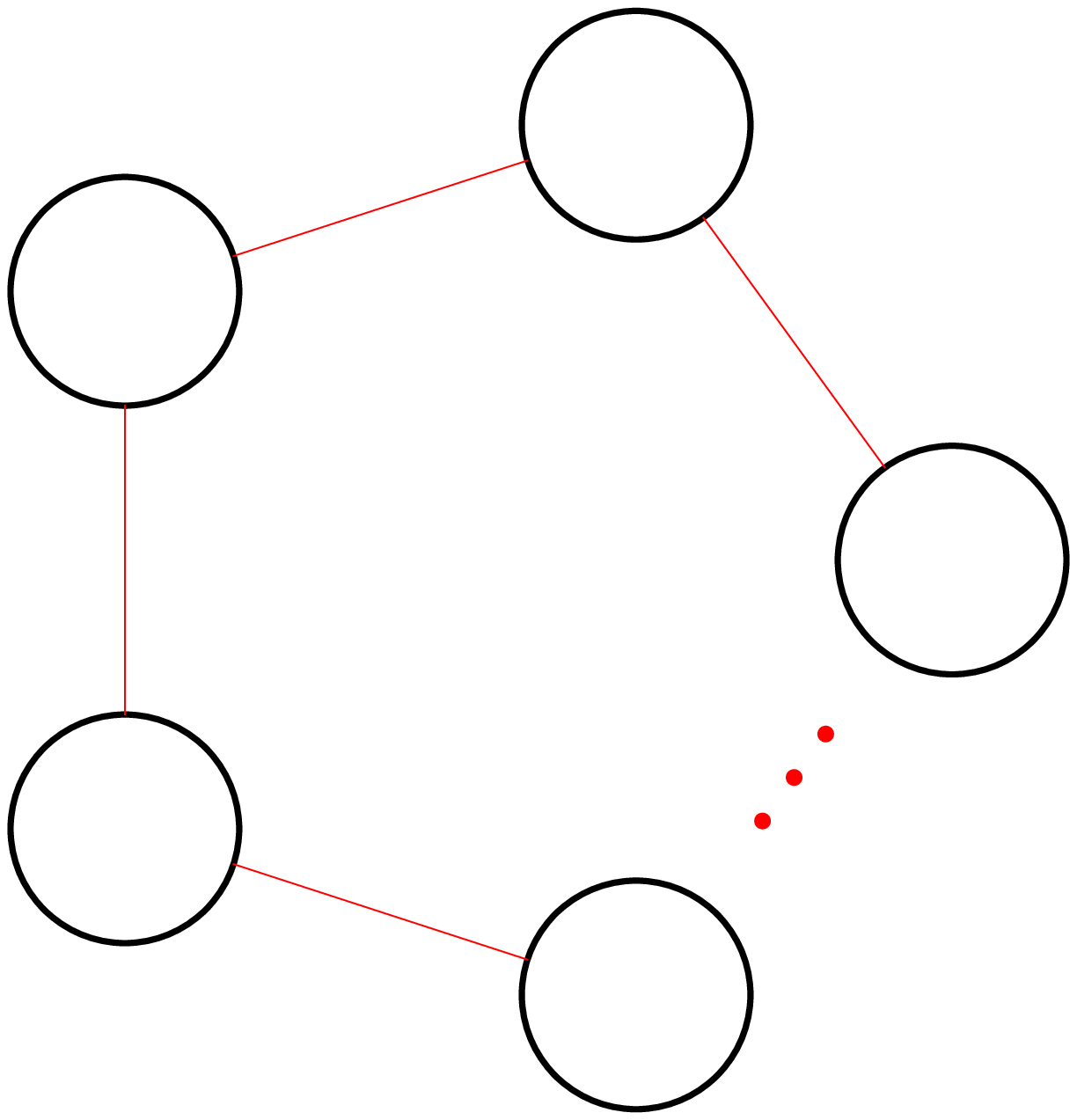}
  \label{GPR1}
\end{figure}
\end{center}
\end{ex}
\vspace{5mm}
 $\clubsuit$
 For the  AF $N$-ring the ${\mathbbm J}$-matrix is given by
 \begin{equation}\label{exR1}
  {\mathbbm J}=\left(\begin{array}{ccccccc}
                0 & \frac{1}{2} & 0 & \ldots &  & 0 & \frac{1}{2} \\
                 \frac{1}{2} &0 & \frac{1}{2} & \ddots &  & \ldots & 0 \\
                 0 &  \ddots &  \ddots & \ddots &  \ddots & & \vdots \\
                 \vdots &  \ddots &  \ddots &  \ddots &  \ddots &  \ddots & \vdots \\
                  &  &  \ddots &  \ddots &  \ddots &  \ddots & 0 \\
                 \vdots &  &  &  \ddots & \frac{1}{2} &0& \frac{1}{2} \\
                 \frac{1}{2} &0 &  & \ldots & 0 & \frac{1}{2} & 0
               \end{array}\right).
 \end{equation}
It is already homogeneously gauged and has the form of a ``circulant" \cite{A01}, i.~e., commutes with the Abelian group
of cyclic shift matrices. Consequently, its eigenvectors ${\mathbf b}^{(\lambda)}$ can be chosen as the Fourier basis
\begin{equation}\label{exR2}
{\mathbf b}^{(\lambda)}_\nu =\frac{1}{\sqrt{N}}\exp\left(
\frac{2\pi\,i\,\lambda\,\nu}{N}
\right),
\;\lambda,\nu=0,\ldots,N-1
\;,
\end{equation}
and the corresponding eigenvalues are
\begin{equation}\label{exR3}
 j_\lambda = \cos\frac{2\pi\lambda}{N}\;.
\end{equation}
It follows that
\begin{eqnarray}\label{exR4a}
 j&=&j_0=1\;,\\
 \label{exR4b}
j_{min}^{(h)}&=& j_{\lfloor \frac{N}{2}\rfloor}=\left\{\begin{array}{r@{\quad:\quad}l}
-1& N\mbox{ even}\\
-\cos\frac{\pi}{N}& N\mbox{ odd}
  \end{array} \right.\;,
\end{eqnarray}
and hence the bounding parabola assumes the form
\begin{equation}\label{exR5}
E_{bound}(\mu)=\left\{\begin{array}{r@{\quad:\quad}l}
-N+\frac{2}{N}\mu^2& N\mbox{ even}\\
-N\cos\frac{\pi}{N}+\frac{1}{N}\left(1+\cos\frac{\pi}{N}\right)\mu^2 & N\mbox{ odd}
  \end{array} \right.\;.
\end{equation}
The absolute ground states of the AF $N$-ring can be identified with $\sqrt{N}\,{\mathbf b}^{(\lambda)}$ for $\lambda=\lfloor \frac{N}{2}\rfloor$,
representing co-planar or Ising spin vectors by complex numbers of absolute value $1$. These ground states have zero magnetization and
a ground state energy  $e_0=E_{bound}(0)$. Hence the AF $N$-rings are parabolic by Theorem \ref{theoremP}.
\hfill$\clubsuit$\\

It follows that the following systems are parabolic and have an umbrella family in the sense of Theorem \ref{theoremP}: The $N$-pantahedron,
the AF spin ring with constant coupling, the uniform AF systems modelled on the
cube, the octahedron, the cuboctahedron and the icosidodecahedron, see \cite{SL03}, and the various Kagome models considered in \cite{S17b}.
Noticeably, these systems have a large symmetry group.
Recall, however, that uniform AF chains are not parabolic for $N>3$, see Example \ref{exCN}.
In view of these numerous examples one might conjecture that systems having only $3$-dimensional ground states cannot be
parabolic and thus Theorem \ref{theoremP} would cover the most general case of parabolic systems. But this is wrong, as the Example \ref{cex}
of the next section will show.

\section{Reduction to the pure Heisenberg ground state problem}\label{sec:RH}
We have reduced the ground state problem for Heisenberg-Zeeman systems to the problem of determination of relative ground states
of the pure Heisenberg Hamiltonian.
This is a minimization problem under additional constraints: We have not only to allow for ${\mathbf s}_\mu\cdot{\mathbf s}_\mu=1,\;\mu=1,\ldots,N$
but also for $M({\mathbf s})={\mathbf S}\cdot{\mathbf e}=\mu$. The latter constraint is equivalent to
\begin{equation}\label{RH1}
 {\mathbf S}\cdot{\mathbf S}=\mu^2
 \;,
\end{equation}
since every spin configuration satisfying (\ref{RH1}) can be suitable rotated in order to satisfy $M({\mathbf s})=\mu$ without changing
its energy $H_0({\mathbf s})$. This additional constraint leads to a modified stationary state equation (SSE) compared with (\ref{D7}).
We have to multiply the term (\ref{RH1}) with a further Lagrange parameter and to add it to the term $H_0({\mathbf s})$ that has
to be minimized. Because of $ {\mathbf S}\cdot{\mathbf S}=N+\sum_{\mu\neq\nu}{\mathbf s}_\mu\cdot{\mathbf s}_\nu$
the resulting SSE has the same form as the original one but with a modified Hamiltonian: $H_0$ is replaced by
\begin{equation}\label{RH2}
 H_\gamma({\mathbf s}) =\sum_{\mu,\nu=1}^N J^{(\gamma)}_{\mu \nu}\,\mathbf{s}_\mu\cdot \mathbf{s}_\nu
 \;,
\end{equation}
where
\begin{equation}\label{RH3}
  J^{(\gamma)}_{\mu \nu}\equiv \left\{\begin{array}{r@{\quad:\quad}l}
J_{\mu\nu}-\gamma& \mu\neq \nu\;,\\
0&\mu =\nu\;,
  \end{array} \right.
\end{equation}
and $\gamma\in{\mathbbm R}$ is the new Lagrange parameter. Note that by setting  $J^{(\gamma)}_{\mu \mu}=0$ we have
neglected the constant term $-\gamma N$ since this term would anyway vanish upon differentiation. In this way we have further
reduced the ground state problem of Heisenberg-Zeeman systems to the ordinary ground state problem for pure Heisenberg systems,
but with a Hamiltonian $H_\gamma({\mathbf s})$ depending linearly on a parameter $\gamma$. The methods developed in
\cite{S17a} -- \cite{S17c} can thus also be applied to the the ground state problem of Heisenberg-Zeeman systems.
We have only to allow for the possibility that the Lagrange parameters $\kappa_\nu$ of the ground states (\ref{D7}) depend on $\gamma$.
Moreover, we have to correct the energy of the Heisenberg system by subtracting the term
$-\gamma\,\sum_{\mu\neq\nu}{\mathbf s}_\mu\cdot{\mathbf s}_\mu$ since its energy is $H_0({\mathbf s})$ and not $H_\gamma({\mathbf s})$.

We will consider a couple of elementary examples.
\pagebreak

\vspace{5mm}
\begin{ex} \label{fex}
 A frustrated spin triangle ($N=3$)\\
\begin{center}
\begin{figure}[h]
  \centering
    \includegraphics[width=0.3\linewidth]{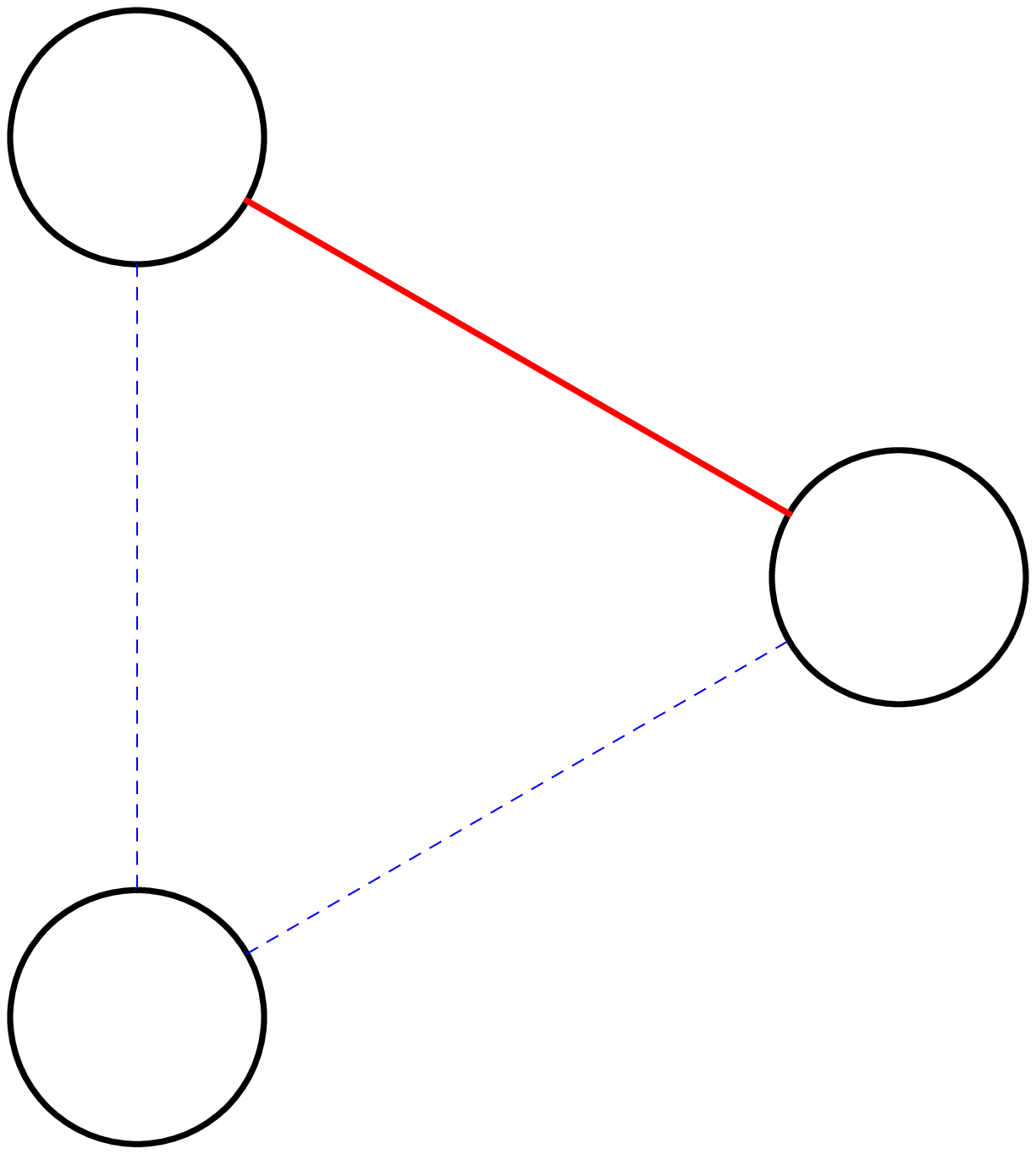}
  \label{GPR3}
\end{figure}
\end{center}
\end{ex}
\vspace{5mm}

\begin{figure}[h]
 \centering
    \includegraphics[width=1.0\linewidth]{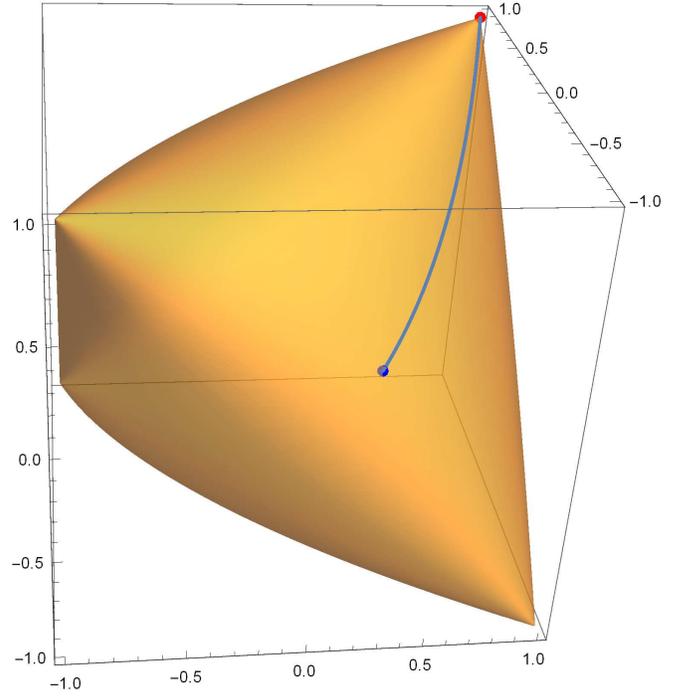}
  \caption[FRH2]
  {The one-parameter family of ground states of $H_\gamma$ according to (\ref{RH6a}) represented as a curve at the boundary of the Gram set
  ${\mathcal G}_3$.   It starts at the absolute ground state $G(\check{\mathbf s})$, given by (\ref{RH5}),
  corresponding to $\gamma=0$ (blue dot), and ends at the ferromagnetic ground state corresponding to $\gamma=1/2$ (red dot).
  }
  \label{FRH2}
\end{figure}

$\clubsuit$ The Heisenberg spin system defined by the Hamiltonian
\begin{equation}\label{RH4}
  H_0=2\,{\mathbf s}_1\cdot {\mathbf s}_2-{\mathbf s}_1\cdot {\mathbf s}_3-{\mathbf s}_2\cdot {\mathbf s}_3
  \;.
\end{equation}
is frustrated  and  has a co-planar absolute ground state $\check{\mathbf s}$ of the form
\begin{equation}\label{RH5}
\check{\mathbf s}_1=\frac{1}{4}  {1 \choose \sqrt{15}},\;
\check{\mathbf s}_2=\frac{1}{4}  {1 \choose -\sqrt{15}},\;
\check{\mathbf s}_3= {1 \choose 0},\;
\end{equation}
with energy $ H_0(\check{\mathbf s})=-\frac{9}{4}$ and magnetization $ M(\check{\mathbf s})=\frac{3}{2}$.
It would be possible to determine the relative ground states with the method used in the Example $3$
but we will rather apply the reduction method outlined above. The modified ${\mathbbm J}$-matrix assumes the form
\begin{eqnarray}\label{RH6a}
{\mathbbm J}^{(\gamma)}&=& \left(
\begin{array}{ccc}
 0 & 1-\gamma  & -\gamma -\frac{1}{2} \\
 1-\gamma  & 0 & -\gamma -\frac{1}{2} \\
 -\gamma -\frac{1}{2} & -\gamma -\frac{1}{2} & 0 \\
\end{array}
\right)\\
&\equiv&
\left(
\begin{array}{ccc}
 0 & J_3(\gamma)  & J_2(\gamma) \\
 J_3(\gamma)  & 0 &J_1(\gamma) \\
 J_2(\gamma) & J_1(\gamma) & 0 \\
\end{array}
\right).
\end{eqnarray}
It turns out that the ground states of ${\mathbbm J}^{(\gamma)}$ are co-planar and hence their Gram matrices can be obtained
explicitly by (\ref{ST7a}) -- (\ref{ST7c}):
\begin{eqnarray}
\label{RH7a}
  u &=& -\frac{4 (\gamma -5) \gamma +7}{8 (\gamma -1)^2}\;, \\
  v &=& w= \frac{2 \gamma +1}{4-4 \gamma }
  \;.
\end{eqnarray}
This defines a one-parameter family of Gram matrices that connects $G(\check{\mathbf s})$ with the Gram matrix of the ferromagnetic ground state,
see Figure \ref{FRH2}. The domain of the Lagrange parameter $\gamma$ is $[0,\frac{1}{2}]$. This follows by calculating the
eigenvalues of the homogeneously gauged matrix ${\mathbbm J}^{(\gamma)(h)}$:
\begin{equation}\label{RH8}
{\mathbbm J}^{(\gamma)(h)}=\left(
\begin{array}{ccc}
 -\frac{1}{2} & 1-\gamma  & -\gamma -\frac{1}{2} \\
 1-\gamma  & -\frac{1}{2} & -\gamma -\frac{1}{2} \\
 -\gamma -\frac{1}{2} & -\gamma -\frac{1}{2} & 1 \\
\end{array}
\right)\;,
\end{equation}
that are of the form $j=-2\gamma,\; j_2=\frac{1}{2} (-3 + 2 \gamma),\;   j_{max}=\frac{1}{2} (3 + 2 \gamma)$.
For $0\le \gamma<\frac{1}{2}$ the lowest eigenvalue is $j_2$, whereas for $\gamma=\frac{1}{2}$  we have
$j_{min}^{(h)}=j=-1$ and the system becomes ferromagnetic, see Section \ref{sec:HZS}.

\begin{figure}[ht]
  \centering
    \includegraphics[width=1.0\linewidth]{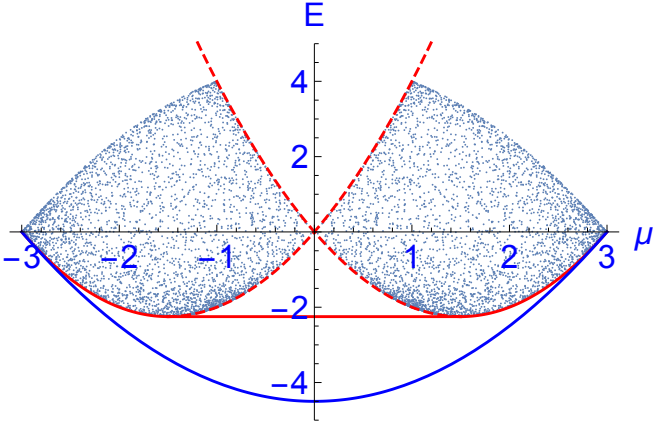}
  \caption[FRH1]
  {The set $\widetilde{E}$ of possible values of $(\mu,E)=\left( M({\mathbf s}),H_0({\mathbf s})\right)$ for the frustrated spin triangle (\ref{RH4}).
  The blue points are determined by randomly chosen spin vectors.
  The red parabolas $E_\pm=\mu(\mu\pm 3)$ locally represent the function $E_{min}$ for $-3\le \mu\le 3$, but only
  the part with $\frac{3}{2}\le |\mu|\le 3$ will correspond to ground states of (\ref{RH4}). The red line
  between the points $(-\frac{3}{2},-\frac{9}{4})$ and $(\frac{3}{2},-\frac{9}{4})$ indicates the convex envelope  $E_{min}^{(co)}$ of $E_{min}$.
  The blue parabola represents the lower bound $E_{bound}$ according to (\ref{ProofB1}). Since $E_{bound}(\mu)<E_{min}(\mu)$ for
  $-3<\mu<3$ the system is not parabolic.
  }
  \label{FRH1}
\end{figure}

\begin{figure}[ht]
  \centering
    \includegraphics[width=1.0\linewidth]{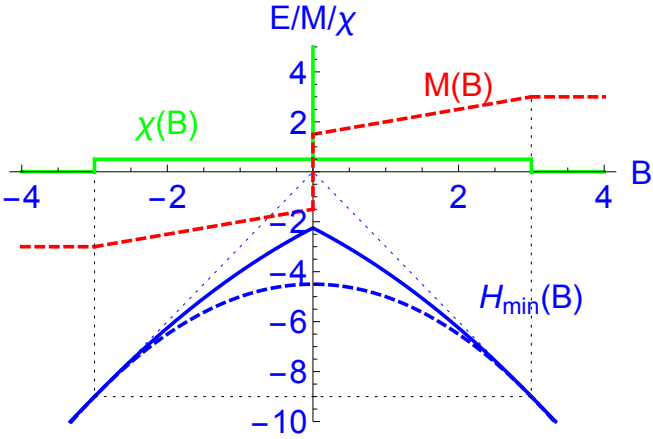}
  \caption[frh3]
  {The three functions $H_{min}(B),{\sf M}(B)$ and $\chi(B)$ for the frustrated spin triangle (\ref{RH4}).
  The lower parabolic bound $H_{bound}(B)$ is indicated by the dashed blue curve.
  }
  \label{FRH3}
\end{figure}

The magnetization of the one-parameter family of ground states is obtained as
\begin{equation}\label{RH8}
 \mu=M(\gamma)=\sqrt{3+2 (u+v+w)}=\frac{3}{2(1- \gamma)}
 \;,
\end{equation}
and the (corrected) ground state energy is given by
\begin{equation}\label{RH9}
 E(\gamma)=2 (J_3 u+J_2 v+J_1 w)+2 \gamma  (u+v+w)=\frac{9 (2 \gamma -1)}{4 (\gamma -1)^2}
 \;.
\end{equation}
(\ref{RH8}) and (\ref{RH9}) define a parametric representation of the parabola
\begin{equation}\label{RH10}
 E=\mu(\mu-3)\;,
\end{equation}
that represents the function $E_{min}$ for $0\le \mu\le 3$.
Since $E_{min}$ is an even function its extension  to all arguments $\mu\in[-N,N]$ is given by
\begin{equation}\label{RH10a}
 E_{min}(\mu)=\left\{\begin{array}{r@{\quad:\quad}l}
\mu(\mu-3)& 0\le\mu\le N;,\\
\mu(\mu+3)& -M\le\mu\le 0
\end{array} \right.\;,
\end{equation}
see Figure \ref{FRH1}.
Similarly as in the Example $3$, $E_{min}$ is not convex and the relative ground states with magnetization $-\frac{3}{2}<\mu<\frac{3}{2}$
are never ground states of the Heisenberg-Zeeman system. The negative Legendre-Fenchel transform of $E_{min}(\mu)$ yields
\begin{equation}\label{RH11}
H_{min}(b)= \left\{\begin{array}{r@{\quad:\quad}l}
-\frac{1}{4}\left(3+|b|\right)^2& |b|\le 3\;,\\
-3|b|& |b|\ge 3
\end{array} \right.\;.
\end{equation}
The complete results for $H_{min}(B),\;{\sf M}(B)$ and $\chi(B)$ are shown in Figure \ref{FRH3}.
The system is not parabolic since the absolute ground state $\check{\mathbf s}$ satisfies $\check{\mu}=\frac{3}{2}$ and $e_0=-\frac{9}{4}$ whereas $j_{min}^{(h)}\,N+\frac{j-j_{min}}{N} \check{\mu}^2=-\frac{27}{8}<-\frac{9}{4}$. Thus (\ref{HZP3b}) is not satisfied, see also
Figure \ref{FRH1}. We point out that $B_{thr}=0$ although the absolute ground state has a non-zero magnetization  $\check{\mu}=3/2$. This is due to
the vanishing slope of the supporting line of $E_{min}$ at $\check{\mu}=3/2$.
\hfill$\clubsuit$\\

$\clubsuit$ It will be instructive to re-consider the above Example \ref{ex3} of the AF $3$-chain.
Although we have already determined its relative ground states we will try to recover them as ground states of the modified Heisenberg
Hamiltonian $H_\gamma$ according to (\ref{RH2}). The modified ${\mathbbm J}$-matrix in the ground state gauge reads
\begin{equation}\label{RH12}
 {\mathbbm J}^{(\gamma)}=\left(
\begin{array}{ccc}
 \frac{1}{6} (4 \gamma -1) & -\gamma  & \frac{1}{2}-\gamma  \\
 -\gamma  & \frac{1}{6} (4 \gamma -1) & \frac{1}{2}-\gamma  \\
 \frac{1}{2}-\gamma  & \frac{1}{2}-\gamma  & \frac{1}{3} (1-4 \gamma ) \\
\end{array}
\right)\;,
\end{equation}
and has the eigenvalues
\begin{eqnarray}
\label{RH13a}
 j_1 &=& \frac{5}{6}-\frac{7 \gamma }{3}\;, \\
 \label{RH13b}
  j_2 &=&\frac{2 (\gamma -1)}{3}\;, \\
  \label{RH13c}
  j_3 &=&\frac{5 \gamma }{3}-\frac{1}{6}\;.
\end{eqnarray}
The eigenvector corresponding to $j_2$ is the Ising state $(-1,-1,1)=\downarrow\downarrow\uparrow$, but this is
the unique absolute ground state of $H_\gamma$ only for $0\le \gamma<\frac{1}{2}$. At $\gamma=\frac{1}{2}$
we have $j_1=j_2=-\frac{1}{3}$ and the matrix ${\mathbbm J}^{(\gamma)}$ assumes the form
\begin{equation}\label{HR14}
 {\mathbbm J}^{(\frac{1}{2})}=\left(
\begin{array}{ccc}
 \frac{1}{6} & -\frac{1}{2} & 0 \\
 -\frac{1}{2} & \frac{1}{6} & 0 \\
 0 & 0 & -\frac{1}{3} \\
\end{array}
\right)\;.
\end{equation}
The eigenspace of ${\mathbbm J}^{(\frac{1}{2})}$ corresponding to the eigenvalue $-\frac{1}{3}$ is spanned by the columns
of the matrix
\begin{equation}\label{HR15}
{\mathbf s}(\alpha)=\left(
\begin{array}{cc}
 \cos \alpha  & \sin \alpha \\
 \cos \alpha  & \sin \alpha  \\
 1 & 0 \\
\end{array}
\right)\;.
\end{equation}
The three rows of ${\mathbf s}(\alpha)$ are identical with the one-parameter family of relative co-planar ground states, see (\ref{HZM2}) with $\alpha=\beta$, and hence the latter
can be recovered as the absolute ground states of $H_{\gamma=\frac{1}{2}}$.
Note that due to the form of the matrix (\ref{HR14}) the Hamiltonian $H_{\gamma=\frac{1}{2}}$
is that of a ferromagnetic dimer such that the third spin is not coupled to the dimer. Hence its ground state can be directly determined
as, say, $\downarrow\downarrow$ and the third spin being arbitrary. This is, up to a rotation, exactly the one-parameter family (\ref{HR15})
with the two limit cases ${\mathbf s}_3=\uparrow$ and ${\mathbf s}_3=\downarrow$ leading to Ising ground states.
In \cite{S17c} we have enumerated the $6$ one-dimensional faces of the Gram set ${\mathcal G}_3$ consisting of line segments
joining two of the $4$ Ising states. The Gram matrices of the ground states of $H_{\gamma=\frac{1}{2}}$ form such a face generated
by $G(\downarrow\downarrow\uparrow)$ and $G(\downarrow\downarrow\downarrow)$.
\hfill$\clubsuit$\\

We have thus encountered two different scenarios in the context of the reduction to the pure Heisenberg ground state problem:
In the Example \ref{fex} the family $H_\gamma$ of modified Hamiltonians has a corresponding continuous family of ground states
${\mathbf s}_\gamma$. This case will be called ``continuous reduction". In contrast to this the Example \ref{ex3} is a case
of ``discontinuous reduction" where the family $H_\gamma$ has the the same ground state ${\mathbf s}_0$ until $\gamma$
reaches a critical value $\gamma_c$ such that all relative ground states of $H_0$, including the ferromagnetic one, are absolute ground states of
$H_{\gamma_c}$. We do not know whether these two cases are the only ones that can occur.

Next we consider the example of a $3$-chain with alternating signs of the coupling constants, that already appeared in \cite{SL03}. Here continuous reduction takes place and $H_{min}$ can  be analytically determined without being a quadratic function in the domain $B_{thr}\le B \le B_{sat}$ as in all other examples of this paper.

\vspace{5mm}
\begin{ex} \label{exAC}
 Alternating $3$-chain $(N=3)$\\
\begin{center}
\unitlength1cm
\begin{picture}(8,1)(0,0)
\thicklines
 \put(0.5,0.5){\circle{0.5}}
  \put(2.5,0.5){\circle{0.5}}
  \put(4.5,0.5){\circle{0.5}}
  \textcolor{red}{ \put(0.75,0.5){\line(1,0){1.5}}}
   \textcolor{blue}{ \put(2.65,0.5){\line(1,0){1.5}}}
\end{picture}
\end{center}
\end{ex}
\vspace{5mm}

$\clubsuit$
\begin{figure}[ht]
  \centering
    \includegraphics[width=1.0\linewidth]{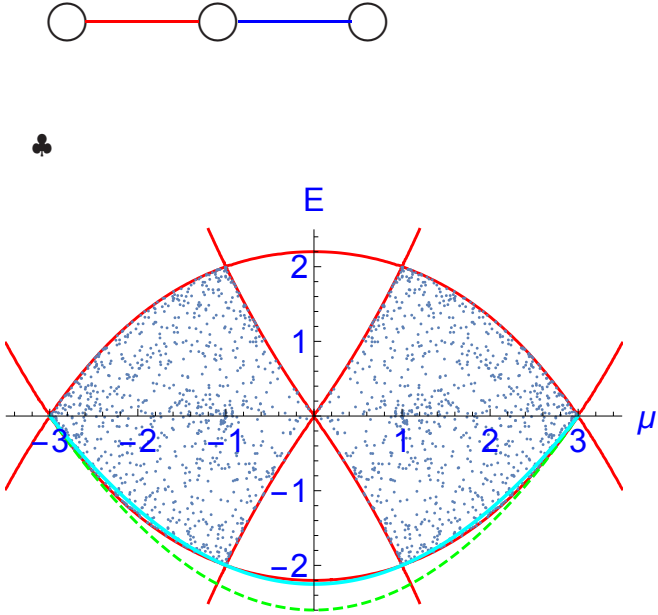}
  \caption[exAC]
  {The Heisenberg energy $E$ of the alternating $3$-chain vs.~the magnetization $\mu$. The blue points correspond to randomly chosen
  co-planar spin configurations, the red curves are solutions of (\ref{AC5}) and locally define the function $E_{min}(\mu)$. The cyan curve $E_{umb}(\mu)$
  represents the energy of an umbrella family joining  the absolute Ising ground state  $\uparrow\downarrow\downarrow$
  with the ferromagnetic ground state $\downarrow\downarrow\downarrow$. The dashed green curve is the lower parabolic bound $E_{bound}(\mu)$. Note that
  for $1<|\mu|<3$ we have $E_{bound}(\mu)<E_{min}(\mu)<E_{umb}(\mu)$.
  }
  \label{FHZP5}
\end{figure}

The alternating $3$-chain is defined by the Hamiltonian
\begin{equation}\label{AC1}
 H_0({\mathbf s})={\mathbf s}_1\cdot{\mathbf s}_2 - {\mathbf s}_2\cdot{\mathbf s}_3\;,
 \end{equation}
and has obviously the absolute Ising ground state $\check{\mathbf s}=\uparrow\downarrow\downarrow$ with magnetization
$M(\check{\mathbf s})=\pm 1$ and ground state energy $e_0=-2$. We will again illustrate the reduction of the Heisenberg-Zeeman
ground state problem to the pure Heisenberg ground state problem for this example and consider the modified ${\mathbbm J}$-matrix
\begin{equation}\label{AC2}
  {\mathbbm J}^{(\gamma)}=\left(\begin{array}{ccc}
 0 & \frac{1}{2}-\gamma  & -\gamma  \\
 \frac{1}{2}-\gamma  & 0 & -\gamma -\frac{1}{2} \\
 -\gamma  & -\gamma -\frac{1}{2} & 0 \\
\end{array}\right)\;.
\end{equation}
Similarly as in the Example \ref{fex} there exists a one-parameter family of co-planar ground states ${\mathbf s}(\gamma)$ of
$H^{(\gamma)}$ given by the Gram matrix elements, see (\ref{ST2}):
\begin{eqnarray}
\label{AC3a}
  u &=& \frac{8 \gamma ^2 (1-2 (\gamma -2) \gamma )-1}{4 (1-2 \gamma )^2 \gamma  (2 \gamma +1)}\;, \\
  \label{AC3b}
  v &=& \frac{1-16 \left(\gamma ^4+\gamma ^2\right)}{8 \gamma ^2 \left(4 \gamma ^2-1\right)}\;, \\
  \label{AC3c}
  w &=& -\frac{8 (2 \gamma  (\gamma +2)-1) \gamma ^2+1}{4 \gamma  (2 \gamma -1) (2 \gamma   +1)^2}\;.
\end{eqnarray}
In this case the parameter domain $(\gamma_1,\gamma_2)$, where $\gamma_1=\frac{1}{2} \left(\sqrt{2}-1\right)=0.2071\ldots$
and $\gamma_2=\frac{1}{2 \sqrt{3}}=0.288675\ldots$, covers the magnetization interval $\mu\in(1,3)$, whereas for smaller values of
$\gamma$, say $0\le \gamma < \gamma_1$ the Hamiltonian $H^{(\gamma)}$ has the unique Ising ground state $\check{\mathbf s}$.
For the domain $(\gamma_1,\gamma_2)$ we obtain
magnetization and Heisenberg energy in the parameter representation
\begin{eqnarray}
\nonumber
  M(\gamma) &=&\sqrt{3+2(u+v+w)}=\frac{\sqrt{48 \gamma ^4+24 \gamma ^2-1}}{2 \gamma  \left(1-4 \gamma ^2\right)}\;, \\
   \label{AC4a}
   &&\\
  \nonumber
  E(\gamma)&=& 2 ({J_3} u+{J_2} v+{J_1} w)+2 \gamma  (u+v+w)\\
   \label{AC4b}
  &=&\frac{48 \gamma ^4+8 \gamma ^2-1}{2 \gamma  \left(1-4 \gamma ^2\right)^2}
  \;.
\end{eqnarray}
It is possible to eliminate $\gamma$ from (\ref{AC4a}) and (\ref{AC4b}) which yields the polynomial equation
\begin{eqnarray}
\nonumber
  0 &=&16 E^6-8 E^4 \left(\mu ^2+3\right)^2 \\
  \nonumber
   &+&E^2 \left(\mu ^8-24 \mu ^6+18 \mu ^4+288 \mu^2-27\right)\\
   \label{AC5}
   &+&\mu ^2 \left(\mu ^4-10 \mu ^2+9\right)^2.
\end{eqnarray}
Its solution defines a one-dimensional algebraic variety in the $(\mu,E)$-plane that includes the boundary of $\widetilde{E}$, the set of
physical $(\mu,E)$-values, and hence locally defines $E_{min}$, see Figure \ref{FHZP5}. Without giving the details
of the straight forward calculation we note that the lower parabolic bound is given by
\begin{equation}\label{AC6}
  E_{bound}(\mu)=-\frac{3 \sqrt{3}}{2}+\frac{\mu ^2}{2 \sqrt{3}}\;,
\end{equation}
and that the system is not parabolic since $E_{bound}(\mu)<E_{min}(\mu)$ for $|\mu|<3$. Further, it is interesting that
the umbrella family joining the two Ising states $\uparrow\downarrow\downarrow$  and $\downarrow\downarrow\downarrow$
yields a curve $E_{umb}(\mu)=\frac{1}{4} \left(\mu ^2-9\right)$ that is slightly above $E_{min}(\mu)$ for $1<|\mu|<3$, see Figure  \ref{FHZP5}.
Although the absolute ground state of the alternating $3$-chain is an Ising state, the reduction of this system is continuous.

\begin{figure}[ht]
  \centering
    \includegraphics[width=1.0\linewidth]{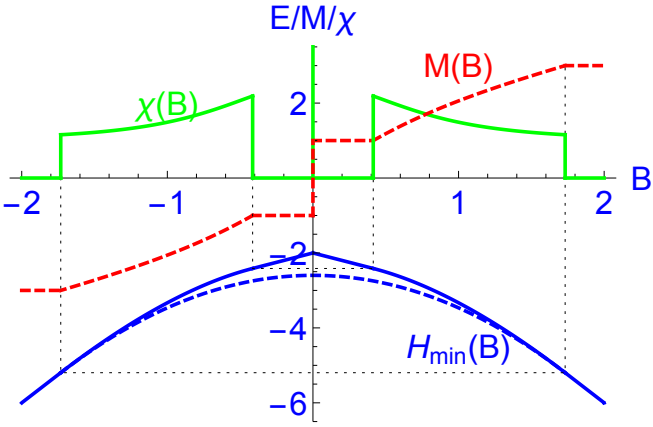}
  \caption[P6]
  {The three functions $H_{min}(B), {\sf M}(B)$ and $\chi(B)$ for the alternating $3$-chain. The dashed blue curve is the lower
  parabolic bound $H_{bound}(B)$.
  }
  \label{FHZP6}
\end{figure}

Finally we give the parametric representations of the magnetic field $B$, $H_{min}$ and susceptibility $\chi$ that can be calculated in a straight forward
manner from (\ref{AC4a}) and (\ref{AC4b}) and hold for $\gamma\in[\gamma_1,\gamma_2]$:
\begin{eqnarray}
\label{AC7a}
  B(\gamma) &=& \frac{\sqrt{48 \gamma ^4+24 \gamma ^2-1}}{1-4 \gamma ^2}\;, \\
  \label{AC7b}
  H_{min}(\gamma) &=& -\frac{8 \gamma }{\left(1-4 \gamma ^2\right)^2}\;, \\
  \label{AC7c}
  \chi(\gamma)&=& \frac{1}{32 \gamma ^3}+\gamma  \left(\frac{16}{12 \gamma
   ^2+1}+\frac{1}{2}\right)-\frac{3}{4 \gamma }\;.
\end{eqnarray}
From this one concludes $B_{thr}=B(\gamma_1)=\sqrt{2}-1$ and $B_{sat}=B(\gamma_2)=\sqrt{3}$.
The complete results for minimal energy $H_{min}$, magnetization ${\sf M}$  and susceptibility $\chi$ are represented in Figure \ref{FHZP6}.
They qualitatively look similar as in the Example \ref{ex3} of the AF $3$-chain, see Figure \ref{FEMchi}, but $\chi(B)$ is not constant
in the interval $B_{thr}<B<B_{sat}$ due to the more complicated form of $H_{min}$.

\hfill$\clubsuit$\\

The last Example \ref{cex} shows discontinuous reduction and, moreover, serves as the counter-example showing that Theorem
\ref{theoremP} does not cover all cases of parabolic systems.
\pagebreak

\vspace{5mm}
\begin{ex} \label{cex}
 Parabolic system without umbrella family $(N=6)$\\
\begin{center}
\begin{figure}[h]
  \centering
    \includegraphics[width=0.3\linewidth]{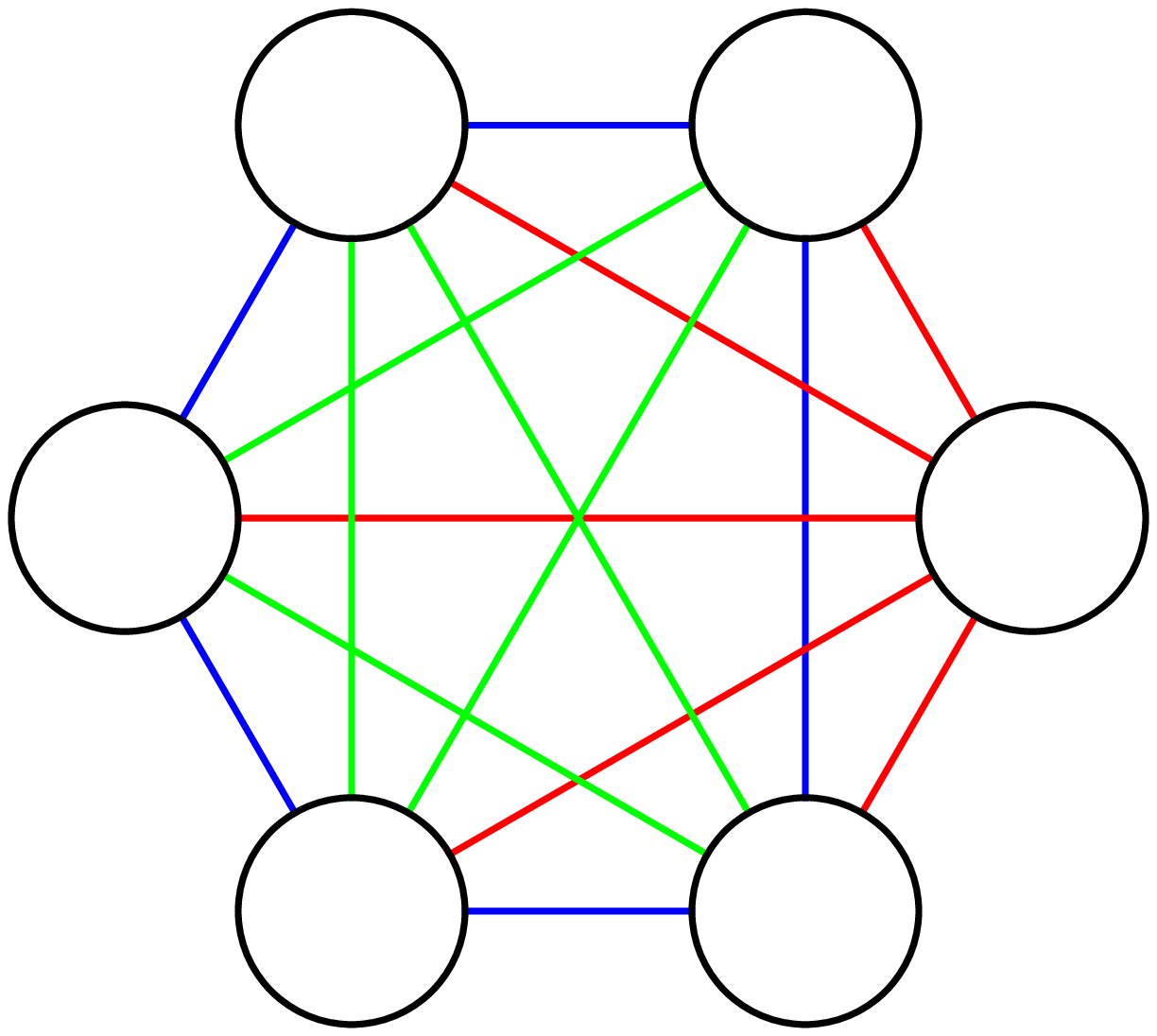}
  \label{GPR6}
\end{figure}
\end{center}
\end{ex}
\vspace{5mm}
$\clubsuit$ 
\begin{figure}[ht]
  \centering
    \includegraphics[width=1.0\linewidth]{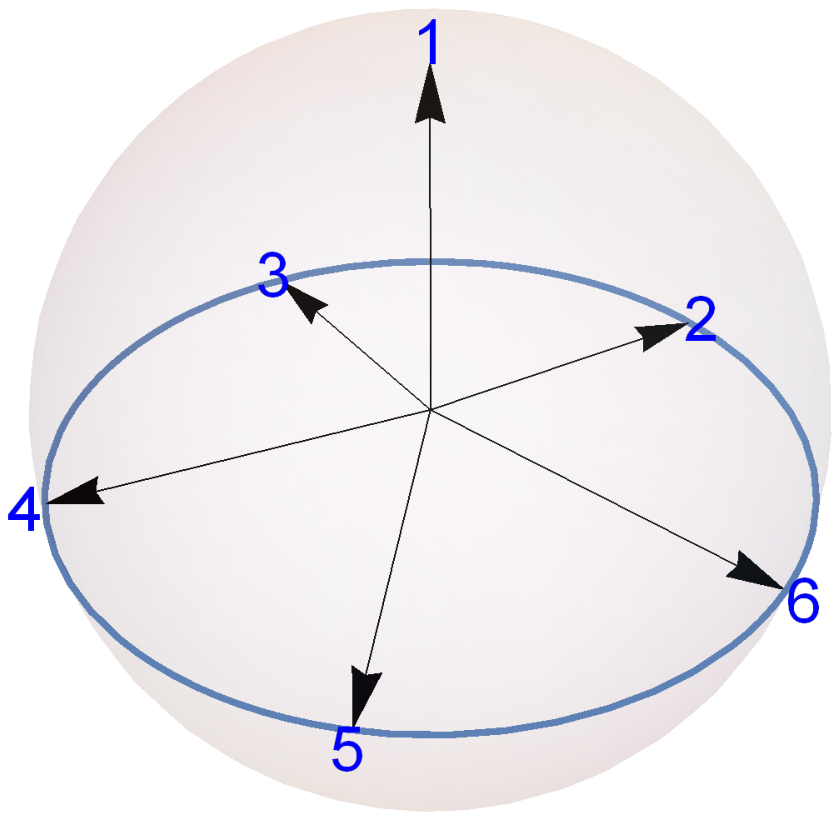}
  \caption[cex]
  {The absolute ground state (\ref{cex3a}), (\ref{cex3b}) of the system given by (\ref{cex1}).
  }
  \label{FHZP4}
\end{figure}

In order to show that not every parabolic Heisenberg-Zeeman system possesses an absolute co-planar or Ising ground state
and hence, according to Theorem \ref{theoremP}, an umbrella family, we consider a system of $N=6$ spins with a Heisenberg Hamiltonian
given by the following ${\mathbbm J}$-matrix:
\begin{equation}\label{cex1}
{\mathbbm J}=\left(
\begin{array}{cccccc}
 -1 & 1 & 1 & 1 & 1 & 1 \\
 1 & \frac{1}{5} & n & p & p & n \\
 1 & n &  \frac{1}{5} & n & p & p \\
 1 & p & n &  \frac{1}{5} & n & p \\
 1 & p & p & n &  \frac{1}{5} & n \\
 1 & n & p & p & n &  \frac{1}{5} \\
\end{array}
\right)
  \;,
\end{equation}
where $p=\frac{7 + 3 \sqrt{5}}{10}=1.37082\ldots$ and $n=\frac{7 - 3 \sqrt{5}}{10}=0.0291796\ldots$. It follows that ${\mathbbm J}$ is already homogeneously gauged and has the eigenvalues
with multiplicities
\begin{eqnarray}\label{cex2a}
j&=&\;\;\;4\quad (\mbox {mult.}=1),\\
\label{cex2b}
j_{min}^{(h)}&=&-2\quad (\mbox{mult.}=3),\\
\label{cex2c}
j_2&=&\;\;\;1\quad  (\mbox{mult.}=2)\;.
\end{eqnarray}
The absolute ground state of the system is $3$-dimensional, essentially unique (i.~e.~unique up to rotational/reflectional degeneracy) and lives on the $3$-dimensional eigenspace of ${\mathbbm J}$ corresponding to $j_{min}^{(h)}=-2$. It has the explicit form
\begin{eqnarray}
\label{cex3a}
 \check{\mathbf s}_1&=& \left( \begin{array}{c}
                           0 \\
                           0 \\
                          1 \end{array}\right)\;, \\
                          \label{cex3b}
  \check{\mathbf s}_\mu &=& \left( \begin{array}{c}
                           \frac{2\sqrt{6}}{5} \cos \frac{2\pi(\mu-1)}{5}\\
                           \frac{2\sqrt{6}}{5}\sin\frac{2\pi(\mu-1)}{5}  \\
                          -\frac{1}{5} \end{array}\right)\;, \mu=2,\ldots,6,
\end{eqnarray}
see Figure \ref{FHZP4}, with
the total magnetization $M(\check{\mathbf s})=0$ and a ground state energy $e_0=j_{min}^{(h)}\,N=-12$.  Since the essential uniqueness
of the absolute ground state is crucial for what follows we will sketch the proof of it using methods from \cite{S17a}.
The eigenspace of  ${\mathbbm J}$ corresponding to the eigenvalue $j_{min}^{(h)}=-2$ is spanned by the three columns of the matrix
\begin{equation}\label{cex4}
W=\left(
\begin{array}{ccc}
 -5-\sqrt{5} & 2 \sqrt{5} & -5-\sqrt{5} \\
 1+\sqrt{5} & -1-\sqrt{5} & 2 \\
 2 & -1-\sqrt{5} & 1+\sqrt{5} \\
 0 & 0 & 2 \\
 0 & 2 & 0 \\
 2 & 0 & 0 \\
\end{array}
\right).
\end{equation}
The corresponding ADE (\ref{D4}) has the unique solution
\begin{equation}\label{cex5}
 \Delta= \frac{1}{100}\left(
\begin{array}{ccc}
 25 & \left(-5+6 \sqrt{5}\right) & \left(-5-6 \sqrt{5}\right) \\
  \left(-5+6 \sqrt{5}\right) & 25 &  \left(-5+6 \sqrt{5}\right) \\
 \left(-5-6 \sqrt{5}\right) &  \left(-5+6 \sqrt{5}\right)   &25 \\
\end{array}
\right)\;,
\end{equation}
and leads to the Gram matrix $G=W\,\Delta\,W^\top$ that is identical with the Gram matrix $G(\check{\mathbf s})$
of the ground state (\ref{cex3a}), (\ref{cex3b}).

The bounding parabola  $E_{bound}$  of the system has the form
\begin{equation}\label{cex3}
E_{bound}(\mu)=j_{min}^{(h)}\,N +\frac{j-j_{min}^{(h)}}{N}\,\mu^2= -12 +\mu^2
 \;.
\end{equation}

We consider a ``partial umbrella family" ${\mathbf s}(\theta)$ of the form
\begin{eqnarray}
\label{cex4a}
 {\mathbf s}_1(\theta)&=& \left( \begin{array}{c}
                           0 \\
                           0 \\
                          1 \end{array}\right)\;, \\
\label{cex4b}
  {\mathbf s}_\mu(\theta) &=& \left( \begin{array}{c}
                           \sin\theta\,\cos \frac{2\pi(\mu-1)}{5}\\
                           \sin\theta\,\sin\frac{2\pi(\mu-1)}{5}  \\
                          \cos\theta \end{array}\right)\;, \mu=2,\ldots,6,
\end{eqnarray}
that interpolates between the Ising states ${\mathbf f}=\uparrow\uparrow\uparrow\uparrow\uparrow\uparrow$
for $\theta=0$ and  ${\mathbf g}=\uparrow\downarrow\downarrow\downarrow\downarrow\downarrow$ for $\theta=\pi$
and contains the absolute ground state (\ref{cex3a}), (\ref{cex3b}) for $\cos\theta =-\frac{1}{5}$.
This family has the magnetization $\mu=M({\mathbf s}(\theta))=5 \cos (\theta )+1$ and the
Heisenberg energy $E=H_0({\mathbf s}(\theta))=\frac{1}{2} (20 \cos (\theta )+25 \cos (2 \theta )+3)$.
Eliminating $\theta$ from the last two equations yields $E=-12+\mu^2$, which coincides with
(\ref{cex3}). The family (\ref{cex4a}), (\ref{cex4b}) covers the part of the bounding parabola (\ref{cex3})
between $\mu=-4$ and $\mu=6$.
Hence this system is parabolic but does not possess an umbrella family in the sense of Theorem \ref{theoremP} since it has
no absolute co-planar or Ising ground states. Interestingly, there exists an umbrella family joining ${\mathbf f}$ and ${\mathbf g}$,
but this family only covers the part of the bounding parabola (\ref{cex3}) between $\mu=4$ and $\mu=6$.

\begin{figure}[ht]
  \centering
    \includegraphics[width=1.0\linewidth]{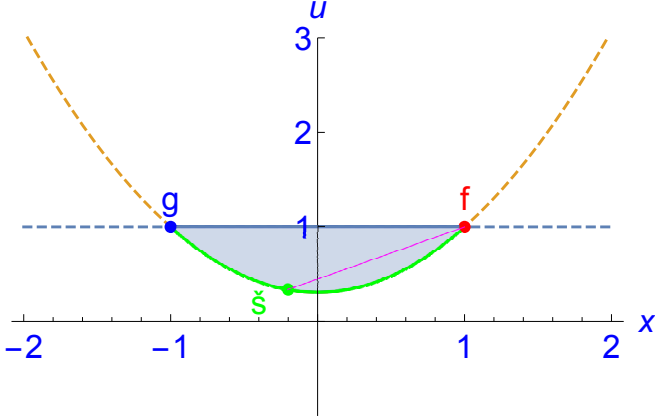}
  \caption[cex]
  {The convex set ${\mathcal S}_{=}$ in the $x,u$-plane the points of which correspond to certain ground states of $H_1$.
  We have displayed three special cases: ${\mathbf g}=\uparrow\downarrow\downarrow\downarrow\downarrow\downarrow$ (blue point),
  ${\mathbf f}=\uparrow\uparrow\uparrow\uparrow\uparrow\uparrow$ (red point ) and $\check{\mathbf s}$ (green point) according to
  (\ref{cex3a}) and (\ref{cex3b}). The blue line segment connecting  ${\mathbf g}$ and  ${\mathbf f}$ corresponds to a $2$-dimensional
  umbrella family, the green parabolic segment connecting  ${\mathbf g}$ and  ${\mathbf f}$ corresponds to the $3$-dimensional
  partial umbrella family (\ref{cex4a}) and (\ref{cex4b}), and the magenta line segment
  connecting  $\check{\mathbf s}$ and  ${\mathbf f}$ corresponds to a $4$-dimensional and hence unphysical umbrella family.
  }
  \label{FHZP3}
\end{figure}

We will further analyze this example in the context of the heading of this section, the reduction to the pure Heisenberg ground state problem.
To this end we consider the modified Hamiltonian $H_\gamma$ according to (\ref{RH2}) and (\ref{RH3}). It turns out that the case
of discontinuous reduction applies, namely that for $0\le \gamma<1$ the Hamiltonian $H_\gamma$ has the essentially unique ground state
$\check{\mathbf s}$. For
$\gamma=1$ the Hamiltonian $H_\gamma$ becomes the first time ferromagnetic
and has both states, $\check{\mathbf s}$ and ${\mathbf f}$ as ground states. We will determine more ground states of $H_1=H_{\gamma=1}$ (including unphysical ones) and recall the fact that the corresponding Gram matrices form a face of the Gram set ${\mathcal G}_6$.
To this end we consider the homogeneously gauged matrix ${\mathbbm J}_1$ corresponding to $H_1$ and its eigenspace corresponding
to the $4$-fold degenerate lowest eigenvalue $-1$. It is spanned by the four columns of the matrix
\begin{equation}\label{cex5}
W=\left(
\begin{array}{cccc}
 0 & 0 & 0 & 1 \\
 \frac{1}{2} \left(1+\sqrt{5}\right) & \frac{1}{2} \left(-1-\sqrt{5}\right) & 1 & 0
   \\
 1 & \frac{1}{2} \left(-1-\sqrt{5}\right) & \frac{1}{2} \left(1+\sqrt{5}\right) & 0
   \\
 0 & 0 & 1 & 0 \\
 0 & 1 & 0 & 0 \\
 1 & 0 & 0 & 0 \\
\end{array}
\right).
\end{equation}
The corresponding ADE has solutions $\Delta\ge0$ depending on $4$ real parameters $u,x,y,z$:
\begin{equation}\nonumber
\Delta(u,x,y,z)=
\end{equation}
\begin{equation}\label{cex6}
\left(
\begin{array}{cccc}
 1 & u & -\frac{-2 \left(2+\sqrt{5}\right) u+\sqrt{5}+3}{1+\sqrt{5}} & x \\
 u & 1 & u & y \\
 -\frac{-2 \left(2+\sqrt{5}\right) u+\sqrt{5}+3}{1+\sqrt{5}} & u & 1 & z \\
 x & y & z & 1 \\
\end{array}
\right).
\end{equation}
We need not investigate the $4$-dimensional convex set ${\mathcal S}_{ADE}$ of solutions $\Delta(u,x,y,z)\ge0$
in  detail and will only consider the intersection ${\mathcal S}_{=}$ of ${\mathcal S}_{ADE}$ with the two-dimensional subspace $x=y=z$.
It is bounded by the line $u=1$ and the parabola $u=\frac{\left(5+\sqrt{5}\right) x^2+\sqrt{5}+1}{2 \left(3+\sqrt{5}\right)}$,
see Figure \ref{FHZP3}. Recall that the points of ${\mathcal S}_{=}$ correspond to certain Gram matrices $G(x,u)$ of
ground states of $H_1$ via $G(x,u)=W\,\Delta(u,x,x,x)\,W^\top$. In Figure \ref{FHZP3} we have displayed three points
corresponding to the Gram matrices of the ground states ${\mathbf g}$, ${\mathbf f}$ and $\check{\mathbf s}$,
namely $G({\mathbf g})=G(-1,1)$, $G({\mathbf f})=G(1,1)$ and $G(\check{\mathbf s})=G(-\frac{1}{5},\frac{1}{25} \left(6 \sqrt{5}-5\right))$.
It turns out that there are two prominent curves connecting ${\mathbf g}$ and ${\mathbf f}$: the straight line segment and the parabolic curve.
The first one corresponds to a $2$-dimensional $1$-parameter umbrella family; the second one corresponds to the family
${\mathbf s}(\theta)$, see (\ref{cex4a}), (\ref{cex4b}), and is given by the correspondence $x(\theta)=\cos\theta$
and $u(\theta)=\frac{1}{4} \left(\sqrt{5}-1\right) \sin ^2(\theta )+\cos ^2(\theta )$. W.~r.~t.~magnetization and energy,
both curves cover certain parts of the bounding parabola (\ref{cex3}). Another line segment connects $\check{\mathbf s}$ with ${\mathbf f}$.

We will dwell upon some details concerning these three curves. The $2$-dimensional umbrella family connecting ${\mathbf g}$ and ${\mathbf f}$
is explicitly given by
\begin{equation}\label{cex7}
  {\mathbf s}_\nu(\alpha)={f_\nu\,\cos\alpha \choose g_\nu\,\sin\alpha}\;\nu=1,\ldots,6
  \;,
\end{equation}
where the $g_\nu$ and $f_\nu$ are the components of the Ising states  ${\mathbf g}$ and ${\mathbf f}$.
Instead of rotating these spin vectors such that the total spin ${\mathbf S}$ points into the direction of the magnetic field
it is easier to directly replace $M({\mathbf s}(\alpha))$ by $\mu(\alpha)=\sqrt{{\mathbf S}\cdot{\mathbf S}}=\sqrt{10 \cos (2 \alpha )+26}$.
Using the result $E(\alpha)=H_0({\mathbf s}(\alpha))=10 \cos (2 \alpha )+14$ it is straight forward to verify $E(\alpha)=-12+\mu^2(\alpha)$, which means
that the $2$-dimensional umbrella family realizes the bounding parabola (\ref{cex3}). However, since $4\le \mu(\alpha)\le 6$ for all $0\le\alpha\le \pi$
only a part of the bounding parabola is covered and this $2$-dimensional umbrella family cannot be obtained by means of Theorem \ref{theoremP}.

This is different for the second partial umbrella family ${\mathbf s}(\theta)$ that covers the part of the bounding parabola given by $-4\le \mu\le 6$
and hence proves the parabolicity of the present system, although ${\mathbf s}(\theta)$ is not an umbrella family in the strict sense
since ${\mathbf s}_1(\theta)$ is constant, see (\ref{cex4a}).

The line segment connecting $\check{\mathbf s}$ with ${\mathbf f}$ can be represented by the $4$-dimensional umbrella family
\begin{equation}\label{cex8}
   {\mathbf s}_\nu(\beta)={\check{\mathbf s}_\nu\,\cos\beta \choose f_\nu\,\sin\beta}\;\nu=1,\ldots,6,
   \; 0\le \beta\le \pi/2.
\end{equation}
It covers the bounding parabola for $0\le \mu \le 6$ but its spin vectors are $4$-dimensional and hence unphysical.
Its Gram matrices $G( {\mathbf s}(\beta))$ satisfy
\begin{equation}\label{cex9}
 G( {\mathbf s}(\beta)) = \cos^2\beta\,G(\check{\mathbf s})+(1-\cos^2\beta)\,G({\mathbf f})
 \;,
\end{equation}
which confirms the statement that the $G( {\mathbf s}(\beta))$ run through the line segment between $G(\check{\mathbf s})$ and
$G({\mathbf f})={\mathbf 1}$. The former results represented in Figure \ref{FHZP3} show that this line segment will \underline{not} be a face
of the Gram set ${\mathcal G}$ and hence the face generated by the two extremal points $G(\check{\mathbf s})$ and $G({\mathbf f})$ must be
larger. However, this face will be contained in the $4$-dimensional face of Gram matrices of ground states of $H_1$.
\hfill$\clubsuit$\\

\section*{Acknowledgment}
I have greatly profited from the long lasting cooperation with Marshall Luban and Christian Schr\"oder including work
on classical ground states that has left its mark on the theory presented here. Especially,  for this article I have used unpublished material 
of a joint project. Moreover, I thank Hugo Touchette for hints concerning the literature on the Legendre-Fenchel transform
and its use in physics.



\begin{thebibliography}{99}
\bibitem{foot2}
There exists a connection of the homogeneous gauge to quantum theory:
The eigenvalues of the Hamiltonian of a system of $N$ spins with quantum number $s$ restricted to the $1$-magnon sector are
$E_\alpha=s^2 N j +2 s(j_\alpha^{(h)} -j)$ where the $j_\alpha^{(h)}$ are the eigenvalues of ${\mathbbm J}^{(h)}$, see \cite{S02}, (24).
\bibitem{foot1}
This is a limit case of the general statement that $\chi(B,T)\ge 0$ for Heisenberg-Zeeman systems (classical or quantum)
and $T>0$ that follows from $\chi(B,T)=\frac{1}{T}\left( \langle {\mathbf S}\cdot{\mathbf e}^2\rangle -\langle {\mathbf S}\cdot{\mathbf e}\rangle^2\right)\ge 0$, where $\langle \ldots \rangle$ denotes the thermal expectation value w.~r.~t.~the canonical ensemble.
\bibitem{foot3}
In quantum theory the saturation field $B_{sat}$ would be defined as the slope $\Delta E/\Delta M$ of the line connecting the points
$(E_{max},M_{max})=(s^2 N j, N s)$ and $(E_1,Ns-1)$ where $E_1$ denotes the minimal energy in the sector $M=N s-1$. According to
\cite{foot2} this amounts to $B_{sat}= 2 s (j-j_{min}^{(h)})$. This is compatible with (\ref{HZS24}) if one takes into account that in the classical
limit $s\longrightarrow \infty$ the energies scale with $s(s+1)$ whereas the magnetization scales with $s$.
\end{thebibliography}
\end{document}